\documentclass[9pt,twoside]{extarticle}
\usepackage[a4paper]{geometry}
\usepackage{lastpage}
\usepackage{fancyhdr}
\fancypagestyle{plain}{%
  \fancyhf{}%
  \fancyfoot[C]{}%
  \fancyhead[HR]{\color{fresnelmiddleblue} \textbf{\thepage}}
  \fancyhead[HL]{\color{fresneldarkblue} \textbf{Electrodynamics of an oscillating particle without cheating} }
}

\usepackage[english]{babel}
\usepackage[scaled]{helvet}
\usepackage{authblk}
\usepackage{graphicx,xcolor}
\definecolor{color0}{RGB}{0,0,0}
\definecolor{red_orange}{RGB}{240,100,100}
\definecolor{guillgreen}{RGB}{50,240,50}
\definecolor{fredred}{RGB}{240,50,50}
\definecolor{fresneldarkblue}{RGB}{44,46,131} % banner ppt charte
\definecolor{fresnelmiddleblue}{RGB}{34,124,192} % banner ppt charte
\definecolor{fresnellightblue}{RGB}{0,165,226} % banner ppt charte
\newcommand{\titlefont}{\color{fresnellightblue} \normalfont\sffamily\bfseries\fontsize{21}{23}\selectfont}

\usepackage{amsmath,amsfonts,amssymb}
\usepackage{booktabs,multirow,tabularx}
\usepackage[colorlinks=true, allcolors=blue]{hyperref}

\usepackage{physics}
\usepackage{lineno}
\usepackage{caption}
\usepackage{subcaption}
\usepackage{amsmath}
\usepackage{mathtools}
\usepackage{amsfonts}
\usepackage{accents}
\usepackage{soul,verbatim} 
\newcommand*{\dt}[1]{%
  \accentset{\mbox{\large\bfseries .}}{#1}}

\newcommand{\w}{\omega}
\newcommand{\rmd}{\mathrm{d}}
\newcommand{\eps}{\epsilon}
\newcommand{\Field}[2][]{\mathbf{#2}_{\mathrm{#1}}}
\newcommand{\Fieldhat}[2][]{\mathbf{\hat{#2}}_{#1}}
\newcommand{\Eq}[1]{Eq.~\eqref{#1}} 
\newcommand{\Curl}[2][]{\mathbf{\nabla} \times \Field[#1]{#2}}

\newcommand{\Div}[2][]{\mathbf{\nabla} \cdot \Field[#1]{#2}}

\newcommand{\PD}[3][]{\frac{\partial^{#1} #2}{\partial #3^{#1}}}

\newcommand{\FTw}[1]{\mathcal{F}_{t \rightarrow \omega} \{ #1 \} }
\newcommand{\FTIw}[1]{\mathcal{F}^{-1}_{\w \rightarrow t} \{ #1 \} }

\newcommand{\INT}[1]{ \int_{#1 \in \mathbb{R}}}

\newcommand{\RE}{\Re \mathrm{e}}
\newcommand{\Z}{\mathbb{Z}}

\newcommand{\eq}[1]{Eq.~(\ref{#1})}

\newcommand{\textsubfigure}{0.45 \textwidth}
\newcommand{\textfigure}{0.9 \textwidth}
\definecolor{SBase03}{RGB}{0,43,54}
\definecolor{SBase02}{RGB}{7,54,66}
\definecolor{SBase01}{RGB}{88,110,117}
\definecolor{SBase00}{RGB}{101,123,131}
\definecolor{SBase0}{RGB}{131,148,150}
\definecolor{SBase1}{RGB}{147,161,161}
\definecolor{SBase2}{RGB}{238,232,213}
\definecolor{SBase3}{RGB}{253,246,227}
\definecolor{SYellow}{RGB}{181,137,0}
\definecolor{SOrange}{RGB}{203,75,22}
\definecolor{SRed}{RGB}{220,50,47}
\definecolor{SMagenta}{RGB}{211,54,130}
\definecolor{SViolet}{RGB}{108,113,196}
\definecolor{SBlue}{RGB}{38,139,210}
\definecolor{SCyan}{RGB}{42,161,152}
\definecolor{SGreen}{RGB}{50,200,50}

\begin{document}

\title{\titlefont Electrodynamics of an oscillating particle without cheating PART I : In vacuo. PART II : Near a dispersive bulk}

\author[1]{M. Garcia-Vergara}
\author[2]{G. Dem\'esy}
\author[2]{A Nicolet}
\author[2*]{F. Zolla}

\affil[1]{Departamento de Física, Facultad de Ciencias, Universidad Nacional Autónoma de México,
Ciudad Universitaria, Av. Universidad 3000, Mexico City, 04510, Mexico}

\affil[2]{Aix Marseille Univ, CNRS, Centrale Marseille, Institut Fresnel, Marseille, France}

\affil[*]{Corresponding author: frederic.zolla@fresnel.fr} %% email address is required

\maketitle

\begin{abstract}
In this paper, the electromagnetic radiation from an oscillating particle placed in the vicinity of an object of size comparable to the wavelength is studied. Although this problem may seem academic at first sight, the details of the calculations are presented throughout without any detail left under the carpet. A polyharmonic decomposition of the radiation sources allows the diffraction problem to be fully characterised while satisfying energy conservation. Finally, the source expressions obtained are suitable for use in a numerical code. A 3D illustration using finite elements is provided.
\end{abstract}

%%%%%%%%%%%%%%%%%%%%%%%%%%  body  %%%%%%%%%%%%%%%%%%%%%%%%%%
 
%--------------------------------------------------------------------------------------------------%
\section{Introduction}\label{Sec:Osc_Intro}

This paper consists in two main parts: In the first one we will address the problem of the electromagnetic (EM) field generated by an oscillating particle \emph{in vacuo} and the second one is devoted to its interaction near to a bulk made of some dispersive material. Albeit the apparently pure academic nature of this problem, many applications to this case of configurations can be found. Starting from the study of antennas made by Hertz and Sommerfeld \cite{harish2007antennas,huang2008antennas, sommerfeld1960partial} to the more recently study of quantum nanoemitters and its interaction with nano spheres \cite{Lassalle:17,klimov_1996}. For this kind of phenomena, it is common to use an approximation to describe the field generated by an oscillating charged particle, the most common of these approximations for the far field is to use a \emph{dipole approximation} as described in \cite{jackson_classical_1998, griffiths1999introduction, novotny2012principles,huang2008antennas}. However, it is important to remark here that \textbf{the dipole approximation is only valid when dealing with the far field}.

It is then necessary to find a better way to describe the field generated by an oscillating charge, the immediate approach is to use the \emph{Liénard-Wiechert} fields, which describe in quite a compact form the fields produced by a moving charge. The problem with this solution is that is not tractable from the practical (\emph{i.e.} numerical) point of view as we will show in Section \ref{Sec:Osc_LW}. Other approaches have been to consider the moving charge as a delta distribution that can be approximated by a Taylor's series \cite{Raimond_PMC_2016} or a harmonic expansion \cite{landau1975classical}. Here we propose a different way for obtaining the harmonic representation of the fields which explicitly depends on time and space. Finally, once sources are obtained in harmonic form we use them as an incident field of the diffraction problem to  obtain the diffracted field by a sphere by using the Finite Elements Method.

\section{Mathematical description of the problem}\label{Sec:Osc_Math_Form}
The mathematical description of the interaction of an EM field with a time dispersive bulk is given by the Maxwell equations:
\begin{align}
&\Curl[\textit{$\textit{$\xi$}$}]{E} = -\partial_t \Field[\textit{$\xi$}]{B}, \\
&\Curl[\textit{$\xi$}]{H} = \partial_t \Field[\textit{$\xi$}]{D} + \Field[]{\boldsymbol{\jmath}} ,\\
&\Div[\textit{$\xi$}]{D}  = \rho,\\
&\Div[\textit{$\xi$}]{B}  = 0
\end{align}
where $\textit{$\xi$} = \{I,II\}$ represents the restriction of the fields, $I$ corresponds to the total field \emph{outside the bulk} and $II$ to the field \emph{inside the bulk}. The constitutive relations are then $\Field[\textit{$\xi$}]{H} = \mu_0^{-1} \Field[\textit{$\xi$}]{B}$ (\emph{i.e.} the dispersive bulk has no magnetization) and $\Field[\textit{$\xi$}]{D} = \eps_0(\eps_{r,\textit{$\xi$}}*\Field[\textit{$\xi$}]{E})$, where 
\begin{equation}
	\eps_{r,\textit{$\xi$}}(t) = \begin{cases}
    					         2\pi \delta(t) &\textrm{ if } \xi = I,  \\
                                  2\pi \delta(t) + \chi(t) &\textrm{ if } \xi = II,
                                 \end{cases}
\end{equation}
and $\chi(t)$ is the causal electric susceptibility with its support within the bulk. The convolution is the one corresponding to the Fourier Transform convention in Appendix A. Thus, our new system of equations in terms of the ($\Field[\xi]{E},\Field[\xi]{B}$) fields reads:
\begin{align}
&\Curl[\textit{$\xi$}]{E} = -\partial_t \Field[\textit{$\xi$}]{B}, \\
&\Curl[\textit{$\xi$}]{B} =  \frac{1}{c^2}\partial_t(\eps_{r,\textit{$\xi$ }}*\Field[\textit{$\xi$}]{E}) + \mu_0\Field[]{\boldsymbol{\jmath}},\\
&\mathbf{\nabla}\cdot(\eps_{r,\textit{$\xi$}}*\Field[\textit{$\xi$}]{E})  = \frac{\rho}{\eps_0},\\
&\Div[\textit{$\xi$}]{B}  = 0.
\end{align}
The sources $\rho$ and $\Field[]{\boldsymbol{\jmath}}$ represent the charge density and current corresponding to an oscillating particle and their explicit expression will be given in Section \ref{Sec:Osc_Harm_Sour}. The fields generated by these sources \emph{in vacuo} are:
\begin{align}
\label{Eq:Faraday_0}
&\Curl[]{E}^0 = -\partial_t \Field[]{B}^0, \\
\label{Eq:Ampere_0}
&\Curl[]{B}^0 = \frac{1}{c^2}\partial_t \Field[]{E}^0 + \mu_0\Field[]{\boldsymbol{\jmath}},\\
\label{Eq:Gauss_E_0}
&\Div[]{E}^0  = \frac{\rho}{\eps_0},\\
\label{Eq:Gauss_M_0}
&\Div[]{B}^0  = 0
\end{align} 
where the constitutive relations $\Field[]{D}^0 = \eps_0 \Field[]{E}^0$ and $\Field[]{H}^0 = \mu_0^{-1} \Field[]{B}^0$, $\xi=\{I,II\}$ have been used. Let us remark here that ($\Field[]{E}^0$,$\Field[]{B}^0$) is the incident field of our diffraction problem and that it is defined evrywhere in space.
The next step is to consider a new set of fields defined as: $\Field[\textit{$\xi$}]{E}^1 := \Field[\textit{$\xi$}]{E}-\Field[]{E}^0$, $\Field[\textit{$\xi$}]{B}^1 := \Field[\textit{$\xi$}]{B}-\Field[]{B}^0$ which, in the present linear case, satisfy the system of so called diffracted fields: 
\begin{align}
\label{Eq:Faraday_1}
&\Curl[\textit{$\xi$}]{E}^1 = -\partial_t \Field[\textit{$\xi$}]{B}^1, \\
\label{Eq:Ampere_1}
&\Curl[\textit{$\xi$}]{B}^1 = \frac{1}{c^2}\partial_t(\eps_{r,\textit{$\xi$}}*\Field[\textit{$\xi$}]{E}^1) + \mu_0\Field[II]{\boldsymbol{\jmath}}^0,\\
\label{Eq:Gauss_E_1}
&\mathbf{\nabla}\cdot(\eps_{r,\textit{$\xi$}}*\Field[\textit{$\xi$}]{E}^1)  = \frac{\rho^0_{II}}{\eps_0},\\
\label{Eq:Gauss_M_1}
&\Div[\textit{$\xi$}]{B}^1  = 0,
\end{align} 
whose  sources are defined as:
\begin{equation}\label{Eq:Sources_0}
\rho^0_{II} := -\eps_0 \mathbf{\nabla}\cdot([\eps_{II}-2\pi\delta]*\Field[]{E}^0),
\end{equation}
and
\begin{equation}
\Field[II]{\boldsymbol{\jmath}}^0 := \eps_0 \partial_t([\eps_{II}-2\pi\delta]*\Field[]{E}^0),
\end{equation}
and from these definitions it is easy to see that they satisfy charge conservation. Notice that the support of these new sources is within the bulk and depends on the electromagnetic field generated by the oscillating particle. However obtaining a handy expression of such field requires a very careful crafting as it will be shown in the next section.
 
%%%%--SECTION--%%%
\section{The search for the EM Field generated by an oscillating charge \emph{in vacuo}}\label{Sec:Osc_EM}
%\section{The EM Field generated by an arbitrary charge distribution}\label{Sec:Osc_EM}
% Guillaume : I replaced panofsky1962classical by panofsky2005classical
The problem of obtaining the electromagnetic field generated by a charged oscillating particle is a very important problem \emph{per se}. This section will start with the very general approach of finding the ($\Field[]{E}$,$\Field[]{B}$) fields generated by an arbitrary pair of charge and current distributions ($\rho$,$\jmath$). Usually these solutions are given by the so-called \emph{Jefimenko's equations}. However, it will be shown how by following a similar method  as in Ref.~\cite{panofsky2005classical} for the far field approximations, it is possible to express the electric field mostly in terms of the current density. Next we will discuss some of the efforts that have been made in order to represent the EM field created by and oscillating charge. 

\subsection{The EM Field generated by an arbitrary charge distribution}

First, we consider the system of equations (\ref{Eq:Faraday_0}-\ref{Eq:Gauss_M_0}) (in the sequel the superscript $0$ denoting the incident field radiated by the oscillating particle in freespace has been removed to alleviate notations). By considering the \emph{Lorentz gauge}, one can obtain the electric and magnetic fields ($\Field[]{E}$, $\Field[]{B}$) using the expressions:
\begin{align}
    \label{Eq:E_partial_A_grad_phi}
    &\Field[]{E}  = - \partial_t\Field[]{A}-\mathbf{\nabla}\phi, \\
    \label{Eq:B_curl_A}
    &\Field[]{B} = \Curl[]{A}
\end{align}
where $\phi$ and $\Field[]{A}$ are the retarded potentials \cite{griffiths1999introduction,jackson_classical_1998}
\begin{align}
	\label{Eq:phi_int}
	 \phi       &= \frac{1}{4\pi \eps_0} \int_{\mathbb{R}^3} \frac{\rho(\Field[]{x}^{\prime}, t_r)}{R} \,\rmd \Field[]{x}^{\prime}, \\
     \label{Eq:A_int}
	\Field[]{A} &= \frac{\mu_0}{4 \pi}  \int_{\mathbb{R}^3} \frac{\Field[]{\boldsymbol{\jmath}}(\Field[]{x}^{\prime}, t_r)}{R} \,\rmd \Field[]{x}^{\prime},
\end{align}

\noindent with the retarded time $t_r = t - \frac{R}{c}$ and $R = |\Field[]{R}|$ with $\Field[]{R} = \Field[]{x}-\Field[]{x}^{\prime}$. The next step consists in  plugging  the retarded potentials Eqs.~(\ref{Eq:phi_int},\ref{Eq:A_int}) into Eqs.~(\ref{Eq:E_partial_A_grad_phi},\ref{Eq:B_curl_A}) and  then take the appropriate derivatives. Nevertheless, this implies to take time derivatives with respect to a function which is in terms of the retarded time $t_r$. In order to avoid this difficulty and carry our calculations further, we propose to use the Fourier transform in time (See Annexe \ref{An:Fourier-T}, $\hat{f}$ denoting the Fourier transform of $f$) and then consider only the spatial derivatives. Starting by transforming the potential $\phi$ one has:
\begin{align}
	\hat{\phi} %=& \frac{1}{4\pi \eps_0} \int_{\mathbb{R}^3} \frac{1}{R} \FTw{\rho(\Field[]{x}^{\prime}, t_r)} \,\rmd\Field[]{x}^{\prime} \\
    	\nonumber
         =& \frac{1}{4\pi \eps_0} \int_{\mathbb{R}^3} \frac{1}{R}\FTw{\rho(\Field[]{x}^{\prime},t - R/c )} \,\rmd\Field[]{x}^{\prime}, \\
         \label{Eq:phi_int_hat}
         =&  \frac{1}{4\pi \eps_0} \int_{\mathbb{R}^3} \frac{1}{R} e^{iR k_0} \hat{\rho}(\Field[]{x}^{\prime},\w) \,\rmd\Field[]{x}^{\prime},
\end{align}
where $k_0 = \frac{\w}{c}$. In a similar way the vector potential in the frequency domains reads:
\begin{equation}
	\label{Eq:A_int_hat}
	\Fieldhat[]{A} = \frac{\mu_0}{4 \pi}  \int_{\mathbb{R}^3} e^{+iR k_0} \frac{\Fieldhat[]{\boldsymbol{\jmath}}(\Field[]{x}^{\prime},\w)}{R} \,\rmd\Field[]{x}^{\prime}.
\end{equation}  
Before proceeding, the following identities are necessary:
\begin{align}
	\label{Eq:Vec_id_1}
	&\mathbf{\nabla} R = \frac{\Field[]{R}}{R}, \\
    \label{Eq:Vec_id_2}
    &\mathbf{\nabla} \frac{1}{R} = -\frac{1}{R^2} \mathbf{\nabla} R=- \frac{\Field[]{R}}{R^3},\\
    \label{Eq:Vec_id_3}
    &\mathbf{\nabla}  e^{iR k_0} = ik_0 e^{iR k_0} \mathbf{\nabla} R = ik_0 e^{iR k_0} \frac{\Field[]{R}}{R},
\end{align} 
and combining equations~(\ref{Eq:Vec_id_2}) and (\ref{Eq:Vec_id_3}) we get:
\begin{equation}
	 \mathbf{\nabla}\bigg(\frac{e^{iR k_0}}{R}\bigg) = \bigg(ik_0-\frac{1}{R}\bigg)\frac{e^{iR k_0}}{R^2}\Field[]{R}.
\end{equation} \label{Eq:Vec_id_4}
Equipped with these tools, it is quite easy to obtain $\Field[]{B}$ by simply taking the curl of \eq{Eq:A_int_hat}:
\begin{align}
	\Fieldhat[]{B} = & \frac{\mu_0}{4 \pi}  \int_{\mathbb{R}^3}  \mathbf{\nabla} \times \bigg( \frac{e^{+iR k_0}}{R} \Fieldhat[]{\boldsymbol{\jmath}}(\Field[]{x}^{\prime},\w)  \bigg) \,\rmd\Field[]{x}^{\prime}\\
    =&\frac{\mu_0}{4 \pi}  \int_{\mathbb{R}^3} \mathbf{\nabla}\bigg(\frac{e^{iR k_0}}{R}\bigg) \times \Fieldhat[]{\boldsymbol{\jmath}}(\Field[]{x}^{\prime},\w)  \,\rmd\Field[]{x}^{\prime}\\
    =& \frac{\mu_0}{4 \pi}  \int_{\mathbb{R}^3} \bigg(ik_0-\frac{1}{R}\bigg)\frac{e^{iR k_0}}{R^2}\Field[]{R} \times \Fieldhat[]{\boldsymbol{\jmath}}(\Field[]{x}^{\prime},\w)  \,\rmd\Field[]{x}^{\prime}
\end{align}
And by taking the inverse Fourier transform, we arrive to the expression:
\begin{equation}
	\Field[]{B} = \Field[int]{B} + \Field[rad]{B},
\end{equation}
where $\Field[int]{B}$ is the intermediate field and $\Field[rad]{B}$ is the radiated field which are given by the integrals:
\begin{align}
	\label{Eq:B_int}
	& \Field[int]{B} =  \frac{\mu_0}{4 \pi}\int_{\mathbb{R}^3} \frac{\Field[]{\boldsymbol{\jmath}}(\Field[]{x}^{\prime}, t_r) \times \Field[]{R}}{R^3}   \,\rmd\Field[]{x}^{\prime},\\
	\label{Eq:B_rad}
	&\Field[rad]{B} = \frac{\mu_0}{4 \pi c}\int_{\mathbb{R}^3} \frac{\partial_t \Field[]{\boldsymbol{\jmath}}(\Field[]{x}^{\prime}, t_r) \times \Field[]{R}}{R^2}   \,\rmd\Field[]{x}^{\prime}.
\end{align}
In order to obtain the electric field, it is necessary to consider the Fourier transform of Eq.~(\ref{Eq:E_partial_A_grad_phi}):
\begin{equation}\label{Eq:E_partial_A_grad_phi_hat}
	\Fieldhat[]{E}  = +i\w\Fieldhat[]{A}-\mathbf{\nabla}\hat{\phi}.
\end{equation}
The first term on the right hand side of Eq.~(\ref{Eq:E_partial_A_grad_phi_hat}) is quite easy to calculate,
\begin{equation}\label{Eq:iwA}
	i\w \Fieldhat[]{A} = \frac{1}{4\pi\eps_0c} \int_{\mathbb{R}^3} i k_0 \Fieldhat[]{\boldsymbol{\jmath}}(\w, \Field[]{x}^{\prime}) \frac{e^{iR k_0}}{R}  \,\rmd\Field[]{x}^{\prime}.
\end{equation}
The second term is much more tricky and its derivation goes as follows:
\begin{align}
	\nonumber
	- \mathbf{\nabla} \hat{\phi} = &\frac{-1}{4\pi \eps_0} \int_{\mathbb{R}^3}\mathbf{\nabla}\bigg(\frac{e^{iR k_0}}{R}\bigg) \hat{\rho}(\Field[]{x}^{\prime},\w) \,\rmd\Field[]{x}^{\prime} \\
    \nonumber
    =&\frac{-1}{4\pi \eps_0} \int_{\mathbb{R}^3}\mathbf{\nabla}\bigg(\frac{e^{iR k_0}}{R}\bigg) \hat{\rho}(\Field[]{x}^{\prime},\w) \,\rmd\Field[]{x}^{\prime},\\
    \nonumber
    %=&\frac{-1}{4\pi \eps_0} \int_{\mathbb{R}^3} \bigg(ik_0\hat{\rho}(\Field[]{x}^{\prime},\w)-\frac{1}{R}\hat{\rho}(\Field[]{x}^{\prime},\w)\bigg)\frac{e^{iR k_0}}{R^2}\Field[]{R}  \,\rmd\Field[]{x}^{\prime}\\
    \nonumber
    =&\frac{-1}{4\pi \eps_0} \int_{\mathbb{R}^3} ik_0\hat{\rho}(\Field[]{x}^{\prime},\w)\frac{e^{iR k_0}}{R^2}\Field[]{R}  \,\rmd\Field[]{x}^{\prime}\\
    \label{Eq:minus_grad_phi}
    &+\frac{1}{4\pi \eps_0} \int_{\mathbb{R}^3}\hat{\rho}(\Field[]{x}^{\prime},\w)\frac{e^{iR k_0}}{R^3}\Field[]{R}  \,\rmd\Field[]{x}^{\prime}
\end{align}
Up to this point most textbooks \emph{e.g.} Refs.~\cite{jackson_classical_1998, griffiths1999introduction,jefimenko1966electricity} simply take the inverse Fourier transform of \eq{Eq:minus_grad_phi} and give the electric field in terms of the derivative of the density of charges and obtain the so called \emph{Jefimenko's equations} \cite{jefimenko1966electricity, griffiths1999introduction}. Despite the straightforward nature of this derivation, we are going to carry the calculations further. The interest of doing so will be shown later. For our purposes, the second integral in \eq{Eq:minus_grad_phi} is already in optimum form, and we will focus our attention on the integral defined within a bounded volume $\Omega$. 
\begin{equation}
	\label{Eq:Int_Omega}
	{\Field[]{I}}_{\Omega}:= \int_{\Omega} ik_0\hat{\rho}(\Field[]{x}^{\prime},\w)\frac{e^{iR k_0}}{R^2}\Field[]{R}  \,\rmd\Field[]{x}^{\prime}.
\end{equation}
Notice that the integral we are looking for, is the limit case when $\Omega \rightarrow \mathbb{R}^3$. Next, we will make use of the continuity equation in the frequency domain:
\begin{equation}
	i\w \hat{\rho}(\Field[]{x}^{\prime}, \w) = \mathbf{\nabla}^{\prime} \cdot \Fieldhat[]{\boldsymbol{\jmath}}  
\end{equation}
and by defining the function: 
\begin{equation}
	F(R) := \frac{e^{iR k_0}}{cR^2},
\end{equation}
the integral ${\Field[]{I}}_{\Omega}$ in \eq{Eq:Int_Omega} can be written in a more compact way (omitting the $\Field[]{x}^{\prime}$, $R$ and $\w$ dependencies) as:
\begin{equation}
	{\Field[]{I}}_{\Omega} := \int_{\Omega} F \mathbf{\nabla}^{\prime} \cdot \Fieldhat[]{\boldsymbol{\jmath}}  \Field[]{R}  \,\rmd\Field[]{x}^{\prime}. 
\end{equation}
It is very important to remark here that \textbf{the divergence is being taken with respect to the primed coordinates (hence the prime superscript in $\mathbf{\nabla}^{\prime}$)}. Due to the fact that we are working with Cartesian coordinates, it is possible to write 
\begin{equation}
	\Field[]{R} = \Field[]{x}-\Field[]{x}^{\prime} = \sum_{\eta = x,y,z} (\eta-\eta^{\prime}) {\Field[]{e}}_{\eta},
\end{equation}
where $(\Field[x]{e},\Field[z]{e},\Field[z]{e})$ denote the cartesian unit vectors,  and then:
\begin{equation}\label{Eq:Sum_I_eta}
	{\Field[]{I}}_{\Omega} = \sum_{\eta = x,y,z} \int_{\Omega} (\eta-\eta^{\prime}) F \mathbf{\nabla}^{\prime} \cdot \Fieldhat[]{\boldsymbol{\jmath}}    \,\rmd\Field[]{x}^{\prime} {\Field[]{e}}_{\text{$\eta$}} = \sum_{\eta = x,y,z} I_{\Omega, \eta} {\Field[]{e}}_{\text{$\eta$}}. 
\end{equation}
Each one of these $\eta$ integrals can be evaluated by means of the identity $ (\eta-\eta^{\prime}) F \mathbf{\nabla}^{\prime} \cdot \Fieldhat[]{\boldsymbol{\jmath}} = \mathbf{\nabla}^{\prime} \cdot \big[  (\eta-\eta^{\prime}) F \Fieldhat[]{\boldsymbol{\jmath}} \big]-\Fieldhat[]{\boldsymbol{\jmath}} \cdot \mathbf{\nabla}^{\prime} \big[  (\eta-\eta^{\prime}) F \big]$ and the Green-Ostrogradsky's theorem \cite{petit_outil_1991,ruiz1995calculo} as follows:
\begin{equation}
	I_{\Omega, \eta} = \int_{\partial \Omega} (\eta-\eta^{\prime}) F \Fieldhat[]{\boldsymbol{\jmath}}\cdot \Field[out]{n} \big|_{\partial \Omega}\,\rmd S^{\prime} - \int_{\Omega} \Fieldhat[]{\boldsymbol{\jmath}} \cdot \mathbf{\nabla}^{\prime} \big[  (\eta-\eta^{\prime}) F \big] \,\rmd\Field[]{x}^{\prime}
\end{equation}
Taking the limit $\Omega \rightarrow \mathbb{R}^3$ and keeping in mind that the boundary term vanishes as $\frac{1}{R}$ we have:
\begin{equation}
	I_{\mathbb{R}^3, \eta} =  - \int_{\mathbb{R}^3} \Fieldhat[]{\boldsymbol{\jmath}} \cdot \mathbf{\nabla}^{\prime} \big[  (\eta-\eta^{\prime}) F \big] \,\rmd\Field[]{x}^{\prime}.
\end{equation}
The gradient (with respect to the primed coordinates) can be computed explicitly:
\begin{align}
	\mathbf{\nabla}^{\prime} \big[  (\eta-\eta^{\prime}) F \big] = &F \mathbf{\nabla}^{\prime} (\eta-\eta^{\prime}) + (\eta-\eta^{\prime}) \mathbf{\nabla}^{\prime} F \\
    = &-F \Field[\text{$\eta$}]{e} -   (\eta-\eta^{\prime}) \Field[]{R}\bigg(\frac{ik_0}{R}-\frac{2}{R^2}\bigg)F
\end{align}
and then
\begin{equation}\label{Eq:I_eta_final}
	\Field[\mathbb{R}^3, \text{$\eta$}]{I} =   \int_{\mathbb{R}^3} F \Fieldhat[]{\boldsymbol{\jmath}} \cdot \Field[\text{$\eta$}]{e} + (\eta-\eta^{\prime}) \Fieldhat[]{\boldsymbol{\jmath}} \cdot \Field[]{R}\bigg(\frac{ik_0}{R}-\frac{2}{R^2}\bigg)F \,\rmd\Field[]{x}^{\prime}.
\end{equation}
Plugging \Eq{Eq:I_eta_final} into \Eq{Eq:Sum_I_eta} we get: 
\begin{equation}\label{Eq:I_final}
	\Field[\mathbb{R}^3]{I} =   \int_{\mathbb{R}^3} \bigg[ \Fieldhat[]{\boldsymbol{\jmath}} + \Field[]{R}(\Fieldhat[]{\boldsymbol{\jmath}} \cdot \Field[]{R}) \bigg(\frac{ik_0}{R}-\frac{2}{R^2}\bigg) \bigg]F \,\rmd\Field[]{x}^{\prime}.
\end{equation}
Before taking the inverse Fourier transform, we will try to express the term between square brackets (which we will call $\Field[]{\boldsymbol{\jmath}}$) in a more illuminating way. First, we rearrange $\Field[]{\boldsymbol{\jmath}}$ as:
\begin{equation}
	   \Field[]{\boldsymbol{\jmath}} = \Fieldhat[]{\boldsymbol{\jmath}}-\Field[]{R}(\Fieldhat[]{\boldsymbol{\jmath}} \cdot \Field[]{R}) \frac{2}{R^2} +  ik_0 \frac{\Field[]{R}(\Fieldhat[]{\boldsymbol{\jmath}} \cdot \Field[]{R})}{R}.
\end{equation}  
Now we consider the vector identity $(\Fieldhat[]{\boldsymbol{\jmath}}\times \Field[]{R})\times \Field[]{R} = \Field[]{R}(\Fieldhat[]{\boldsymbol{\jmath}} \cdot \Field[]{R})-R^2 \Fieldhat[]{\boldsymbol{\jmath}}$ \cite{ruiz1995calculo}, and from this we have:
\begin{align}
	&\frac{\Field[]{R}(\Fieldhat[]{\boldsymbol{\jmath}} \cdot \Field[]{R})}{R} = \frac{(\Fieldhat[]{\boldsymbol{\jmath}}\times \Field[]{R})\times \Field[]{R}}{R}+R \Fieldhat[]{\boldsymbol{\jmath}},\\
    & \Fieldhat[]{\boldsymbol{\jmath}}-2\frac{\Field[]{R}(\Fieldhat[]{\boldsymbol{\jmath}} \cdot \Field[]{R})}{R^2} = -\frac{(\Fieldhat[]{\boldsymbol{\jmath}}\times \Field[]{R})\times \Field[]{R}}{R^2}-\frac{\Field[]{R}(\Fieldhat[]{\boldsymbol{\jmath}} \cdot \Field[]{R})}{R^2}.
\end{align}
Then $\Field[]{\boldsymbol{\jmath}}$ can be seen as:
\begin{equation}
	   \Field[]{\boldsymbol{\jmath}} =  - \bigg[\frac{(\Fieldhat[]{\boldsymbol{\jmath}}\times \Field[]{R})\times \Field[]{R}}{R^2}+\frac{\Field[]{R}(\Fieldhat[]{\boldsymbol{\jmath}} \cdot \Field[]{R})}{R^2}\bigg] +  ik_0 \bigg[\frac{(\Fieldhat[]{\boldsymbol{\jmath}}\times \Field[]{R})\times \Field[]{R}}{R}+R \Fieldhat[]{\boldsymbol{\jmath}} \bigg].
\end{equation}
Substituting this result into \Eq{Eq:I_final} and then plugging that new integral into \Eq{Eq:minus_grad_phi} we finally arrive to the expression: 
\begin{align}
	\nonumber
	- \mathbf{\nabla} \hat{\phi} =&\frac{1}{4\pi \eps_0} \int_{\mathbb{R}^3}\hat{\rho}\frac{e^{iR k_0}}{R^3}\Field[]{R}  \,\rmd\Field[]{x}^{\prime}
 \\
    \nonumber
    &-\frac{1}{4\pi \eps_0 c} \int_{\mathbb{R}^3} i k_0 \frac{(\Fieldhat[]{\boldsymbol{\jmath}}\times \Field[]{R})\times \Field[]{R}}{R^3} e^{iR k_0} \,\rmd\Field[]{x}^{\prime},\\
    \nonumber
    &+\frac{1}{4\pi \eps_0 c} \int_{\mathbb{R}^3}  \bigg[\frac{(\Fieldhat[]{\boldsymbol{\jmath}}\times \Field[]{R})\times \Field[]{R}}{R^4}+\frac{\Field[]{R}(\Fieldhat[]{\boldsymbol{\jmath}} \cdot \Field[]{R})}{R^4}\bigg]e^{iR k_0} \,\rmd\Field[]{x}^{\prime}\\
    &- \frac{1}{4\pi\eps_0c} \int_{\mathbb{R}^3} i k_0 \Fieldhat[]{\boldsymbol{\jmath}}(\w, \Field[]{x}^{\prime}) \frac{e^{iR k_0}}{R}  \,\rmd\Field[]{x}^{\prime}
\end{align}
From \Eq{Eq:iwA} we recognize the last integral as $-i\w \Fieldhat[]{A}$ and then, after taking the inverse Fourier transform, we get:
\begin{equation}
  \Field[]{E} = \Field[c]{E} + \Field[int]{E} + \Field[rad]{E},
\end{equation}
where $\Field[c]{E}$ is the Coulomb field, $\Field[int]{E}$ the intermediate field and $\Field[rad]{E}$ the radiated field given by the integrals:
\begin{align}
	\label{Eq:E_c}
 	& \Field[c]{E} =  \frac{1}{4\pi \eps_0} \int_{\mathbb{R}^3}\frac{\rho(\Field[]{x}^{\prime},t_r)}{R^3}\Field[]{R}  \,\rmd\Field[]{x}^{\prime}, \\
    \label{Eq:E_int}
 	&\Field[int]{E} = \frac{1}{4\pi \eps_0 c} \int_{\mathbb{R}^3}  \bigg[\frac{(\Field[]{\boldsymbol{\jmath}}(\Field[]{x}^{\prime},t_r)\times \Field[]{R})\times \Field[]{R}}{R^4}+\frac{(\Field[]{\boldsymbol{\jmath}}(\Field[]{x}^{\prime},t_r) \cdot \Field[]{R})}{R^4}\Field[]{R}\bigg] \,\rmd\Field[]{x}^{\prime}, \\
    \label{Eq:E_rad}
	&\Field[rad]{E} = \frac{1}{4\pi \eps_0 c^2} \int_{\mathbb{R}^3} \frac{(\partial_t \Field[]{\boldsymbol{\jmath}}(\Field[]{x}^{\prime},t_r)\times \Field[]{R})\times \Field[]{R}}{R^3} \,\rmd\Field[]{x}^{\prime}.
\end{align}
At this point the reader may be wondering the reason for why we have made all these extra steps when the Jefimenko's equations already provide an explicit expression for the electric field. This is because when dealing with the Jefimenko's equations, the magnetic field is terms of the electric current $\Field[]{\boldsymbol{\jmath}}$ and the electric field is in terms of the distribution of charge $\rho$ \cite{jefimenko1966electricity, griffiths1999introduction}. This point of view albeit intuitively is very clear, makes it difficult to compare the terms corresponding to the radiation field. By making the manipulations described above, we have ensured the fact that the intermediate and radiated electric and magnetic fields  are all expressed in terms of the electric current solely. The utility of this approach will be shown later.

%%%%------------------------------------------%%%
\subsection{The different approximations to the oscillating source problem}\label{Sec:Osc_Approx}

%%%%------------------------------------------%%%
\subsubsection{The Liénard-Wiechert's Field approach}\label{Sec:Osc_LW}

The academic problem of describing the $\Field[]{E}$ and $\Field[]{B}$ fields generated by a charged particle that moves along a given trajectory $\Field[]{u}(t)$, which are called the \emph{Liénard-Wiechert} fields, has been studied in many books \cite{jackson_classical_1998, griffiths1999introduction,landau1975classical,spohn2004dynamics, heald2012classical,panofsky2005classical}. The basic idea is to consider a charge density $\rho$ and an electric current $\Field[]{j}$ given by:  
\begin{align}
	\label{Eq:rho_delta}
	& \rho(\Field[]{x}, t)        = q\delta(\Field[]{x} - \Field[]{u}(t)),\\ 
    \label{Eq:j_delta}
    &\Field[]{\boldsymbol{\jmath}}(\Field[]{x}, t) = q\delta(\Field[]{x} - \Field[]{u}(t))\Field[]{v}(t),
\end{align}
where $\delta$ is a Delta distribution and $\Field[]{v}(t)  = \frac{d \Field[]{u}}{dt}$. And from here there are many ways to tackle the problem of obtaining the $\Field[]{E}$ and $\Field[]{B}$ generated fields: Jackson \cite{jackson_classical_1998} and Landau \cite{landau1975classical} consider an elegant formalism using quadrivector approach. Panofsky \cite{panofsky2005classical} and  Heald \cite{spohn2004dynamics} use the so called Liénard-Wiechert potentials which can be obtained by direct substitution on equations (\ref{Eq:phi_int}-\ref{Eq:A_int}) and then carrying all the necessary derivatives. The deduction of the fields $\Field[]{E}$ and $\Field[]{B}$ following this procedure can be seen in Ref.~\cite{griffiths1999introduction}. For this section we have decided not to follow any of these approaches, but rather to proceed by direct substitution of the sources (\ref{Eq:rho_delta}) and (\ref{Eq:j_delta}) into equations (\ref{Eq:B_int}-\ref{Eq:B_rad}) and (\ref{Eq:E_c}-\ref{Eq:E_rad}), that is: 

\begin{align}
	\label{Eq:B_int_delta_int}
	&\Field[int]{B} =  \frac{qc \mu_0}{4 \pi}\int_{\mathbb{R}^3} \delta(\Field[]{x}^{\prime} - \Field[]{u}) \frac{\boldsymbol{\beta} \times \Field[]{n}}{R^2}   \,\rmd\Field[]{x}^{\prime},\\
	\label{Eq:B_rad_delta_int}
	&\Field[rad]{B} = \frac{qc \mu_0}{4 \pi}\partial_t \int_{\mathbb{R}^3} \delta(\Field[]{x}^{\prime} - \Field[]{u}) \frac{ \boldsymbol{\beta} \times \Field[]{n}}{c R}   \,\rmd\Field[]{x}^{\prime},\\
 	&\Field[c]{E} =  \frac{q}{4\pi \eps_0} \int_{\mathbb{R}^3} \frac{\delta(\Field[]{x}^{\prime} - \Field[]{u})}{R^2}\Field[]{n}  \,\rmd\Field[]{x}^{\prime} \\
    \label{Eq:E_int_delta_int}
 	&\Field[int]{E} = \frac{q}{4\pi \eps_0} \int_{\mathbb{R}^3}  \frac{\delta(\Field[]{x}^{\prime} - \Field[]{u})}{R^2} \big[(\boldsymbol{\beta}\times \Field[]{n})\times \Field[]{n}+(\boldsymbol{\beta} \cdot \Field[]{n})\Field[]{n}\big] \,\rmd\Field[]{x}^{\prime}, \\
    \label{Eq:E_rad_delta_int}
	&\Field[rad]{E} = \frac{q}{4\pi \eps_0 } \partial_t \int_{\mathbb{R}^3} \frac{\delta(\Field[]{x}^{\prime} - \Field[]{u})}{c R} (\boldsymbol{\beta}\times \Field[]{n})\times \Field[]{n} \,\rmd\Field[]{x}^{\prime},
\end{align}
where we have introduced the ususal following short hand conventions:
\begin{equation}
	\boldsymbol{\beta} := \frac{\Field[]{v}}{c}, \qquad  \Field[]{n} := \frac{\Field[]{R}}{R}.
\end{equation}
The procedure that is shown in Annexe 2, follows the ideas expressed by Heald and Marion in Ref.~\cite{heald2012classical}, the main difference is that while Heald and Marion consider the Jefimenko's equations, we are going to use equations (\ref{Eq:B_int_delta_int}-\ref{Eq:E_rad_delta_int}). \textbf{We have decided to include the full deduction of the Liénard-Wiechert fields because as far as we have seen this result is quoted but the steps towards its obtention are not shown. Griffiths just states that the deduction is \emph{very difficult} and Heald and Marion say that it is necessary to perform \emph{heroic algebra}. Therefore, we believe that it is important to show, as best as we can, how to obtain one of the main results in classical electrodynamics.} 

Once the fields $\Field[]{E}$ and $\Field[]{B}$ are given by \Eq{Eq:E_LW} and \Eq{Eq:B_LW}, namely
\begin{equation*}
	 \Field[]{B} = \frac{qc\mu_0}{4 \pi} \bigg[\frac{\Field[]{a}\times \Field[]{n}}{K^2Rc^2}+\frac{\Field[]{a}\cdot \Field[]{n}(\boldsymbol{\beta}\times \Field[]{n})}{c^2K^3R} +\frac{(1-\beta^2)(\boldsymbol{\beta}\times \Field[]{n})}{K^3R^2} \bigg]_{t_r}
\end{equation*}
and
\begin{equation*}
 \Field[]{E} = \frac{q}{4 \pi \eps_0}\bigg[\frac{(1-\beta^2)( \Field[]{n}-\boldsymbol{\beta})}{K^3R^2} + \frac{\Field[]{a}\times(\Field[]{n}-\boldsymbol{\beta})\times \Field[]{n} }{c^2K^3R} \bigg]_{t_r},
\end{equation*}
with $\Field[]{a} = Kc\boldsymbol{\dt{\beta}}$, it would be easy to think that the fields produced by an oscillating particle could be retrieved by considering the specific trajectory:
\begin{equation}
	\Field[]{u}(t_r) = a \cos(\w_0 t_r) \Field[z]{e},
\end{equation}
where $a$ and $\w_0$ are the oscillation amplitude and frequency respectively. Nevertheless, as pointed by Spohn in \cite{spohn2004dynamics}, the \emph{Liénard Wiechert} fields are less explicit than they appear to be. This is due to the fact that \Eq{Eq:E_LW} and \Eq{Eq:B_LW} depend on the retarded time which is still itself a solution of a (in general non trivial) transcendental equation, namely:
\begin{equation}
	t_r= t - \frac{|\Field[]{x}-a\cos(\w_0 t_r) \Field[z]{e}|}{c}.
\end{equation}
Let us notice here that if the particle is at constant speed with a straight trajectory, the \emph{Liénard Wiechert} fields can be almost straightforwardly \cite{griffiths1999introduction}. However, the solution of this problem, when dealing with an oscillating charge, in this case the retarded time is a function of the present time and the position ($t_r := t_r(t, \Field[]{x})$). 
%%%%------------------------------------------%%%
\subsubsection{The Landau's Spectral resolution approach}\label{Sec:Osc_Landau}

In \emph{The classical theory of fields}, Landau \emph{et al.} \cite{landau1975classical} consider that the fields produced by moving charges can be expanded as a superpostion of monochromatic waves. Assuming that $\rho(\Field[]{x},t)$ and $\Field[]{\jmath}(\Field[]{x},t)$ have a Fourier integral representation, we can write 
\begin{align}
    \label{Eq:rho_fourier}
    \rho(\Field[]{x},t) &= \INT{\w} \hat{\rho}(\Field[]{x},\w) e^{-i\w t} \,\mathrm{d}\w,\\
    \label{Eq:j_Fourier}
    \Field[]{\jmath}(\Field[]{x},t) &= \INT{\w} \Fieldhat[]{\jmath}(\Field[]{x},\w) e^{-i\w t} \,\mathrm{d}\w.
\end{align}
According to Landau: \emph{It is clear that each Fourier component of $\rho(\Field[]{x},t)$ and $\Field[]{\jmath}(\Field[]{x},t)$ is responsible for the creation of the corresponding monochromatic component of the field.} 
Thus, it is natural to consider the following Fourier integral representations of the potentials $\phi(\Field[]{x},t)$ and $\Field[]{A}(\Field[]{x},t)$:
\begin{align}
    \label{Eq:phi_Fourier}
    \phi(\Field[]{x},t) &= \INT{\w} \hat{\phi}(\Field[]{x},\w) e^{-i\w t} \,\mathrm{d}\w, \\
    \label{Eq:A_Fourier}
    \Field[]{A}(\Field[]{x},t) &= \INT{\w} \Fieldhat[]{A}(\Field[]{x},\w) e^{-i\w t} \,\mathrm{d}\w.
\end{align}
For the sequel, we will only work with $\phi(\Field[]{x},t)$, because all the discussion applies also to $\Field[]{A}(\Field[]{x},t)$.
and substituting \eq{Eq:rho_fourier} into \eq{Eq:phi_int} we get
\begin{align}
    \phi(\Field[]{x},t) 
    &= \frac{1}{4\pi \eps_0} \int_{\mathbb{R}^3} \INT{\w} \hat{\rho}(\Field[]{x},\w) e^{-i\w t_r} \,\mathrm{d}\w \frac{1}{R} \,\rmd\Field[]{x}^{\prime}, \\
    &= \INT{\w} e^{-i\w t} \bigg( \frac{1}{4\pi \eps_0} \int_{\mathbb{R}^3} \hat{\rho}(\Field[]{x}^{\prime},\w) \frac{e^{i\w R / c}}{R}   \,\rmd\Field[]{x}^{\prime} \bigg) \,\mathrm{d}\w.     
\end{align}
Comparing this last equality with \eq{Eq:phi_Fourier}, one gets that
\begin{equation}
    \label{Eq:phi_Fourier_component}
    \hat{\phi}(\Field[]{x},\w) = \frac{1}{4\pi \eps_0} \int_{\mathbb{R}^3} \hat{\rho}(\Field[]{x}^{\prime},\w) \frac{e^{i\w R / c}}{R}   \,\rmd\Field[]{x}^{\prime}.
\end{equation}
Remembering that $\hat{\rho}(\Field[]{x},\w)$ is the Fourier transform of $\rho(\Field[]{x}, \tau)$  and after some manipulations $\hat{\phi}(\Field[]{x},\w)$ can be written as per
\begin{equation}
    \label{Eq:phi_Fourier_int_tau_int_space}
    \hat{\phi}(\Field[]{x},\w) = \frac{1}{4\pi \eps_0} \frac{1}{2\pi} \INT{\tau} \int_{\mathbb{R}^3} \rho(\Field[]{x}^{\prime},\tau) \frac{e^{i\w (t+|\Field[]{x}-\Field[]{x}^{\prime}|/c) }}{|\Field[]{x}-\Field[]{x}^{\prime}|}   \,\rmd\Field[]{x}^{\prime}  \,\rmd\tau.
\end{equation}
Now, we consider the singular charge distribution as in \eq{Eq:rho_delta} to obtain
\begin{equation}
    \label{Eq:phi_Fourier_int_tau_int_space_for delta}
    \hat{\phi}(\Field[]{x},\w) = \frac{q}{4\pi \eps_0} \frac{1}{2\pi} \INT{\tau}\frac{e^{i\w (\tau+|\Field[]{x}-\Field[]{u}(\tau)|/c) }}{|\Field[]{x}-\Field[]{u}(\tau)|}   \,\rmd\tau.
\end{equation}
Upon substitution of this expression into \Eq{Eq:phi_Fourier} we arrive to 
\begin{equation}
    \label{Eq:phi_Landau}
    \phi(\Field[]{x},t) =  \frac{q}{4\pi \eps_0} \frac{1}{2\pi} \INT{\w}\INT{\tau}\frac{e^{i\w (\tau+|\Field[]{x}-\Field[]{u}(\tau)|/c-t) }}{|\Field[]{x}-\Field[]{u}(\tau)|}   \,\rmd\tau  \,\mathrm{d}\w,
\end{equation}
and similarly for $\Field[]{A}(\Field[]{x},t)$
\begin{equation}
    \label{Eq:A_Landau}
   \Field[]{A}(\Field[]{x},t) =  \frac{q \mu_0}{4\pi} \frac{1}{2\pi} \INT{\w}\INT{\tau} \Field[]{v}(\tau) \frac{e^{i\w (\tau+|\Field[]{x}-\Field[]{u}(\tau)|/c-t) }}{|\Field[]{x}-\Field[]{u}(\tau)|}   \,\rmd\tau  \,\mathrm{d}\w.
\end{equation}

From the above expressions, it's evident that the right-hand side of the equation explicitly depends on the present time. This feature avoids issues related to retarded time, which is a notable problem in the Liénard-Wiechert fields. However, difficulties emerge when attempting to explicitly compute these integrals. This arises due to the fact that the particle's trajectory, denoted by $\mathbf{u}$, is, in principle, unrestricted in its choice of any argument $\tau$. Moreover, the exponential function in equations \eqref{Eq:phi_Landau} and \eqref{Eq:A_Landau} necessitates a sweep over all the $(\omega, \tau)$ values in $\mathbb{R}^2$.

\subsubsection{The Raimond's Taylor series expansion approach}\label{Sec:Raimond_Ty}

In \cite{Raimond_PMC_2016} J.M. Raimond proposes another way to represent the charge density $\rho(\Field[]{x},t)=q \delta(\Field[]{x}-a\cos(\w_0 t)\Field[z]{e})$ by using a Taylor series expansión of the Dirac delta in the sense of distributions: 
\begin{equation}
    \delta(\Field[]{x}-g(t)) = \delta(\Field[]{x})-g(t)\Field[z]{e}\cdot\nabla \delta(\Field[]{x})+\frac{g^2(t)}{2}\Field[z]{e}\cdot\nabla [\Field[z]{e}\cdot\nabla\delta(\Field[]{x})] +\dots ,
\end{equation}
where $g(t)=a\cos(\w_0t)$. In this way one can rewrite the charge density:
\begin{equation}
    \rho(\Field[]{x},t) = \sum_{n=0}^{\infty} \rho_n(\Field[]{x},t) = \sum_{n=0}^{\infty} (g(t))^n \varrho_n(\Field[]{x}), 
\end{equation}
where the $\rho_n(\Field[]{x},t)$ is the $n$-th charge density and $\varrho_n(\Field[]{x})$ is given by
\begin{equation}\label{Eq:rho_n}
    \varrho_n(\Field[]{x}) = \frac{q[-\Field[z]{e}\cdot \nabla]^n \delta(\Field[]{x})}{n!} = \frac{q(-1)^n }{n!} \PD[n]{ \delta(\Field[]{x})}{z}.  
\end{equation}
The reader may identify $\varrho_n(\Field[]{x})$ as the singular charge distribution of a $2^n$-pole (\emph{i.e.} $n=0$ monopole, $n=1$ dipole, $n=2$ quadrupole, etc.) \cite{jentschura2017advanced,stratton1941electromagnetic}.
Now, by considering the charge conservation we can write:
\begin{align}
    \PD[]{\rho}{t}
    % =& \sum_{n=0}^{\infty}\PD[]{}{t} \rho_n(\Field[]{x},t)\\
    % =& \sum_{n=0}^{\infty} n(g(t))^{n-1}g^{\prime}(t) \varrho_n(\Field[]{x})\\
    % =& \sum_{n=0}^{\infty} n(g(t))^{n-1}g^{\prime}(t) \frac{q(-1)^n }{n!} \PD[n]{ \delta(\Field[]{x})}{z}\\
    \nonumber
    =& \sum_{n=0}^{\infty} n(g(t))^{n-1}g^{\prime}(t) \frac{q(-1)^n }{n!} \PD[n]{ \delta(\Field[]{x})}{z}\\
    \nonumber
    =& 0 -\PD[]{}{z} \sum_{n=1}^{\infty} (g(t))^{n-1}g^{\prime}(t) \frac{q(-1)^{n-1} }{(n-1)!} \PD[n-1]{ \delta(\Field[]{x})}{z}\\
    \nonumber
    =& 0 -\PD[]{}{z} \sum_{n=1}^{\infty} (g(t))^{n-1}g^{\prime}(t) \varrho_{n-1}(\Field[]{x})\\
    =& 0 -\PD[]{}{z} \sum_{n=1}^{\infty} g^{\prime}(t) \rho_{n-1}(\Field[]{x},t).
\end{align}
From this last expression, we can see that the current distribution $\Field[]{\jmath}(\Field[]{x},t)$ can be written as:
\begin{equation}
    \Field[]{\jmath}(\Field[]{x},t) = \sum_{n=0}^{\infty} \Field[n]{\jmath}(\Field[]{x},t), %= \Field[]{0}+ \sum_{n=1}^{\infty}  g^{\prime}(t)  \rho_{n-1}(\Field[]{x})\Field[z]{e}. 
\end{equation}
where $\Field[0]{\jmath}(\Field[]{x},t)=\Field[]{0}$ and $\Field[n]{\jmath}(\Field[]{x},t) = g^{\prime}(t)  \rho_{n-1}(\Field[]{x})\Field[z]{e}$.
 From the above expressions it is easy to see that there is a conservation of charge between the $n$-th current density $\Field[n]{\jmath}(\Field[]{x},t)$  and the $n$-th charge density $\rho_n(\Field[]{x},t)$ for $n\geq0$. %The proof is as follows
% \begin{align}
%     \nabla \cdot \Field[n-1]{\jmath}(\Field[]{x},t) & = g^{\prime}(t) (g(t))^{n-1} \PD[]{\rho_{n-1}(\Field[]{x})}{z}, \\
%     & = g^{\prime}(t) (g(t))^{n-1} \PD[]{}{z}\bigg( \frac{q(-1)^{n-1} }{(n-1)!} \PD[n-1]{ \delta(\Field[]{x})}{z}\bigg), \\
%     & = - \frac{1}{n}\PD[]{(g(t))^n}{t} \frac{q(-1)^{n} }{(n-1)!} \PD[n]{ \delta(\Field[]{x})}{z}, \\
%     & =  - \PD[]{}{t} \bigg( (g(t))^n\frac{q(-1)^{n} }{n!} \PD[n]{ \delta(\Field[]{x})}{z} \bigg), \\
%     & = - - \partial_t \rho_n(\Field[]{x},t).
% \end{align}
Thus one can consider the charge distribution pairs $(\rho_n(\Field[]{x},t),\Field[n]{\jmath}(\Field[]{x},t))$ in order to obtain the fields generated by an oscillating charge. In this case the retarded potentials are:
\begin{align}
    \label{Eq:phi_Raimond}
    \nonumber
    \phi(\Field[]{x},t) 
    & = \frac{q}{4\pi \eps_0} \sum_{n=0}^{\infty}  \frac{(-1)^n }{n!} \expval{\frac{\partial^{n} \delta(\Field[]{x}^{\prime})}{\partial z^{\prime n}}  ,\frac{(g(t-c^{-1}|| \Field[]{x} - \Field[]{x}^{\prime}||) )^n}{|| \Field[]{x} - \Field[]{x}^{\prime}||}  }, \\  
    & = \frac{q}{4\pi \eps_0} \sum_{n=0}^{\infty}  \frac{1}{n!}  \frac{\partial^{n} }{\partial z^{\prime n}}\bigg[\frac{(a\cos(\w t-k|| \Field[]{x} - \Field[]{x}^{\prime}||) )^n}{|| \Field[]{x} - \Field[]{x}^{\prime}||}  \bigg]_{\Field[]{x}^{\prime}=\Field[]{0}}.
\end{align}
In the same fashion, one can write the vectorial magnetic potential as:
\begin{align}
    \label{Eq:A_Raimond}
    \nonumber
    \Field[]{A}(\Field[]{x},t)
    & = \frac{q \mu_0}{4\pi} \sum_{n=1}^{\infty}  \frac{(-1)^{n-1} }{(n-1)!}  \expval{\frac{\partial^{n-1} \delta(\Field[]{x}^{\prime})}{\partial z^{\prime (n-1) }}    ,\frac{g^{\prime}(t-c^{-1}|| \Field[]{x} - \Field[]{x}^{\prime}||)(g(t-c^{-1}|| \Field[]{x} - \Field[]{x}^{\prime}||) )^n}{|| \Field[]{x} - \Field[]{x}^{\prime}||}} \Field[z]{e}\\
    & = \frac{q \mu_0}{4\pi} \sum_{n=1}^{\infty}  \frac{-\w a^2}{(n-1)!} \frac{\partial^{n-1} }{\partial z^{\prime (n-1) }} \bigg[\frac{\sin(\w t-k|| \Field[]{x} - \Field[]{x}^{\prime}||)(\cos(\w t-k|| \Field[]{x} - \Field[]{x}^{\prime}||) )^n}{|| \Field[]{x} - \Field[]{x}^{\prime}||} \bigg]_{\Field[]{x}^{\prime}=\Field[]{0}} \Field[z]{e}.
\end{align}
The potentials in equations \eqref{Eq:phi_Raimond} and \eqref{Eq:A_Raimond} are also expressed in terms of the present time $t$. However, the multiple derivatives that must be computed render the expressions impractical. One could argue that for a large value of $||\mathbf{x}||$, only a few terms are necessary to obtain a good approximation, as demonstrated in \cite{Raimond_PMC_2016} by retaining up to the dipole term. However, this approach would essentially involve a far-field approximation once again \cite{jackson_classical_1998}. In the following section, we will present a more practical and elegant method for representing an oscillating charge.

%%%%------------------------------------------%%%
\section{Harmonic decomposition of the sources}\label{Sec:Osc_Harm_Sour}

As we saw in the previous section, the \emph{Liénard-Wiechert} fields are not the best way to obtain the fields produced by an oscillating particle. The main problem is that the source terms depend on the trajectory that describes the charge. For this reason it will be convenient to find a way to decompose the source terms in a polyharmonic way. This idea has been previously considered by Landau \cite{landau1975classical} and, as evident from equations \eqref{Eq:phi_Raimond} and \eqref{Eq:A_Raimond}, also by Raimond \cite{Raimond_PMC_2016}, albeit indirectly. However, as demonstrated in the previous section, the expressions derived from Landau's and Raimond's ideas are challenging to implement in practice. Thus, a new approach for describing the sources is necessary.

In this section we propose another way inspired in quantum mechanics, which can be summarized as: \emph{The superposition of waves spread in a certain domain can be seen as a particle}. Physically, this means that a very localized source can be seen as the interference of a certain kind of waves. Mathematically speaking, we are looking to find a sequence such that, we can have convergence in the sense of distributions to a Dirac delta \cite{kreyszig1989introductory,reddy2013introductory, saichev1997distributions}. It is worth noting that a similar concept is employed in references \cite{luo2003cerenkov, lin2018controlling} for the case of a charged particle in uniform motion. 

\subsection{Two Fourier expansions for the sources}

Let us start our analysis by having the charge density $\rho$ and its current density $\Field[]{\boldsymbol{\jmath}}$ simply given by:
\begin{equation}\label{Eq:rho_def}
	 \rho(\Field[]{x}, t) = q \delta(\Field[\bot]{x})\otimes \varrho(z,t),\qquad \Field[\bot]{x} = (x,y)
\end{equation}
with $\varrho(z,t) := \delta(z-a\cos(\w_0 t))$ and
\begin{equation}\label{Eq:j_vec_def}
\Field[]{\boldsymbol{\jmath}}(\Field[]{x}, t) = q\delta(\Field[\bot]{x})\otimes \jmath(z,t)\Field[z]{e}, 
\end{equation}
with $\jmath(z,t) := -a\w_0 \sin(\w_0 t) \varrho(z,t)$ out of charge conservation. Notice that $\varrho$ and $\jmath$ are not multiplied by the charge. It then turns out that the charge density is not harmonic despite the harmonic motion of the particle. In other words, no complex function $\underline{\varrho}(z)$ can be found in such a way that $\varrho(z,t) = \RE \{\underline{\varrho}(z) e^{i\w t} \}$. \textbf{This is quite important to remark because in references as \cite{jackson_classical_1998,jentschura2017advanced} this is the starting point when representing and oscillating dipole}.
Nevertheless it is apropos to notice that the distribution $\varrho(z, \cdot)$ is a $\frac{2 \pi}{\w_0}$-periodic distribution. Thus, $\varrho$ as well as $\jmath(z,t)$ can be expanded as a Fourier series
(See Annex \ref{An:Compute-cl})~:
\begin{equation}
	\varrho(z,t) = \sum_{l \in \Z }\varrho_l^F(z,t), \quad \jmath(z,t) = \sum_{l \in \Z } \jmath_l^F(z,t),  
\end{equation}
with 
\begin{equation}
	\label{Eq:rho_F}
	\varrho_l^F(z,t) = w(z) T_l\bigg(\frac{z}{a}\bigg)e^{+il\w_0 t} 
\end{equation}
and
\begin{equation}
	\label{Eq:j_F}
	\boldsymbol{\jmath}_l^F(z,t) = -a\w_0 \sin(\w_0 t) w(z) T_l\bigg(\frac{z}{a}\bigg)e^{+il\w_0 t},
\end{equation}	
where $T_l$ are the Chebyshev polynomials of the first kind \cite{arfken2012mathematical} and $w(z)$ is the weight function 
\begin{equation}
	w(z):= \frac{1}{\pi \sqrt{a^2-z^2}} \chi_{[-a,a]}(z) \, 
\end{equation}
with $\chi_{[-a,a]}(z)$ a characteristic function.
%%%---Subsection---%%%
\subsection{Continuity equation for the harmonic components of the sources}

 In the previous paragraph, an expansion for $\rho$ and $\Field[]{\boldsymbol{\jmath}}$ are linked by the so-called charge conservation, namely: $\Div[]{\boldsymbol{\jmath}} + \dt{\rho} = 0$. Notice that for moving point particles this equation has to be understood in the sense of distributions \cite{petit_outil_1991,kreyszig1989introductory,reddy2013introductory, saichev1997distributions}. What about the different components $\varrho_l^F$ and $\jmath_l^F$? In other words, is there any transference of energy, between the different waves oscillating with the different frequencies at stake $\w_0$, $2\w_0$, etc$\dots$ ? To answer this question, we have to care much more about the notation referring this $l$.
 While $\varrho_{l}^{F}$ oscillates with frequency $l\w_0$, the scalar function $\jmath_l^F$ is a mix of two oscillations with different frequencies namely $(l-1)\w_0$ and $(l+1)\w_0$ due to the presence of $\sin(\w_0 t)$ term in \Eq{Eq:j_F}.
 Making use of the complex representation of $\sin(\w_0 t)$ it follows:
 \begin{equation}
 	\jmath(z,t) = \sum_{l \in \Z }\boldsymbol{\jmath}_l^F(z,t) = \sum_{l \in \Z } \frac{a \w_0}{2i} w(z) T_l(\frac{z}{a}) [e^{i(l-1)\w_0 t} - e^{i(l+1)\w_0 t}].
 \end{equation}
 As we can see, this representation is not in a convenient form, instead we would like to see each term in the series oscillating at frequency $l \omega_0$ (where $l$ is a dummy index), namely: 
\begin{equation}\label{Eq:Final_j}
   	\jmath(z,t) = \sum_{l \in \Z} 	\jmath_l^T(z) e^{+il\w_0 t}.	
\end{equation}
This can be easily achieved after renaming indices ($l+1 \rightarrow l$ and $l-1 \rightarrow l$) for the corresponding terms $e^{i(l-1)\w_0 t}$ and $e^{i(l+1)\w_0 t}$ which allows to obtain: 
\begin{equation}
	\label{Eq:J_l_T}
	 \boldsymbol{\jmath}_l^T(z) := \frac{a \w_0}{2i}[ \textit{$\xi$}_{l+1}(z) - \textit{$\xi$}_{l-1}(z)],
\end{equation}
where $\textit{$\xi$}_l(z)= w(z) T_l(\frac{z}{a})$. Analogously for $\rho(z,t)$ we get: 
\begin{equation}\label{Eq:rho_l_T}
	\rho(z,t) = \sum_{l \in \Z}\varrho_l^F(z,t) = \sum_{l \in \Z}\varrho_l^T(z) e^{+il\w_0 t} = \sum_{l \in \Z}\textit{$\xi$}_l(z) e^{+il\w_0 t}.	
\end{equation}
The arcane meaning of the superscripts $T$ and $F$ is therefore clear: $T$ (resp. $F$) means true (resp. false) in the sense that each spatial coefficient corresponds to only one $l \in \Z $.
Correspondingly the conservation of charge can be now formulated for each multiple of the frequency $\w_0$.
By the definition of $\rho(\Field[]{x},t)$ and $\Field[]{\boldsymbol{\jmath}}(\Field[]{x},t)$ The conservation of charge implies:
\begin{equation}
      \partial_t \varrho(z,t) + \partial_z \jmath(z,t) = 0, 
\end{equation}
and plugging the harmonic expansions of $\varrho(z,t)$ and $\jmath(z,t)$ one gets:
\begin{equation}
      \sum_{l \in \Z} \bigg[ il\w_0 \textit{$\xi$}_l(z) + \frac{a \w_0}{2i} \partial_z( \textit{$\xi$}_{l+1}(z) - \textit{$\xi$}_{l-1}(z)) \bigg]e^{+il\w_0 t} = 0. 
\end{equation}
Given the fact that $e^{+il\w_0 t}$ with $l \in \Z$ is a basis \cite{kreyszig1989introductory, reddy2013introductory} for our polyharmonic decomposition, all the terms between square brackets are equal to zero. Thus the following identity is obtained:
\begin{equation}
 l \textit{$\xi$}_l(z) =  \frac{a}{2} \partial_z( \textit{$\xi$}_{l+1}(z) - \textit{$\xi$}_{l-1}(z)). 
\end{equation}
And the conservation of charge for each $l$-th term in the expressions \Eq{Eq:Final_j} and \Eq{Eq:rho_l_T} follows. Upon demonstrating that there is no transfer among the distinct harmonic components of the charge density $\rho(\Field[]{x},t)$ and the current density $\Field[]{\boldsymbol{\jmath}}(\Field[]{x},t)$, we may proceed to employ these Fourier expansions to derive the electric and magnetic induction fields for an oscillating charged particle.This will be show in the following section. 
%
%------------------------------------------------------------------------------------------------%
\section{The building of \textbf{E} and \textbf{B} via polyharmonic computations}\label{Sec:Osc_Polyharm}
The main consequence of the harmonic decomposition of the sources and the conservation of charge term by term is that the electric and magnetic induction fields, respectively $\Field[]{E}$ and $\Field[]{B}$, can be seen as a superposition of fields $\Field[l]{E}(\Field[]{x},t)$ and $\Field[l]{B}(\Field[]{x},t)$, respectively. That is, fields generated by the harmonic densities of charge $\varrho_l^T(z) e^{+il\w_0 t}$ and current $\jmath_l^T(z)e^{+il\w_0 t}$. Each one of these fields oscillates in terms of multiples of the fundamental frequency $\w_0$. This section is devoted to this issue starting to work with equations (\ref{Eq:B_int}-\ref{Eq:B_rad}) and (\ref{Eq:E_c}-\ref{Eq:E_rad}), considering $\Field[]{R} = \Field[]{x}-\Field[]{x^{\prime}}$, $R = |\Field[]{R}|$ and $t_r = t-\frac{R}{c}$ the retarded time. It is important to remark here that \textbf{in this case the retarded time is not in terms of a transcendental equations but rather explicitly given in terms of the present time $t$}.

\subsection{The Fourier expansion of the Fields \textbf{E} and \textbf{B}}

\textst{After} We start by applying the delta distribution $\delta(\Field[\bot]{x}^{\prime})$ from equations \Eq{Eq:rho_def} and \Eq{Eq:j_vec_def} into the expressions (\ref{Eq:B_int}-\ref{Eq:B_rad}) and (\ref{Eq:E_c}-\ref{Eq:E_rad}), which implies \textst{we get} that $\Field[]{E}$ and $\Field[]{B}$ are given by:
\begin{align}
	\nonumber
	\Field[]{E}(\Field[]{x},t) =& \frac{1}{4\pi\epsilon_0} \int_{\mathbb{R}} \frac{q}{c \tilde{R}^3} \bigg[ \frac{\jmath(z^{\prime}, \tilde{t}_r)}{\tilde{R}} + \frac{\partial_t \jmath(z^{\prime}, \tilde{t}_r)}{c}  \bigg](\Field[z]{e}\times\Field[]{\tilde{R}})\times\Field[]{\tilde{R}}  d z^{\prime}\\	
	 &+\frac{1}{4\pi\epsilon_0} \int_{\mathbb{R}} \frac{q}{\tilde{R}^3}\bigg[\varrho(z^{\prime}, \tilde{t}_r) + \jmath(z^{\prime}, \tilde{t}_r) \frac{\Field[z]{e}\cdot \Field[]{\tilde{R}} }{\tilde{R} c} \bigg] \Field[]{\tilde{R} }  dz^{\prime},\\
	\Field[]{B}(\Field[]{x},t) =& \frac{\mu_0}{4\pi} \int_{\mathbb{R}} \frac{q}{\tilde{R}^2} \bigg[ \frac{\jmath(z^{\prime}, \tilde{t}_r)}{\tilde{R}} + \frac{\partial_t \jmath(z^{\prime}, \tilde{t}_r)}{c}  \bigg]\Field[z]{e}\times\Field[]{\tilde{R}}  d z^{\prime},
\end{align}
where $\Field[]{\tilde{R}} = \Field[]{x}-z^{\prime}\Field[z]{e}$, $\tilde{R} = |\Field[]{\tilde{R}}|$ and $\tilde{t}_r = t-\frac{\tilde{R}}{c}$. The next step is to define the functions:
\begin{align}
	\label{Eq:Q_harm}
	Q(z^{\prime},\Field[]{\tilde{R}}, \tilde{t}_r) &:= \varrho(z^{\prime}, \tilde{t}_r) + \boldsymbol{\jmath}(z^{\prime}, \tilde{t}_r) \frac{\Field[z]{e}\cdot \Field[]{\tilde{R}} }{\tilde{R} c},\\
	\label{Eq:K_harm}
	K(z^{\prime},\Field[]{\tilde{R}}, \tilde{t}_r) &:= \frac{\jmath(z^{\prime}, \tilde{t}_r)}{\tilde{R}} + \frac{\partial_t \jmath(z^{\prime}, \tilde{t}_r)}{c}.
\end{align}
And then the electric and magnetic fields can be written in a more compact way as:
\begin{align}
	\Field[]{E}(\Field[]{x},t) =& \frac{1}{4\pi\epsilon_0} \int_{\mathbb{R}} \frac{q}{c\tilde{R}^3} \, \mathbf{F}_{\mathbf{E}}(\mathbf{x},\mathbf{x}') d z^{\prime},
\end{align}
with
\begin{equation}
\mathbf{F}_{\mathbf{E}}(\mathbf{x},\mathbf{x}') := \bigg( cQ(z^{\prime},\tilde{R}, \tilde{t}_r)  \Field[]{\tilde{R}} + K(z^{\prime},\tilde{R}, \tilde{t}_r)(\Field[z]{e}\times\Field[]{\tilde{R}})\times\Field[]{\tilde{R}} \bigg)
\end{equation}
and
\begin{align}
	\Field[]{B}(\Field[]{x},t) =& \frac{\mu_0}{4\pi} \int_{\mathbb{R}} \frac{q}{\tilde{R}^2} K(z^{\prime},\tilde{R}, \tilde{t}_r) \Field[z]{e} \times \Field[]{\tilde{R}} d z^{\prime}.
\end{align}
From the definitions of $\varrho(z,t)$ in \ref{Eq:Final_varrho} and $\boldsymbol{\jmath}(z,t)$ in \ref{Eq:Final_j} we have that equations (\ref{Eq:Q_harm}) and (\ref{Eq:K_harm}) read:
\begin{align}
	Q(z^{\prime},\Field[]{\tilde{R}}, \tilde{t}_r) &= \sum_{l \in \Z } e^{+il \w_0 t}  Q_l(z^{\prime},\Field[]{\tilde{R}}),\\
	K(z^{\prime},\Field[]{\tilde{R}}, \tilde{t}_r) &= \sum_{l \in \Z } e^{+il \w_0 t}  K_l(z^{\prime},\Field[]{\tilde{R}}),
\end{align}
where
\begin{align}
	\label{Eq:Q_l_harm}
	&  Q_l(z^{\prime},\Field[]{\tilde{R}}) :=   \bigg[\varrho_l^T(z^{\prime}) + \jmath_l^T(z^{\prime}) \frac{\Field[z]{e}\cdot \Field[]{\tilde{R}}}{\tilde{R} c}\bigg] e^{-ilk_0\tilde{R}}, \qquad k_0= \frac{\w_0}{c}\\
	\label{Eq:K_l_harm}
	&  K_l(z^{\prime},\Field[]{\tilde{R}}) :=   \bigg[\frac{1}{\tilde{R}} + \frac{il\w_0}{c} \bigg]\jmath_l^T(z^{\prime}) e^{-ilk_0\tilde{R}}. 
\end{align}
Notice that in these expressions there a phase shift $e^{-ilk_0\tilde{R}}$ due to the retarded time $t_r$. Therefore $\Field[]{E}(\Field[]{x},t)$ and $\Field[]{B}(\Field[]{x},t)$ can be seen as a superposition of elementary harmonic terms, \emph{i.e.}:  
\begin{align}
	\label{Eq:E_0_sum}
	&  \Field[]{E}(\Field[]{x},t)=\sum_{l \in \Z } \Field[l]{E}(\Field[]{x},t) = \sum_{l \in \Z }e^{+il \w_0 t} \Field[l]{E}(\Field[]{x}), \\
	\label{Eq:B_0_sum}
	& \Field[]{B}(\Field[]{x},t) =\sum_{l \in \Z } \Field[l]{B}(\Field[]{x},t) =\sum_{l \in \Z }e^{+il \w_0 t} \Field[l]{B}(\Field[]{x}), 
\end{align}
with spatially dependent coefficients given by:
\begin{align}
	\label{Eq:E_harm_final}
	& \Field[l]{E}(\Field[]{x}) = \frac{1}{4\pi\epsilon_0} \int_{[-a,a]} \frac{q}{c\tilde{R}^3}\bigg( cQ_l(z^{\prime},\tilde{R})  \Field[]{\tilde{R}} + K_l(z^{\prime},\tilde{R})(\Field[z]{e}\times\Field[]{\tilde{R}})\times\Field[]{\tilde{R}} \bigg)d z^{\prime},\\
	\label{Eq:B_harm_final}
	&  \Field[l]{B}(\Field[]{x}) = \frac{\mu_0}{4\pi} \int_{[-a,a]} \frac{q}{\tilde{R}^2} K_l(z^{\prime},\tilde{R}) \Field[z]{e} \times \Field[]{\tilde{R}} d z^{\prime}.
\end{align}
where we have used the fact that the support of $\varrho_l^T(z^{\prime})$ and $\jmath_l^T(z^{\prime})$ is within the interval $[-a,a]$. These coefficients can be computed numerically, and thus the building of the $\Field[]{E}$ and $\Field[]{B}$ fields is complete.

\subsection{A geometrical description of the fields}

\begin{figure}[htbp]
\centering
\begin{subfigure}{\textsubfigure}
\includegraphics[width=\linewidth]{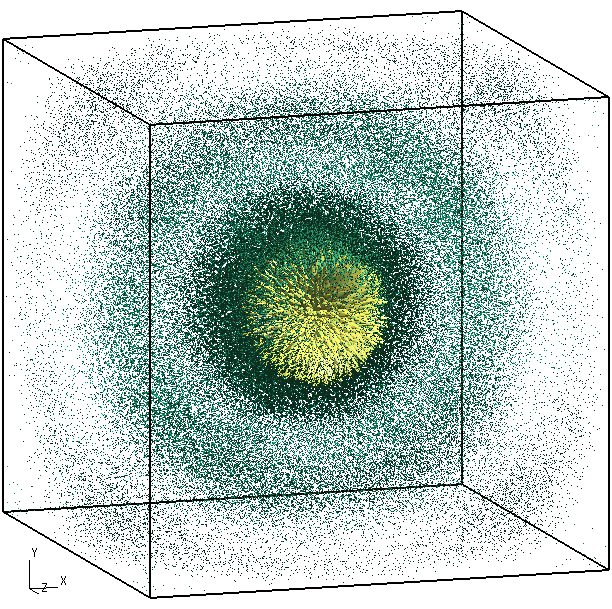}
\captionsetup{justification=centering}
\caption{$\w_0$} \label{fig:0a}
\end{subfigure}
\hspace*{\fill} % separation between the subfigures
\begin{subfigure}{\textsubfigure}
\includegraphics[width=\linewidth]{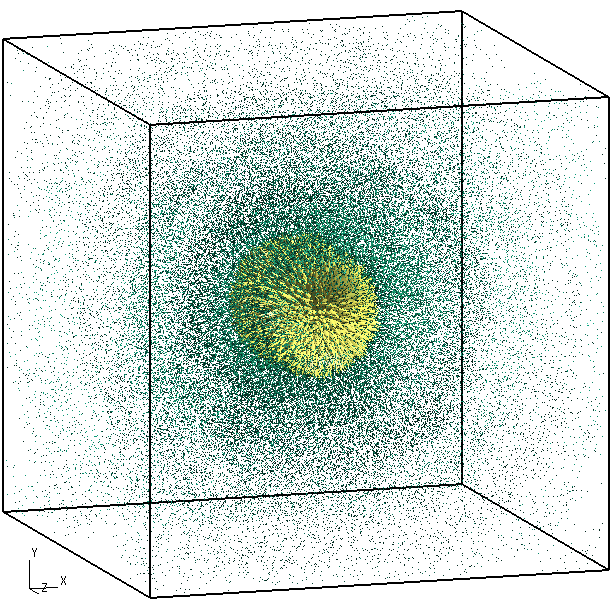}
\captionsetup{justification=centering}
\caption{$2\w_0$} \label{fig:0b}
\end{subfigure}

\centering
\begin{subfigure}{\textsubfigure}
\includegraphics[width=\linewidth]{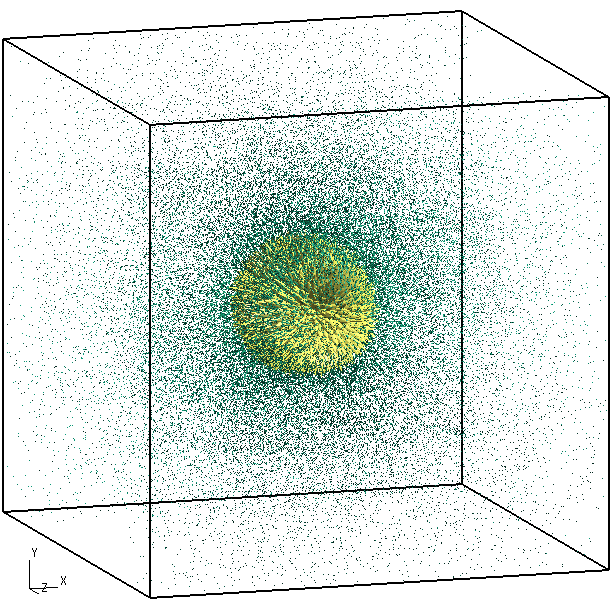}
\captionsetup{justification=centering}
\caption{$3\w_0$} \label{fig:0c}
\end{subfigure}
\hspace*{\fill} % separation between the subfigures
\centering
\begin{subfigure}{\textsubfigure}
\includegraphics[width=\linewidth]{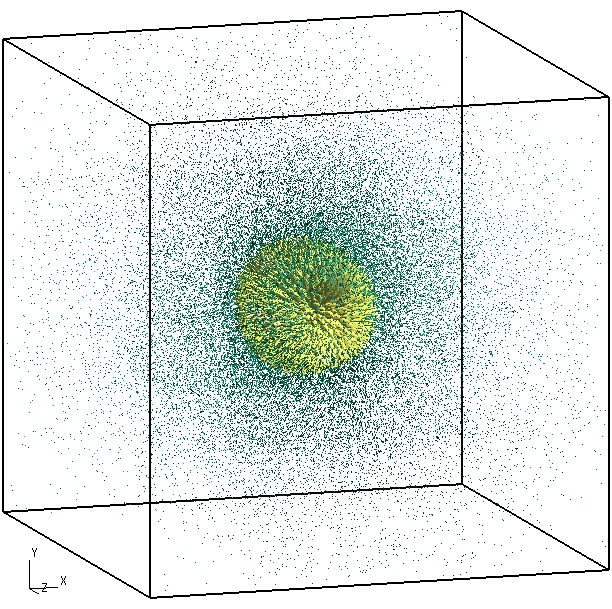}
\captionsetup{justification=centering}
\caption{$4\w_0$} \label{fig:0d}
\end{subfigure}
\captionsetup{justification=centering}
\caption{Harmonic components of the electric field generated by an oscillating particle, $\Field[l]{E}$ for l = $1,2,3,4$} \label{fig:0}
\end{figure}

\begin{figure}[htbp]
\centering
\begin{subfigure}{\textsubfigure}
\includegraphics[width=\linewidth]{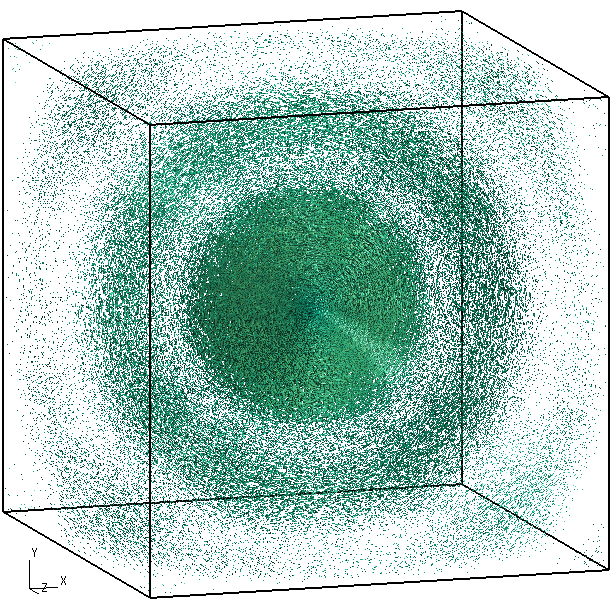}
\captionsetup{justification=centering}
\caption{$\w_0$} \label{fig:1a}
\end{subfigure}
\hspace*{\fill} % separation between the subfigures
\begin{subfigure}{\textsubfigure}
\includegraphics[width=\linewidth]{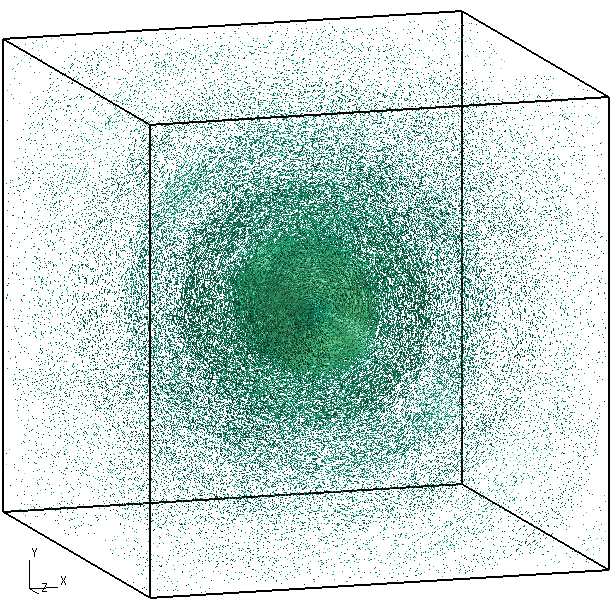}
\captionsetup{justification=centering}
\caption{$2\w_0$} \label{fig:1b}
\end{subfigure}

\centering
\begin{subfigure}{\textsubfigure}
\includegraphics[width=\linewidth]{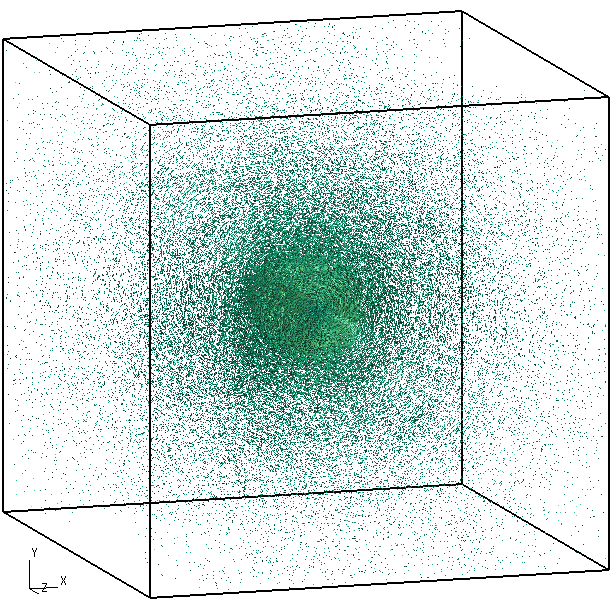}
\captionsetup{justification=centering}
\caption{$3\w_0$} \label{fig:1c}
\end{subfigure}
\hspace*{\fill} % separation between the subfigures
\centering
\begin{subfigure}{\textsubfigure}
\includegraphics[width=\linewidth]{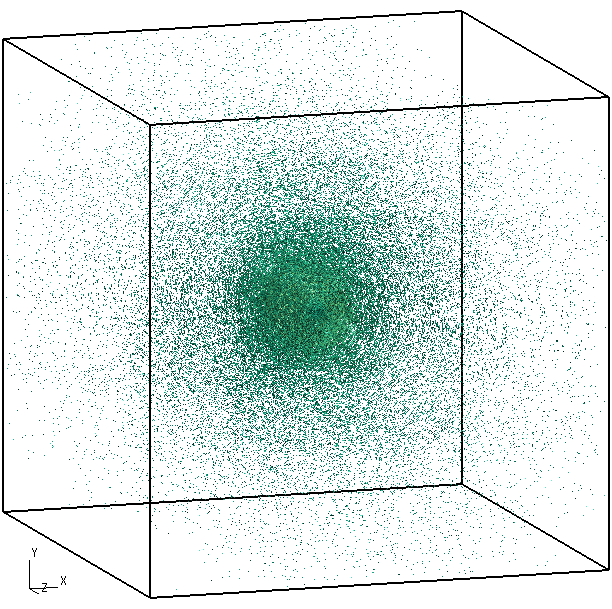}
\captionsetup{justification=centering}
\caption{$4\w_0$} \label{fig:1d}
\end{subfigure}
\captionsetup{justification=centering}
\caption{Harmonic components of the real part of magnetic induction field generated by an oscillating particle, $\Field[l]{B}$ for l = $1,2,3,4$ (cut in canonical planes view)} \label{fig:1}
\end{figure}

\begin{figure}[htbp]
\centering
\begin{subfigure}{\textsubfigure}
\includegraphics[width=\linewidth]{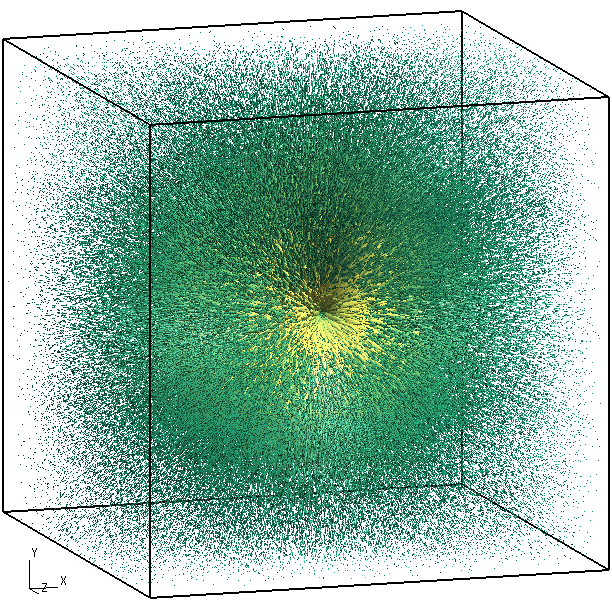}
\captionsetup{justification=centering}
\caption{$\w_0$} \label{fig:2a}
\end{subfigure}
\hspace*{\fill} % separation between the subfigures
\begin{subfigure}{\textsubfigure}
\includegraphics[width=\linewidth]{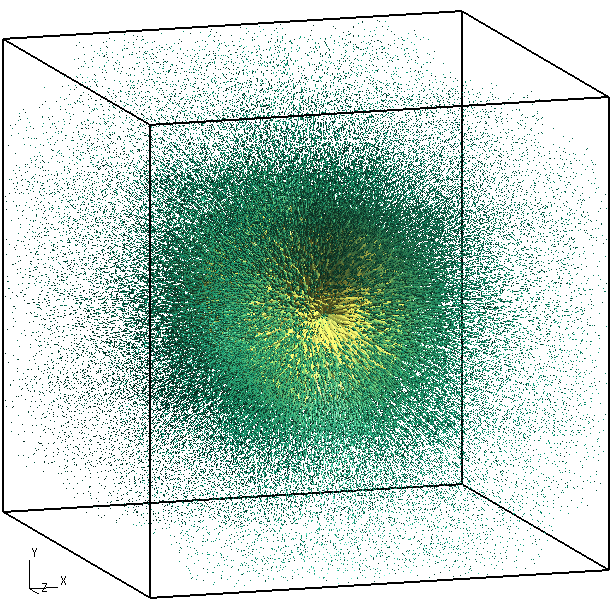}
\captionsetup{justification=centering}
\caption{$2\w_0$} \label{fig:2b}
\end{subfigure}

\centering
\begin{subfigure}{\textsubfigure}
\includegraphics[width=\linewidth]{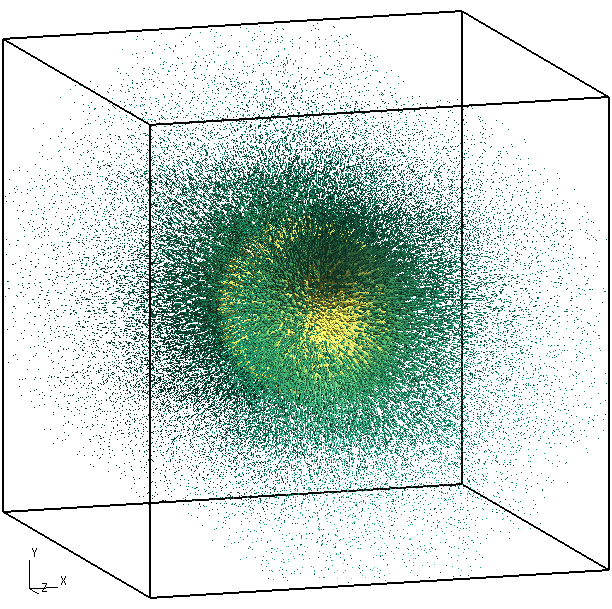}
\captionsetup{justification=centering}
\caption{$3\w_0$} \label{fig:2c}
\end{subfigure}
\hspace*{\fill} % separation between the subfigures
\centering
\begin{subfigure}{\textsubfigure}
\includegraphics[width=\linewidth]{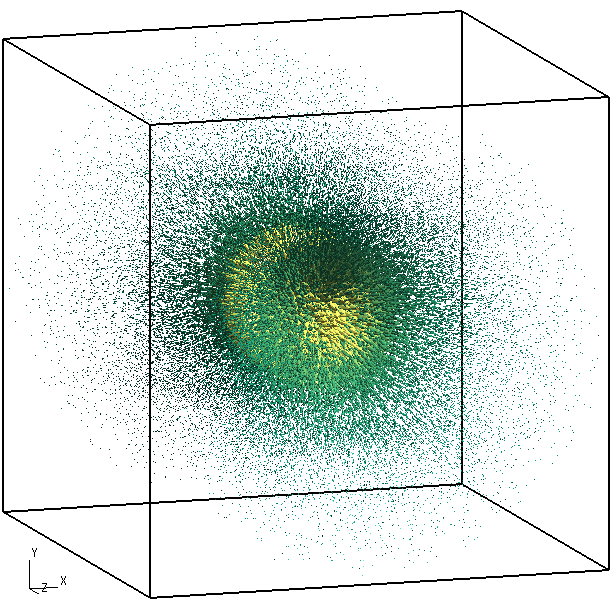}
\captionsetup{justification=centering}
\caption{$4\w_0$} \label{fig:2d}
\end{subfigure}
\captionsetup{justification=centering}
\caption{Harmonic components of the real part of Poynting vector field generated by an oscillating particle, $\Field[l]{S}$ for l = $1,2,3,4$ (cut in canonical planes view)} \label{fig:2}
\end{figure}

\begin{figure}[htbp]
\centering
\begin{subfigure}{\textsubfigure}
\includegraphics[width=\linewidth]{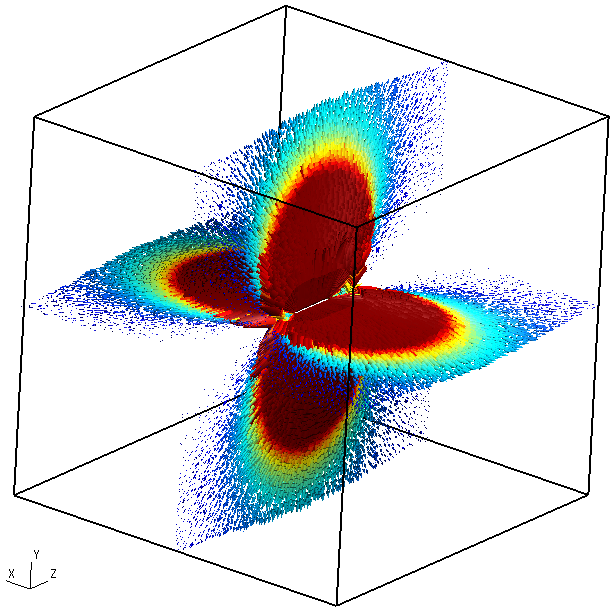}
\captionsetup{justification=centering}
\caption{$\w_0$} \label{fig:22a}
\end{subfigure}
\hspace*{\fill} % separation between the subfigures
\begin{subfigure}{\textsubfigure}
\includegraphics[width=\linewidth]{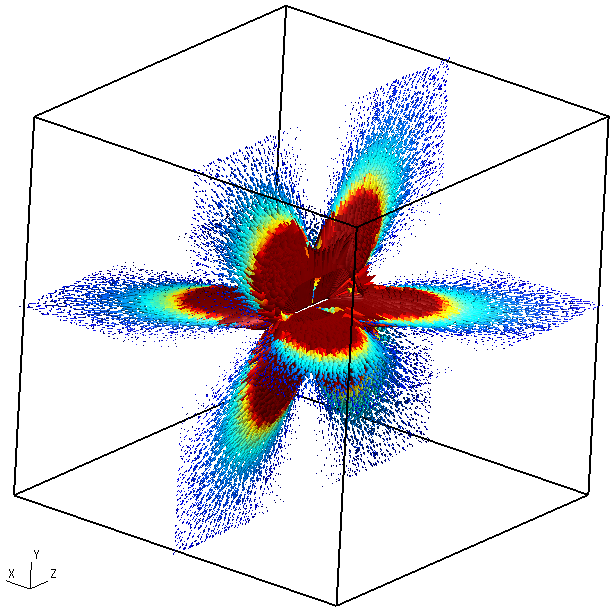}
\captionsetup{justification=centering}
\caption{$2\w_0$} \label{fig:22b}
\end{subfigure}

\centering
\begin{subfigure}{\textsubfigure}
\includegraphics[width=\linewidth]{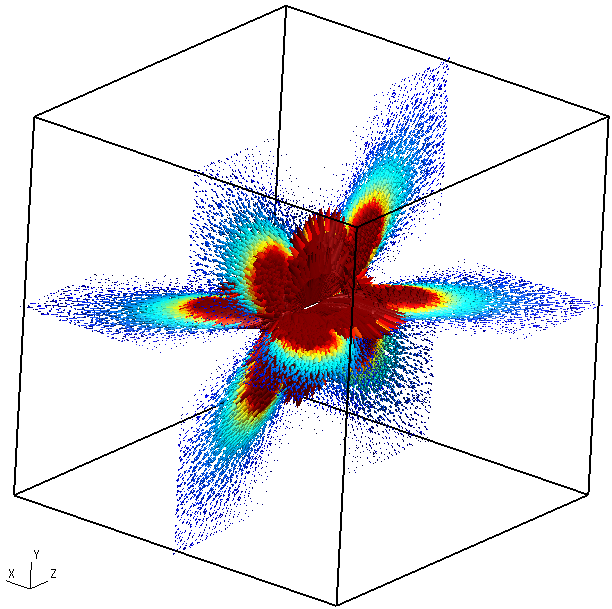}
\captionsetup{justification=centering}
\caption{$3\w_0$} \label{fig:22c}
\end{subfigure}
\hspace*{\fill} % separation between the subfigures
\centering
\begin{subfigure}{\textsubfigure}
\includegraphics[width=\linewidth]{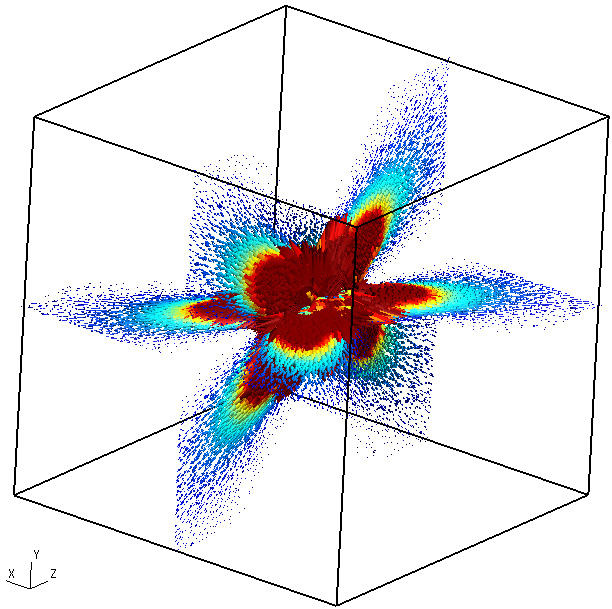}
\captionsetup{justification=centering}
\caption{$4\w_0$} \label{fig:22d}
\end{subfigure}
\captionsetup{justification=centering}
\caption{Harmonic components of the real part of Poynting vector field generated by an oscillating particle, $\Field[l]{S}$ for l = $1,2,3,4$ (cut in canonical planes view)} \label{fig:22}
\end{figure}

Despite the complicated appearance of equations (\ref{Eq:E_harm_final}-\ref{Eq:B_harm_final}) it is posible to extract some \emph{a priori} information about the geometric behaviour of the ($\Field[]{E}$,$\Field[]{B}$) fields. Starting by using the vector identity $(\Field[z]{e}\times\Field[]{\tilde{R}})\times\Field[]{\tilde{R}} = (\Field[z]{e}\cdot \Field[]{\tilde{R}})\Field[]{\tilde{R}}-\tilde{R}^2 \Field[z]{e}$ we can rewrite the spatial coefficients as per:
\begin{align}
	\nonumber
	\Field[l]{E}(\Field[]{x}) =& \displaystyle \frac{1}{4\pi\epsilon_0} 
	\bigg\{ \int_{\mathbb{R}} \frac{q}{c \tilde{R}^3} \Phi_l^{sph}(z',\Field[]{x}) \,  d z^{\prime}  \Field[]{x}\\
	&- \int_{\mathbb{R}} \frac{q}{c \tilde{R}^3} \Phi_l^{flat}(z',\Field[]{x}) \,  d z^{\prime} \Field[z]{e} \bigg\}, 
\end{align}
where $\Field[]{E}^{sph}$ and $\Field[]{E}^{flat}$ are defined by
\begin{equation}
\Phi_l^{flat}(z',\Field[]{x}) := -   z^{\prime}\big(c Q_l(z^{\prime},\tilde{R}) + (z-z^{\prime}) K_l(z^{\prime},\tilde{R})\big) + K_l(z^{\prime},\tilde{R}) \tilde{R}^2 
\end{equation}
and
\begin{equation}
\Phi_l^{sph}(z',\Field[]{x}) :=cQ_l(z^{\prime},\tilde{R}) + (z-z^{\prime}) K_l(z^{\prime},\tilde{R}) 
\end{equation}
\begin{align}
	\Field[l]{B}(\Field[]{x}) =& \frac{\mu_0}{4\pi} \int_{\mathbb{R}} \frac{q}{\tilde{R}^2} K_l(z^{\prime},\tilde{R}) \Field[z]{e} \times \Field[\perp]{x} d z^{\prime},
\end{align}
From this representation it is easy to see that the first integral term of $\Field[]{E}$, which will be called $\Field[]{E}^{sph}$, is a field with spherical symmetry. However due to the action of the second integral term, in the sequel $\Field[]{E}^{flat}$, the total field is flattened out in the direction perpendicular to the motion. On the other hand, the field lines of $\Field[]{B}$ circle around the $z$-axis and, as expected, are perpendicular to the field lines of the electric field. Figure \ref{fig:0} and ( resp.\ref{fig:1}) show imaginary (resp. real) part of the the harmonic field components $\Field[l]{E}$  (resp. $\Field[l]{B}$) for $l=1,2,3,4$. Whereas figure \ref{fig:2} represents the real part of the Poynting vector $\Field[l]{S} = \frac{1}{2 \mu_0}\Field[l]{E} \times \Field[l]{B} $. Albeit the electric and magnetic fields seem to be more or less the same, the figures that show the fields at the canonical planes $y=0$ and $x=0$ reveal a quite different behavior, that is: the electric field shows the expected geometrical behavior (this is more evident for fig. \ref{fig:0d}) and the magnetic field circles around the $z$-axis (see for instance fig. \ref{fig:1a} ). Finally, the projection of the Poynting vector field on the canonical planes is shown in figure \ref{fig:22}.
%----------------------Figure-----------%
%----------------------Figure-----------%
\begin{figure}[htbp]
\centering
\begin{subfigure}{\textsubfigure}
\includegraphics[width=\linewidth]{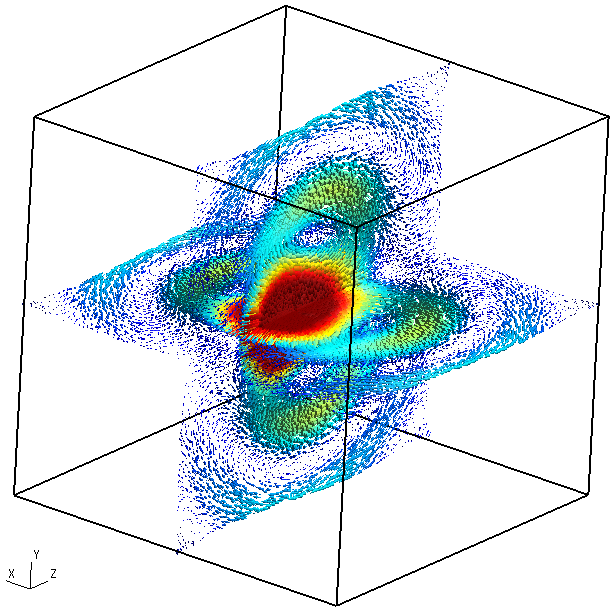}
\captionsetup{justification=centering}
\caption{$\w_0$} \label{fig:00a}
\end{subfigure}
\hspace*{\fill} % separation between the subfigures
\begin{subfigure}{\textsubfigure}
\includegraphics[width=\linewidth]{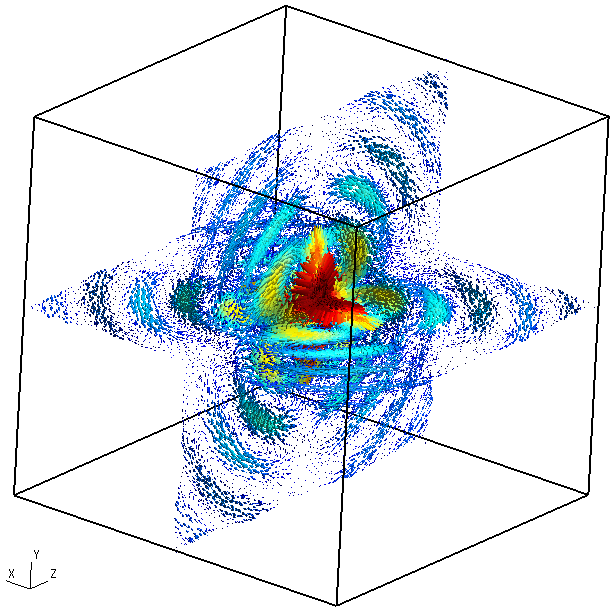}
\captionsetup{justification=centering}
\caption{$2\w_0$} \label{fig:00b}
\end{subfigure}

\centering
\begin{subfigure}{\textsubfigure}
\includegraphics[width=\linewidth]{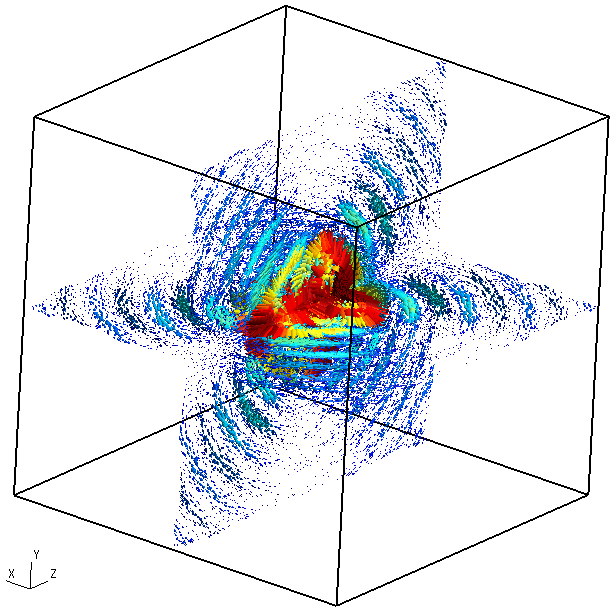}
\captionsetup{justification=centering}
\caption{$3\w_0$} \label{fig:00c}
\end{subfigure}
\hspace*{\fill} % separation between the subfigures
\centering
\begin{subfigure}{\textsubfigure}
\includegraphics[width=\linewidth]{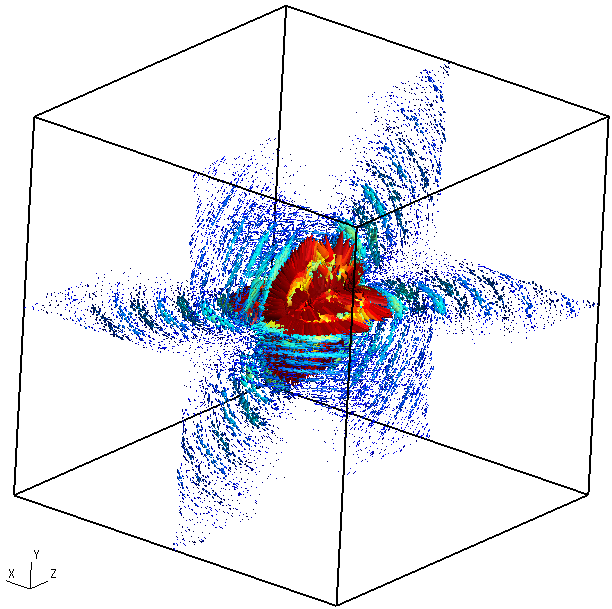}
\captionsetup{justification=centering}
\caption{$4\w_0$} \label{fig:00d}
\end{subfigure}
\captionsetup{justification=centering}
\caption{Harmonic components of the electric field generated by an oscillating particle, $\Field[l]{E}$ for l = $1,2,3,4$ (cut in canonical planes view)} \label{fig:00}
\end{figure}

%----------------------Figure-----------%

%----------------------Figure-----------%
\begin{figure}[htbp]
\centering
\begin{subfigure}{\textsubfigure}
\includegraphics[width=\linewidth]{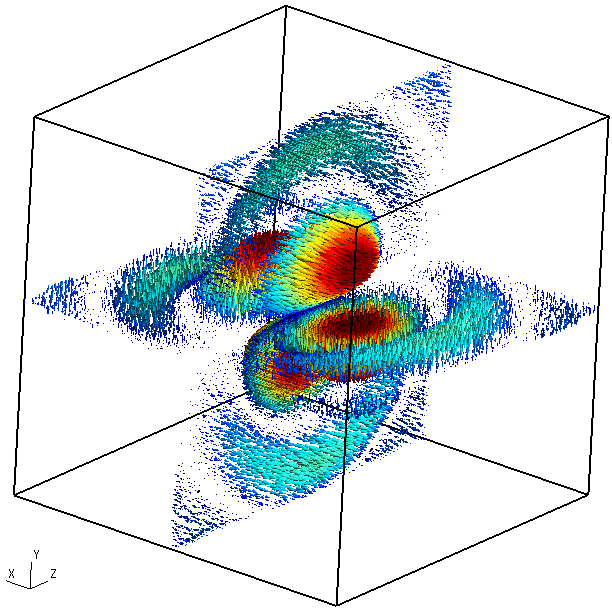}
\captionsetup{justification=centering}
\caption{$\w_0$} \label{fig:11a}
\end{subfigure}
\hspace*{\fill} % separation between the subfigures
\begin{subfigure}{\textsubfigure}
\includegraphics[width=\linewidth]{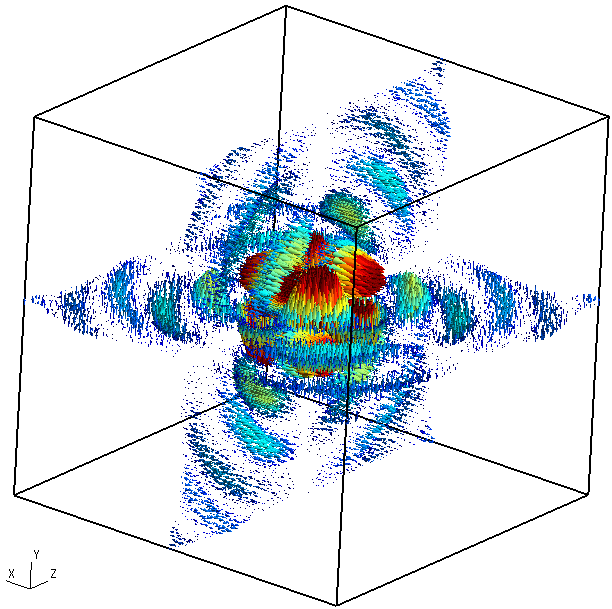}
\captionsetup{justification=centering}
\caption{$2\w_0$} \label{fig:11b}
\end{subfigure}

\centering
\begin{subfigure}{\textsubfigure}
\includegraphics[width=\linewidth]{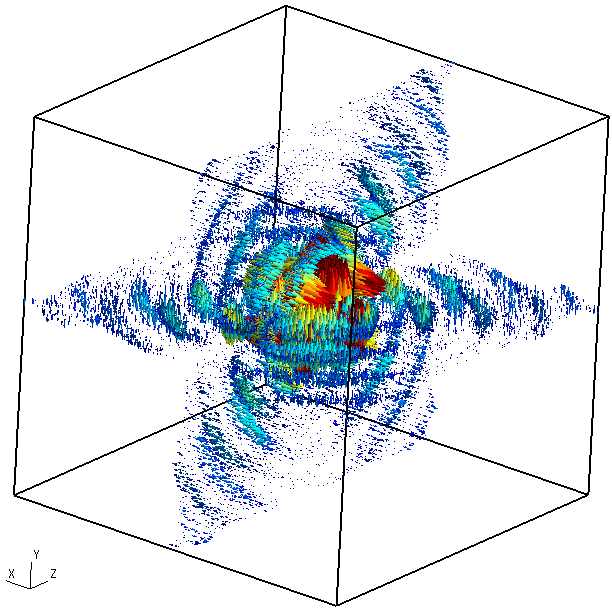}
\captionsetup{justification=centering}
\caption{$3\w_0$} \label{fig:11c}
\end{subfigure}
\hspace*{\fill} % separation between the subfigures
\centering
\begin{subfigure}{\textsubfigure}
\includegraphics[width=\linewidth]{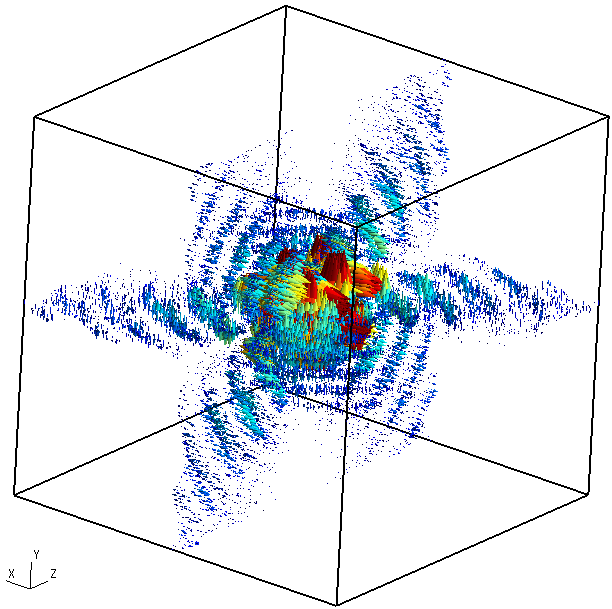}
\captionsetup{justification=centering}
\caption{$4\w_0$} \label{fig:11d}
\end{subfigure}
\captionsetup{justification=centering}
\caption{Harmonic components of the real part of magnetic induction field generated by an oscillating particle, $\Field[l]{B}$ for l = $1,2,3,4$ (cut in canonical planes view)} \label{fig:11}
\end{figure}
%----------------------Figure-----------%
%----------------------Figure-----------%
%%%%%%SECTION%%%%%%%%%%%%%%%%%%%%%%%%%%%%%%%%%%%%%%%%%%%%%%%%%%%%%%%%%%%%%%%%%%%%%%%%%%%%%%%%%%%%%%%%%%%%%%%%%%%%%%%%%%%
%\newpage
\section{Far Field radiated power}\label{Sec:Far_Field_Power}

As it was mentioned before, the main approach towards the study of the EM field generated by an oscillating charge requires to consider the \emph{far field approximation}. In this section we will demonstrate how, by using the polyharmonic expression from previous sections, it is possible to retrive the same results as reported in the literature \cite{jackson_classical_1998}.

\subsection{The polyharmonic representation of the EM far field approximations}

As we know, the radiated electric and magnetic induction fields are given by:
\begin{align}
	&\Field[rad]{B} = q\frac{\mu_0}{4 \pi c}\int_{\mathbb{R}^3} \frac{\partial_t \Field[]{\boldsymbol{\jmath}}(\Field[]{x}^{\prime}, t_r) \times \Field[]{n}}{R}   \,\rmd\Field[]{x}^{\prime},\\
	&\Field[rad]{E} = q\frac{1}{4\pi \eps_0 c^2} \int_{\mathbb{R}^3} \frac{(\partial_t \Field[]{\boldsymbol{\jmath}}(\Field[]{x}^{\prime},t_r)\times \Field[]{n})\times \Field[]{n}}{R} \,\rmd\Field[]{x}^{\prime}.
\end{align}
where $\Field[]{n} = \frac{\Field[]{R}}{R}$. Using the fact that the electric current is defined as $\Field[]{\boldsymbol{\jmath}}(\Field[]{x}^{\prime}, t_r) := q\delta(\Field[\bot]{x})\otimes \boldsymbol{\jmath}(z,t_r)\Field[z]{e}$, the fields read:
\begin{align} 
	\label{Eq:radiated_B}
	&\Field[rad]{B} = q\frac{\mu_0}{4 \pi c}\int_{\mathbb{R}} \frac{\partial_t \boldsymbol{\jmath}(\Field[]{x}^{\prime}, \tilde{t}_r) \Field[z]{e}\times \Field[]{\tilde{n} }}{\tilde{R}}   dz^{\prime},\\
	\label{Eq:radiated_E}
	&\Field[rad]{E} = q\frac{1}{4\pi \eps_0 c^2} \int_{\mathbb{R}} \frac{(\partial_t \Field[]{\boldsymbol{\jmath}}(\Field[]{x}^{\prime},\tilde{t}_r)\times \Field[]{\tilde{n} })\times \Field[]{\tilde{n} }}{\tilde{R}} dz^{\prime}.
\end{align}
and as we did before $\Field[]{\tilde{R}} = \Field[]{x}-z^{\prime}\Field[z]{e}$, $\Field[]{\tilde{n}} = \frac{\Field[]{\tilde{R}}}{\tilde{R}}$,  and  $\tilde{t}_r = t - \frac{\tilde{R}}{c}$. Now, in order to compute the fields radiated \emph{at infinitum} we make the following approximations for when $|\Field[]{x}^{\prime}| \ll |\Field[]{x}|$:
\begin{align}
	& \Field[]{\tilde{n}} \simeq \frac{\Field[]{x}}{|\Field[]{x}|} = \boldsymbol{\nu},\\
	& \frac{1}{\tilde{R}}  \simeq \frac{1}{|\Field[]{x}|},\\
    \label{Eq:tau_r}
	& \tilde{t}_r         \simeq t - \frac{|\Field[]{x}|}{c} + \bigg(\frac{z}{|\Field[]{x}|}\bigg) \bigg(\frac{z^{\prime}}{c}\bigg) = \tau_r.
\end{align}
By plugging these approximations into equations (\ref{Eq:radiated_B}) and (\ref{Eq:radiated_E}), one arrives to the expressions:
\begin{align} 
	\label{Eq:inf_B}
	\Field[rad]{B} \simeq \Field[\infty]{B} =& q\frac{\mu_0}{4 \pi c |\Field[]{x}|}\int_{\mathbb{R}} \partial_t \boldsymbol{\jmath}(z^{\prime}, \tau_r)  dz^{\prime} \Field[z]{e}\times \boldsymbol{\nu}  ,\\
	\label{Eq:inf_E}
	\nonumber
	\Field[rad]{E} \simeq \Field[\infty]{E} =& q\frac{1}{4\pi \eps_0 c^2 |\Field[]{x}|} \int_{\mathbb{R}} \partial_t \boldsymbol{\jmath}(z^{\prime}, \tau_r)  dz^{\prime} (\Field[z]{e}\times \boldsymbol{\nu})\times \boldsymbol{\nu}, \\
		=& c \Field[\infty]{B} \times \boldsymbol{\nu}.
\end{align}
From this last equation it is clear that we can focus our attention just in $\Field[\infty]{B}$. Therefore, the next step is to consider its polyharmonic representation. Remembering the definition of $\boldsymbol{\jmath}(z^{\prime}, \tau_r)$ in equation (\ref{Eq:Final_j}) and $\tau_r$ in (\ref{Eq:tau_r}) we have that 
\begin{equation}\label{Eq:B_infty}
	\Field[\infty]{B} = \sum_{l \in \Z } e^{il\w_0 t} \Field[\infty]{B}^l,
\end{equation}
with the spatial coefficient $\Field[\infty]{B}^l$ defined as:
\begin{equation}
	\Field[\infty]{B}^l  := il\w_0 \frac{q\mu_0}{4 \pi c |\Field[]{x}|} e^{-ilk_0 |\Field[]{x}|} \int_{\mathbb{R}}  \boldsymbol{\jmath}_l^T(z^{\prime}) \exp\bigg[ il k_0 \frac{z}{|\Field[]{x}|} z^{\prime} \bigg] dz^{\prime} (\Field[z]{e}\times \boldsymbol{\nu}).
\end{equation}
Here it is important to emphasize that $\Field[\infty]{B}^0=0$. Thus the results obtained in this section are for $l\neq0$.
Notice that the integral term resembles to a spatial Fourier transform that goes from $z'\rightarrow \eta_l$ with the new variable 
\begin{equation}
	\eta_l := l k_0 \frac{z}{|\Field[]{x}|}. 
\end{equation}
Thus
\begin{equation}\label{Eq:B_infty_l}
	\Field[\infty]{B}^l = il\w_0 \frac{q\mu_0}{4 \pi c |\Field[]{x}|} e^{-ilk_0 |\Field[]{x}|} 2\pi \mathcal{F}_{z'\rightarrow \eta_l} \big\{ \boldsymbol{\jmath}_l^T(z^{\prime}) \big\} (\Field[z]{e}\times \boldsymbol{\nu}). 
\end{equation}
This Fourier transform can be easily computed by using the definition in Eq.(\ref{Eq:J_l_T}):
\begin{equation}
	2\pi \mathcal{F}_{z'\rightarrow \eta_l} \big\{ \boldsymbol{\jmath}_l^T(z^{\prime}) \big\} = 2 \pi \bigg[\mathcal{F}_{z'\rightarrow \eta_l} \big\{ \textit{$\xi$}_{l+1}(z^{\prime})-\mathcal{F}_{z'\rightarrow \eta_l} \big\{ \textit{$\xi$}_{l-1}(z^{\prime}) \big\}  \bigg] \frac{a \w_0}{2i}
\end{equation}
Fortunately, the Fourier transform of $\textit{$\xi$}_l$ is given in equation (\ref{Eq:FT_zk_Bessel}) and after using identity (\ref{Eq:Identity_1}) one gets: 
\begin{equation}
	2\pi \mathcal{F}_{z'\rightarrow \eta_l} \big\{ \boldsymbol{\jmath}_l^T(z^{\prime}) \big\} =  a i^l l \w_0 \frac{\mathrm{J}_l(\eta_l a)}{\eta_l a },
\end{equation}
which implies that:
\begin{align}
\Field[\infty]{B}^l 
 &= i^{l+1} \frac{q\mu_0}{4 \pi c} (l\w_0)^2 \frac{a}{|\Field[]{x}|} e^{-ilk_0 |\Field[]{x}|} \frac{\mathrm{J}_l(\eta_l a)}{\eta_l a }     (\Field[z]{e}\times \boldsymbol{\nu}),\\
 & = i^{l+1} \frac{\mu_0}{4\pi} c(lk_0)^2 qa\frac{e^{-ilk_0 |\Field[]{x}|}}{|\Field[]{x}|} \frac{\mathrm{J}_l(\eta_l a)}{\eta_l a }     (\Field[z]{e}\times \boldsymbol{\nu}),\\
 & =  \frac{\mu_0}{4\pi} c(lk_0)^2 \frac{e^{-ilk_0 |\Field[]{x}|}}{|\Field[]{x}|}(\boldsymbol{\nu}\times \Field[z]{p})\bigg( i^{l-1} \frac{\mathrm{J}_l(\eta_l a)}{\eta_l a } \bigg),
\end{align}
where we have defined the dipole moment vector $\Field[z]{p}= qa\Field[z]{e}$. Notice that this last expression is very similar to the one in \cite{jackson_classical_1998} for the \emph{magnetic dipole field} in the radiation zone.   
Calling 
\begin{equation}
	 A_l = i^{l-1} \frac{\mu_0}{4 \pi} c(lk_0)^2  e^{-ilk_0 |\Field[]{x}|},
\end{equation} 
we get:
\begin{equation}
	\Field[\infty]{B}^l = A_l \frac{\mathrm{J}_l(\eta_l a)}{\eta_l a }    \frac{(\boldsymbol{\nu}\times \Field[z]{p})}{|\Field[]{x}|} ,
\end{equation}
and its square norm is given by:
\begin{align}
	|\Field[\infty]{B}^l|^2 =& \frac{|A_l|^2}{|\Field[]{x}|^2}  \bigg( \frac{\mathrm{J}_l(\eta_l a)}{\eta_l a } \bigg)^2 \bigg[1-\bigg(\frac{z}{|\Field[]{x}|}\bigg)^2\bigg] |\Field[z]{p}|^2,\\
		=& \frac{|A_l|^2}{|\Field[]{x}|^2} \bigg( \frac{\mathrm{J}_l(\eta_l a)}{\eta_l a } \bigg)^2 \bigg[1-\bigg(\frac{a\eta_l}{a lk_0}\bigg)^2\bigg] |\Field[z]{p}|^2,
\end{align}
where it has been used the fact that: $\eta_l  = lk_0 \frac{z}{|\Field[]{x}|}$. This last result will be used in the study of the radiated power. 
\subsection{Radiated power}

The expression for the far field radiated power for an oscillating will be derived. First for the relativistic case and later for the non relativistic one.

\subsubsection{Relativistic case}
We know from the definition of the Poynting vector and (\ref{Eq:inf_E}) that:
\begin{equation}
	\label{Eq:inf_S}
	\Field[\infty]{S} = \Field[\infty]{E}\times \Field[\infty]{H}  = \frac{c}{\mu_0} (\Field[\infty]{B}\times \boldsymbol{\nu})\times \Field[\infty]{B} = \frac{c}{\mu_0} | \Field[\infty]{B} |^2 \boldsymbol{\nu}.
\end{equation}
Remembering  the Fourier expansion of $\Field[\infty]{B}$ in \eq{Eq:B_infty} we have that:
\begin{equation}
	| \Field[\infty]{B} |^2 = \sum_{\substack{l \in \Z \\ l\neq 0 }}\sum_{\substack{m \in \Z \\ m\neq 0 }}  e^{i(l-m)\w_0 t} \Field[\infty]{B}^l \overline{\Field[]{B}}^m_{\infty}.
\end{equation}
Now, we consider the time-averaged Poynting vector, given the fact that $\Field[\infty]{B}$ is $\frac{2\pi}{\w_0}$-periodic we integrate \eq{Eq:inf_S} from $-\frac{\pi}{\w_0}$ to $\frac{\pi}{\w_0}$ to have:
\begin{equation}
	\braket{\Field[\infty]{S}}  = \frac{c}{\mu_0}  \braket{|\Field[\infty]{B}|^2}\boldsymbol{\nu},
\end{equation}
where the quantities between brackets are time averaged. By the orthogonality of the complex exponential we have:
\begin{equation}
	\braket{|\Field[\infty]{B}|^2} 	
	= \sum_{\substack{l \in \Z \\ l\neq 0 }}  \frac{|A_l|^2}{|\Field[]{x}|^2} \bigg( \frac{\mathrm{J}_l(\eta_l a)}{\eta_l a } \bigg)^2 \bigg[1-\bigg(\frac{a\eta_l}{a lk_0}\bigg)^2\bigg]|\Field[z]{p}|^2.
\end{equation}
For obtaining the power it is customary to  calculate the flux of the averaged Poynting vector through a spherical surface of radius $R_0$ (it could be any surface encompassing the EM sources, but the spherical surface is the one that allows the simplest calculations). That is: 
\begin{align}
	\int_{0}^{2 \pi} \int_{0}^{\pi} \braket{\Field[\infty]{S}} R_0^2 \sin \theta\,\rmd\theta\,\rmd\phi 
    & = \frac{c}{\mu_0} \int_{0}^{2 \pi} \int_{0}^{\pi} \braket{|\Field[\infty]{B}|^2} R_0^2 \sin \theta\, \rmd\theta\,\rmd \phi,\\
    & =  \sum_{\substack{l \in \Z \\ l\neq 0 }} \frac{2\pi c}{\mu_0} |A_l|^2 \int_{0}^{\pi} \bigg( \frac{\mathrm{J}_l(\eta_l a)}{\eta_l a } \bigg)^2 \bigg[1-\bigg(\frac{a\eta_l}{a lk_0}\bigg)^2\bigg] |\Field[z]{p}|^2 \sin \theta\,\rmd\theta .
\end{align}
Next, we perform the change of variable:
\begin{equation}
	u = a \eta_l = a lk_0 \frac{z}{R_0} = a lk_0 \cos \theta 
\end{equation}
then the differential $du$ reads
\begin{equation}
	\rmd u =  - a lk_0 \sin \theta\,\rmd\theta.
\end{equation}
After these calculations, the flux of the averaged Poynting vector reads:
\begin{equation}    
\begin{aligned}
	\int_{0}^{2 \pi} \int_{0}^{\pi} \braket{\Field[\infty]{S}} R_0^2 \sin \theta d\theta d \phi  = &\\
	&\sum_{\substack{l \in \Z \\ l\neq 0 }}  \frac{2 \pi c}{\mu_0 alk_0} |A_l|^2  \int_{-alk_0}^{alk_0} \bigg( \frac{\mathrm{J}_l(u)}{u} \bigg)^2 \bigg[1-\bigg(\frac{u}{a lk_0}\bigg)^2\bigg] |\Field[z]{p}|^2 du. 
\end{aligned}
\label{Eq:Flux_sum}
\end{equation}
Calling: 
\begin{equation}
	D_l :=  \frac{c (2 \pi)^2}{\w_0 \mu_0 alk_0} |A_l|^2 = \frac{ 2 \pi c }{\mu_0 al k_0} \bigg(\frac{\mu_0}{4\pi}c (l k_0)^2\bigg)^2 = \frac{Z_0 c^2 (lk_0)^4}{8\pi (alk_0) }, 
\end{equation}
and because $D_{-l} = - D_l$ one has that the averaged energy flux can be seen as:
\begin{align}
	\int_{0}^{2 \pi} \int_{0}^{\pi} \braket{\Field[\infty]{S}} R_0^2 \sin \theta\,\rmd\theta\,\rmd \phi 
     & = \sum_{l=1}^\infty 2 D_l \int_{-alk_0}^{alk_0} \bigg( \frac{\mathrm{J}_l(u)}{u} \bigg)^2 \bigg[1-\bigg(\frac{u}{a lk_0}\bigg)^2\bigg]  \,\rmd u |\Field[z]{p}|^2,\\
      &= \sum_{l=1}^\infty P_l
\end{align}
with $P_l$ the $l$-th contribution given by: 
\begin{align}
	\label{Eq:Pl_eq}
	 P_l &= 4 D_l  \int_{0}^{alk_0} \bigg( \frac{\mathrm{J}_l(u)}{u} \bigg)^2 \bigg[1-\bigg(\frac{u}{a lk_0}\bigg)^2\bigg] \,\rmd u |\Field[z]{p}|^2, \qquad l\geq1,\\
    &= 4 D_l \bigg[\int_{0}^{alk_0} \frac{\mathrm{J}_l^2(u)}{u^2}  \,\rmd u-\frac{1}{(a lk_0)^2} \int_{0}^{alk_0}  \mathrm{J}_l^2(u)   \,\rmd u \bigg]  |\Field[z]{p}|^2
\end{align}
where  by means of the identities obtained in Annexe \ref{An:Bessel identities} one can see that the integrals are given by:
\begin{align}
    \label{Eq:I_S_l_geq_1}
    &\mathcal{I}_l^{S}(alk_0) = \int_{0}^{alk_0} \frac{\mathrm{J}_l^2(u)}{u^2}\,\rmd u=\frac{alk_0}{2l} \bigg[\frac{\mathrm{J}_{l}\mathrm{J}^{\prime}_{l+1}- \mathrm{J}_{l+1}\mathrm{J}^{\prime}_{l} }{1+2l} + \frac{\mathrm{J}_{l}\mathrm{J}^{\prime}_{l-1}- \mathrm{J}_{l-1}\mathrm{J}^{\prime}_{l} }{1-2l}\bigg]_{alk_0}\quad l\geq1,\\
    \label{Eq:I_R_l_g_1}
    &\mathcal{I}_l^{R}(alk_0) = \int_{0}^{alk_0} \mathrm{J}_l^2(u)\,\rmd u = \bigg[ \mathcal{I}_{l-2}^{R}-\frac{2(l-1)}{alk_0}\mathrm{J}_{l-1}^2-2(l-1)\mathcal{I}_{l-1}^{S}\bigg]_{alk_0}\quad l>1, \\
    \label{Eq:I_R_l_eq_1}
    &\mathcal{I}_1^{R}(ak_0) = \int_{0}^{ak_0} \mathrm{J}_1^2(u)\,\rmd u = \big[ \mathcal{I}_0^{R} - ak_0(\mathrm{J}_{1}^2 - \mathrm{J}_{0}^2)\big]_{ak_0}\quad l=1,
\end{align}
where the function $\mathcal{I}^R_0$ is defined trough an integral (see \ref{Eq:I_R_0}). By employing equations (\ref{Eq:I_S_l_geq_1})-(\ref{Eq:I_R_l_eq_1}) it is possible  to compute in a semi analytical manner the radiated power flux. \textst{Let us} In particular, if we focus our attention to the case when $l = 1$:
\begin{equation}
	P_1 = 4 D_1 \int_{0}^{ak_0} \bigg( \frac{\mathrm{J}_1(u)}{u} \bigg)^2 \bigg[1-\bigg(\frac{u}{a k_0}\bigg)^2\bigg]  \,\rmd u |\Field[z]{p}|^2,
\end{equation}
% From the Appendix \ref{An:Bessel identities} we know that 
% \begin{align}
% 	&\int_{0}^{ak_0} \bigg( \frac{\mathrm{J}_1(u)}{u} \bigg)^2 \,\rmd u =  \frac{ak_0}{2}\bigg[\frac{\mathrm{J}_1J_2^{\prime}-\mathrm{J}_2J_1^{\prime}}{3} + \mathrm{J}_1J_0^{\prime}-\mathrm{J}_0J_1^{\prime} \bigg]_{ak_0},\\	
% %
% 	&\int_{0}^{ak_0} \mathrm{J}_1^2(u)  \,\rmd u =  \mathcal{I}_0^R (ak_0)_- ak_0[\mathrm{J}_{1}^2(ak_0) + \mathrm{J}_{0}^2(ak_0)],
% \end{align}
%  Thus
or in a more explicit fashion
\begin{multline}
	P_1 = 4 D_l\bigg\{\frac{ak_0}{2}\bigg[\frac{\mathrm{J}_1\mathrm{J}_2^{\prime}-\mathrm{J}_2\mathrm{J}_1^{\prime}}{3} + \mathrm{J}_1\mathrm{J}_0^{\prime}-\mathrm{J}_0\mathrm{J}_1^{\prime} \bigg]_{ak_0} + \\
	\frac{\mathcal{I}_0^R(ak_0)}{ak_0} - \frac{[\mathrm{J}_{1}^2(ak_0) + \mathrm{J}_{0}^2(ak_0)]}{(ak_0)^2} \bigg\} |\Field[z]{p}|^2.
\end{multline}
It is important to notice that in this case we have made no restrictions regarding the velocity of the charged particle, albeit it can not be faster than $c$ (this will be studied in a more deeper way in subsection \ref{SubSec:Supraluminal}). Thus, the expression derived here are valid even for charges whith speed near $c$.
%%%%%%%%%%%%%
\subsubsection{Non-relativistic case}
\begin{figure}[htbp]
    \centering
    \graphicspath{{Figures/} } 
    \includegraphics[width=\textfigure]{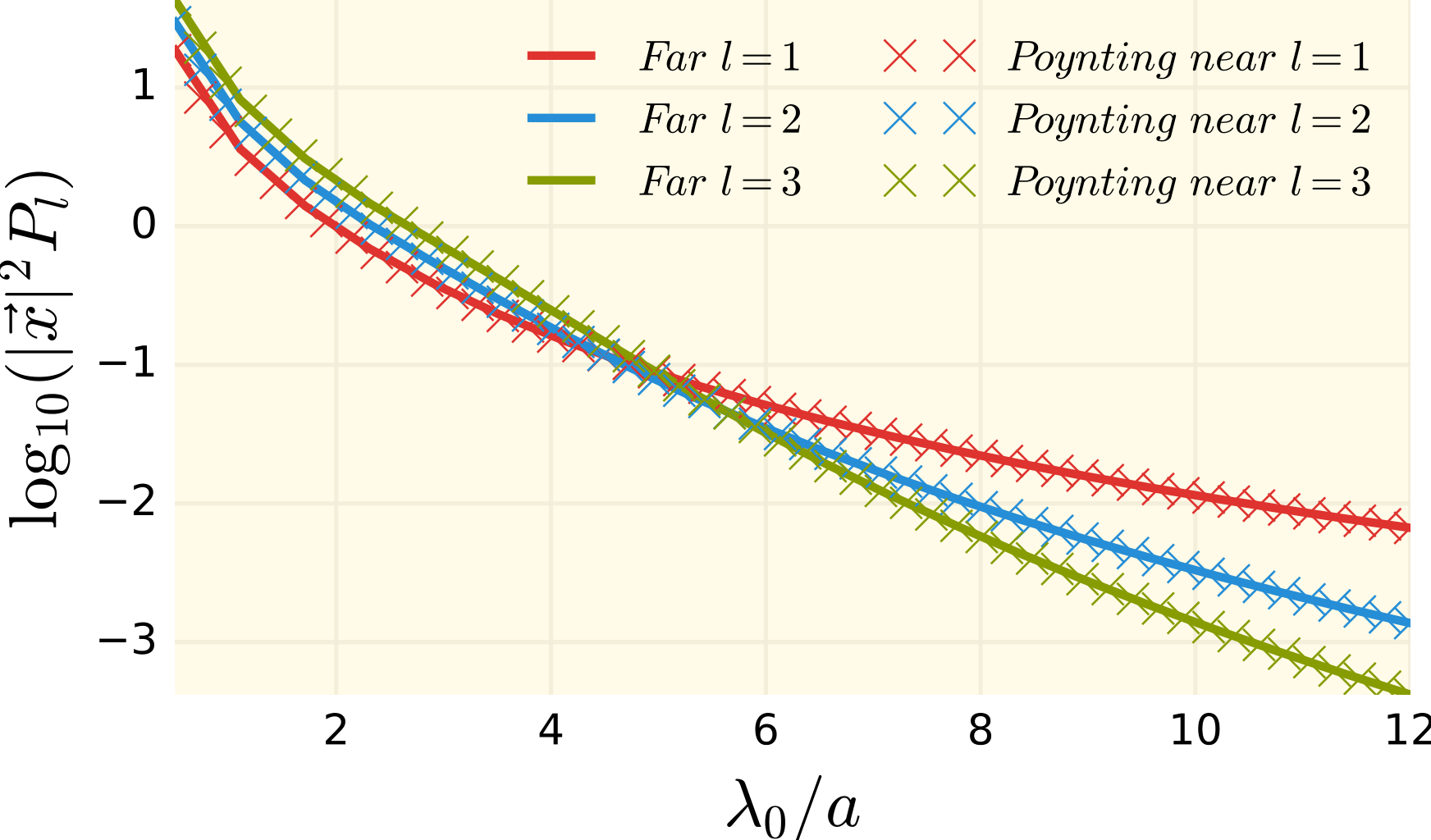}
    \captionsetup{justification=centering}
    \caption{Comparison of the powers $P_l$ ($l=1,2,3$) obtained analytically and numerically. }
    \label{fig:Pl_comp}
\end{figure}

We are then in the case of $alk_0 \ll 1$. For $l=1$, we have $0<u< a k_0 \ll 1$. In that case, the behavior of $\mathrm{J}_1$ near the origin is well known namely $\mathrm{J}_1(u) \sim \frac{u}{2}$ and as a result:
\begin{equation}
\frac{\mathrm{J}_1^2(u)}{u^2} \sim \frac{1}{4}
\end{equation}
and we obtain
\begin{equation}
P_1 \sim  D_1\,|\Field[z]{p}|^2 \int_{0}^{ak_0} \left(1-\frac{u^2}{(ak_0)^2} \right) \, \,\rmd u
\label{eq:P03-beta}
\end{equation}
i.e.
\begin{equation}
P_1 \sim  D_1\frac{2}{3}  ak_0 |\Field[z]{p}|^2 =  \frac{Z_0 c^2 (k_0)^4}{8\pi (ak_0) } \frac{2}{3}  ak_0|\Field[z]{p}|^2 = \frac{Z_0 c^2 (k_0)^4}{12\pi  }|\Field[z]{p}|^2.
\end{equation}
The reader may recognize this last equality as the expression for the total power radiated in \cite{jackson_classical_1998}. Finally, figure \ref{fig:Pl_comp} shows a comparison between the power $P_l$ (for $l = 1,2,3$) obtained analytically by means of \ref{Eq:Pl_eq} when using the far field approximation and the power numerically computed by considering the Poynting vector flux across the surface of a pill box that encloses the trajectory of the charged particle. The Poynting vector is obtained via equations (\ref{Eq:E_harm_final}-\ref{Eq:B_harm_final}) and the numerical integrations is performed by using the solver GetDP \cite{Geuzaine_getDP}. 
As we can see, both approaches fit perfectly.

\subsection{The particle cannot be supraluminal}\label{SubSec:Supraluminal}
It is common sense that for the $l-th$ component of the magnetic induction field in \ref{Eq:inf_B}, namely $\mathbf{B}_{\infty}^l$, the $l$-th averaged Poynting vector $\braket{\boldsymbol{S}_{\infty}^l} : = \frac{c}{\mu_0}|\Field[\infty]{B}^l|^2$ (see \ref{Eq:inf_E}) as well as for the power $P_l $ (given by equations (\ref{Eq:Flux_sum}-\ref{Eq:Pl_eq})) in the value converges with growing $l$. A divergence would result in the total value of infinity. Due to the connection between $\mathbf{B}_{\infty}^l$, $\braket{\boldsymbol{S}_{\infty}^l}$ and $\mathcal{P}_l$ it follows, that if the magnetic field diverges, the two others diverge, too.\\
The $l^{th}$ magnetic field depends on $l$ only in
\begin{equation}
\mathbf{B}_{\infty}^l \sim l\cdot \mathrm{J}_l(l\cdot ak_0\cos\theta)
\label{eq:bdiv}
\end{equation}
For large $l$, $\mathrm{J}_l(l\beta)$ can be written as follows \cite{abramowitz1964handbook}:
\begin{equation}
\mathrm{J}_l(l~ak_0\cos\theta) \sim \frac{e^{l\, (\tanh (a_0)-a_0)}}{\sqrt{\frac{1}{2}\pi\, l\, \tanh (a_0)}}\ \ ,\ \text{with } a_0=\mathrm{arccosh} \left(\frac{1}{ak_0\cos\theta} \right)
\label{eq:Jdiv}
\end{equation}
With this, $\mathbf{B}_{\infty}^l$ goes with
\begin{equation}
\mathbf{B}_{\infty}^l \sim \sqrt{\frac{2l}{\pi\sqrt{1-(ak_0\cos\theta)^2}}}e^{l(\sqrt{1-(ak_0\cos\theta)^2}-\mathrm{arcosh}(\frac{1}{ak_0\cos\theta}))}
\label{eq:Bdiv2}
\end{equation}
This only converges  with $l\rightarrow\infty$ if the argument of the exponential function is negative. That leads in the form of Eq. \ref{eq:Jdiv} to the following inequalities:
\begin{eqnarray}
a_0 > \mathrm{tanh}(a_0) \nonumber \\
\Leftrightarrow a_0 > 0 \nonumber \\
%\Leftrightarrow 0 < \beta < 1 \nonumber \\
\leftrightarrow 0 < ak_0\cos\theta < 1 \nonumber \\
\Leftrightarrow ak_0 < 1 \nonumber \\
\Leftrightarrow v:=a\omega_0 < c
\end{eqnarray}
With $v=a\omega_0$ the maximum velocity of our point charge in his sinusoidal movement, the condition for convergence is matched in the physically sense of Albert Einstein's postulate that nothing is faster than light \cite{griffiths1999introduction, jackson_classical_1998}. It is important to remark here the fact that this result holds irrespective of whether the particle has got a mass.  
%%%%%%SECTION
%
\section{Obtaining the diffracted field}\label{Sec:Osc_Diffracted}

After all this work, we have that the fields $\Field[]{E}^0$ and $\Field[]{B}^0$ of the system of equations (\ref{Eq:Faraday_0}-\ref{Eq:Gauss_M_0})
can be seen as a superposition of waves in the form of \Eq{Eq:E_harm_final} and \Eq{Eq:B_harm_final}. Even more, the analysis fo the radiated energy from previous section, show that from our expressions of $\Field[]{E}^0$ and $\Field[]{B}^0$ one can retrieve the classical results when the non relativistic far field approximations are considered. The second part of this \textst{chapter} work (which is going to be considerably shorter that the first one) can be tackled in a straight forward fashion. First, the sources in \Eq{Eq:Sources_0} can be easily retrieved  by noticing that:
\begin{align}
	\nonumber
	[\eps_{r,II}-2\pi\delta]*\Field[]{E}^0 
	=& \frac{1}{2\pi} \int_{\tau \in \mathbb{R}} [\eps_{r,II}(\tau)-2\pi \delta] \sum_{l \in \Z }e^{+il \w_0 (t-\tau)} \Field[l]{E}^0(\Field[]{x})  d\tau, \\
	\nonumber	
	=& \sum_{l \in \Z }e^{+il \w_0 t}\Field[l]{E}^0(\Field[]{x}) \frac{1}{2\pi} \int_{\tau \in \mathbb{R}} [\eps_{r,II}(\tau)-2\pi \delta] e^{-il \w_0 \tau} d\tau,\\
	=& \sum_{l \in \Z }e^{+il \w_0 t}\Field[l]{E}^0(\Field[]{x}) [\overline{\hat{\eps}}_{r,II}(l\w_0)-1].
\end{align} 
Then:
\begin{align}\label{Eq:Sources_0_harm}
\rho^0_{II} =& -\eps_0 \sum_{l \in \Z }  [\overline{\hat{\eps}}_{r,II}(l\w_0)-1] e^{+il \w_0 t}\mathbf{\nabla}\cdot \Field[l]{E}^0(\Field[]{x}),\\
\Field[II]{\boldsymbol{\jmath}}^0 =& \eps_0 \sum_{l \in \Z } il \w_0 [\overline{\hat{\eps}}_{r,II}(l\w_0)-1]  e^{+il \w_0 t}\Field[l]{E}^0(\Field[]{x}).
\end{align}
It is then natural to propose as solutions of the system (\ref{Eq:Faraday_0}-\ref{Eq:Gauss_M_0}):
\begin{align}
	\label{Eq:E_1_sum}
	\Field[\textit{$\xi$}]{E}(\Field[]{x},t) =& \sum_{l \in \Z }e^{+il \w_0 t} \Field[\textit{$\xi$},l]{E}(\Field[]{x}),\\
	\label{Eq:B_1_sum}
	\Field[\textit{$\xi$}]{B}(\Field[]{x},t) =& \sum_{l \in \Z }e^{+il \w_0 t} \Field[\textit{$\xi$},l]{B}(\Field[]{x}),
\end{align}
Plugging these solutions into equations (\ref{Eq:Faraday_1}-\ref{Eq:Gauss_M_1}) (and recalling that $\Field[\textit{$\xi$},l]{E}^1 := \Field[\textit{$\xi$},l]{E}-\Field[]{E}^0$), we arrive to the following system which must be satisfied for each $l \in \Z $.
\begin{align}
\label{Eq:Faraday_1_l}
&\Curl[\textit{$\xi$},l]{E}^1 = -il\w_0 \Field[\textit{$\xi$},l]{B}^1, \\
\label{Eq:Ampere_1_l}
&\Curl[\textit{$\xi$},l]{B}^1 = \frac{il\w_0}{c^2}\overline{\hat{\eps}}_{r, II}(l\w_0)\Field[\textit{$\xi$},l]{E}^1 + \frac{il\w_0}{c^2}[\overline{\hat{\eps}}_{r,II}(l\w_0)-1]\Field[l]{E}^0(\Field[]{x}),\\
\label{Eq:Gauss_E_1_l}
&\mathbf{\nabla}\cdot\Field[\textit{$\xi$},l]{E}^1  = -\frac{[\overline{\hat{\eps}}_{r,II}(l\w_0)-1]}{\overline{\hat{\eps}}_{r, II}(l\w_0)} \mathbf{\nabla}\cdot \Field[l]{E}^0(\Field[]{x}) ,\\
\label{Eq:Gauss_B_1_l}
&\Div[\textit{$\xi$},l]{B}^1 = 0
\end{align}
Taking the curl on Faraday's Law we get:
\begin{equation}\label{Eq:Driffracted_field_eq}
\mathbf{\nabla} \times \Curl[\textit{$\xi$},l]{E}^1 - \bigg(\frac{l \w_0 }{c}\bigg)^2\overline{\hat{\eps}}_{r,II}(l\w_0) \Field[\textit{$\xi$},l]{E}^1=  \bigg(\frac{l \w_0 }{c}\bigg)^2 [\overline{\hat{\eps}}_{r,II}(l\w_0)-1] \Field[l]{E}^0,
\end{equation}
where the ring hand side of this expression is a source term. Moreover equation can be solved for instance, by using Finite Elements Method as it is explained in Refs.~\cite{zolla_foundations_2005,demesy2010all}.
\section{Numerical illustration}\label{Sec:Osc_Results}

\begin{figure}[htbp]
\centering
\begin{subfigure}{\textsubfigure}
\includegraphics[width=\linewidth]{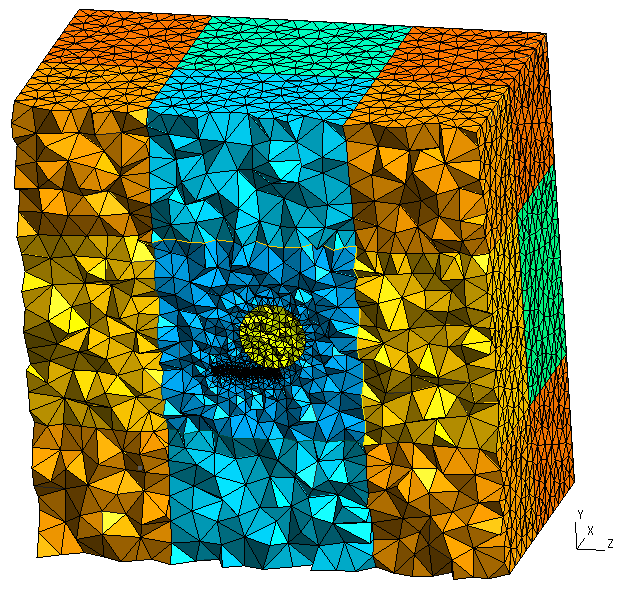}
\captionsetup{justification=centering}
\caption{Mesh showing the different regions of integration} \label{fig:3a}
\end{subfigure}
\hspace*{\fill} % separation between the subfigures
\centering
\begin{subfigure}{\textsubfigure}
\includegraphics[width=\linewidth]{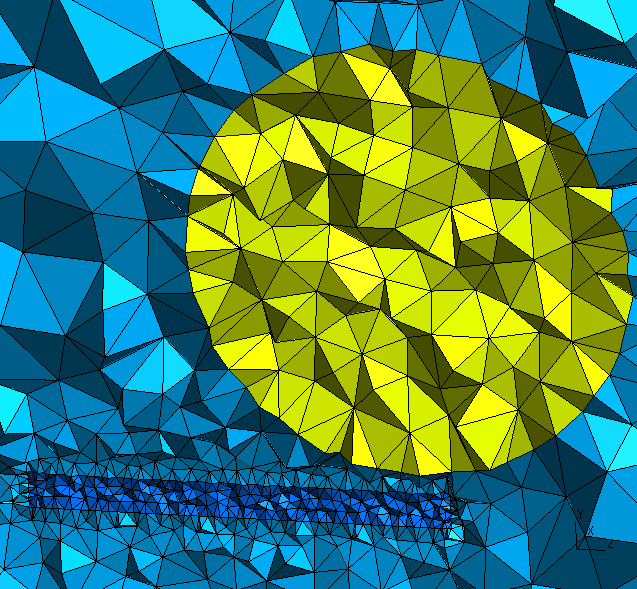}
\captionsetup{justification=centering}
\caption{Zoom enclosing the sphere and the region where the particle oscillates.} \label{fig:3b}
\end{subfigure}
\captionsetup{justification=centering}
\caption{Mesh showing the regions where the different fields are obtained via FEM.} \label{fig:3}
\end{figure}

\begin{figure}[htbp]
    \centering
    \begin{subfigure}{\textsubfigure}
        \includegraphics[width=\linewidth]{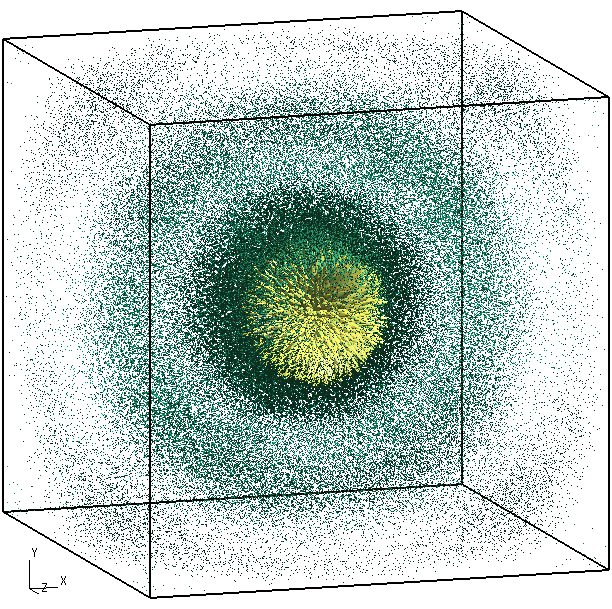}
        \captionsetup{justification=centering}
        \caption{$\w_0$} \label{fig:4a}
        \end{subfigure}
    \hspace*{\fill} % separation between the subfigures
    \begin{subfigure}{\textsubfigure}
        \includegraphics[width=\linewidth]{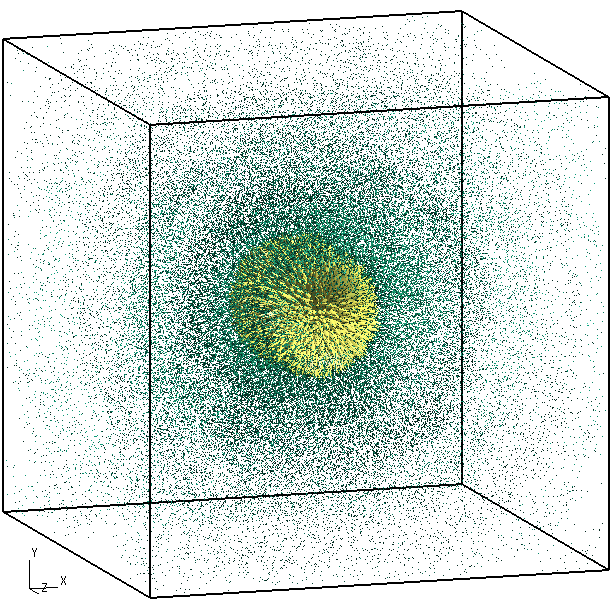}
        \captionsetup{justification=centering}
        \caption{$2\w_0$} \label{fig:4b}
    \end{subfigure}
% \centering
    \begin{subfigure}{\textsubfigure}
        \includegraphics[width=\linewidth]{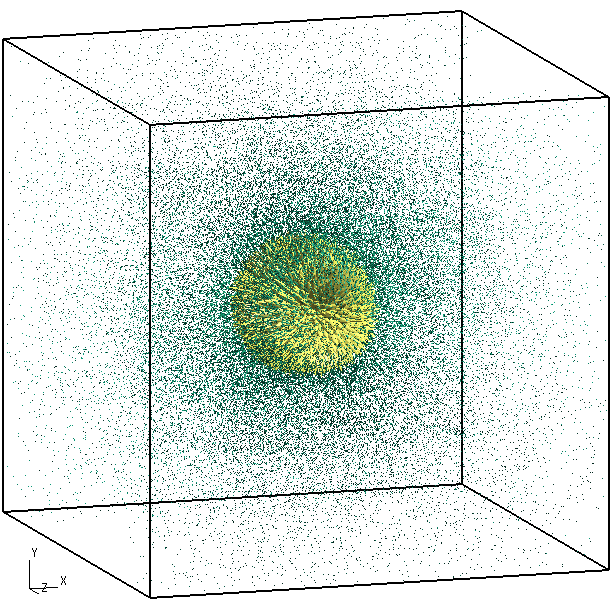}
        \captionsetup{justification=centering}
        \caption{$3\w_0$} \label{fig:4c}
    \end{subfigure}
    \hspace*{\fill} % separation between the subfigures
    \centering
    \begin{subfigure}{\textsubfigure}
        \includegraphics[width=\linewidth]{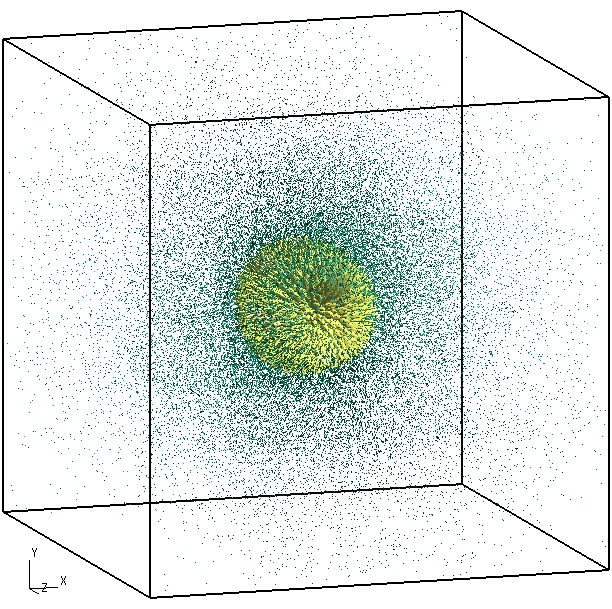}
        \captionsetup{justification=centering}
        \caption{$4\w_0$} \label{fig:4d1}
    \end{subfigure}
    \captionsetup{justification=centering}
    \caption{Harmonic components of the imaginary part of the incident electric field $\Field[l]{E}^0$ for l = $1,2,3,4$} \label{fig:4}
\end{figure}

\begin{figure}[htbp]
    \centering
    \begin{subfigure}{\textsubfigure}
        \includegraphics[width=\linewidth]{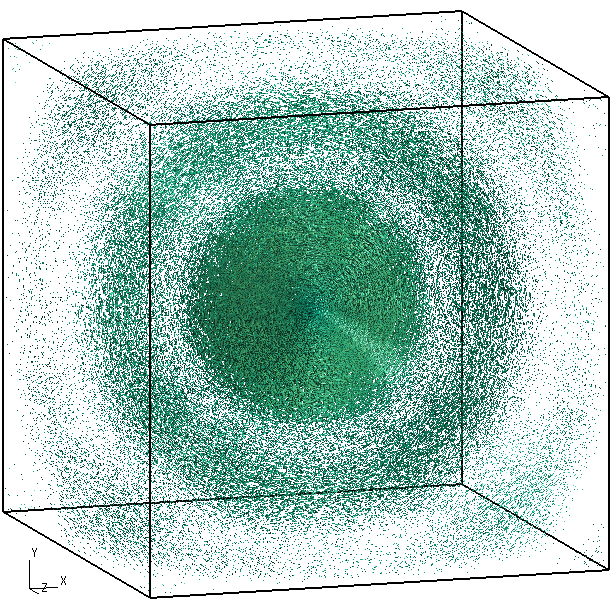}
        \captionsetup{justification=centering}
        \caption{$\w_0$} \label{fig:5a}
        \end{subfigure}
    \hspace*{\fill} % separation between the subfigures
    \begin{subfigure}{\textsubfigure}
        \includegraphics[width=\linewidth]{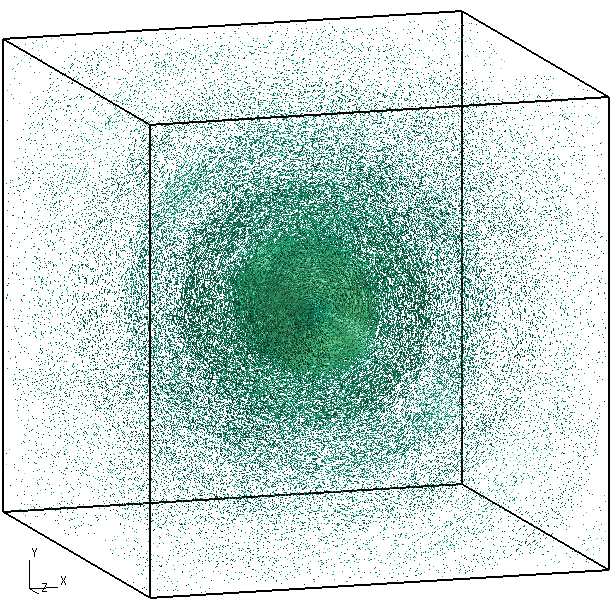}
        \captionsetup{justification=centering}
        \caption{$2\w_0$} \label{fig:5b}
    \end{subfigure}
    % %\hspace*{\fill} % separation between the subfigures
    
    \centering
    \begin{subfigure}{\textsubfigure}
        \includegraphics[width=\linewidth]{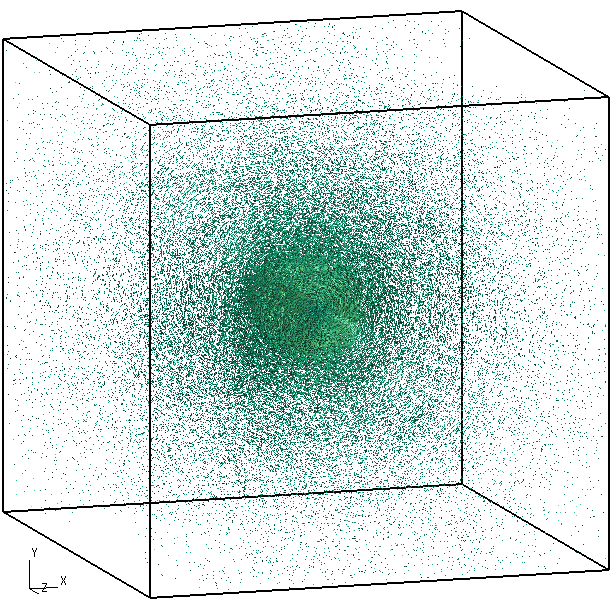}
        \captionsetup{justification=centering}
        \caption{$3\w_0$} \label{fig:5c}
    \end{subfigure}
    \hspace*{\fill} % separation between the subfigures
    \centering
    \begin{subfigure}{\textsubfigure}
        \includegraphics[width=\linewidth]{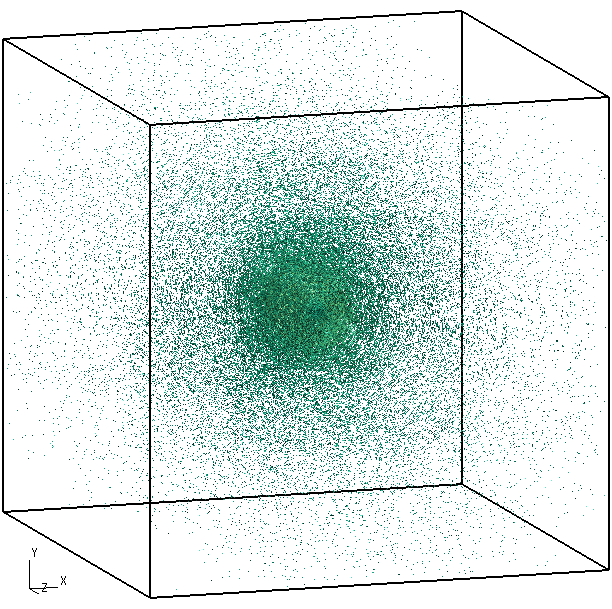}
        \captionsetup{justification=centering}
        \caption{$4\w_0$} \label{fig:4d}
    \end{subfigure}
    \captionsetup{justification=centering}
    \caption{Harmonic components of the real part of the incident magnetic induction field $\Field[l]{B}^0$ for l = $1,2,3,4$} \label{fig:5}
\end{figure}

\begin{figure}[htbp]
    \centering
    \begin{subfigure}{\textsubfigure}
        \includegraphics[width=\linewidth]{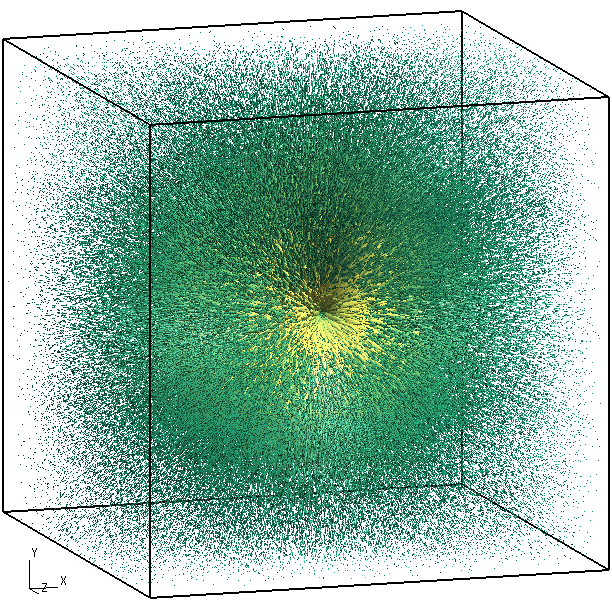}
        \captionsetup{justification=centering}
        \caption{$\w_0$} \label{fig:9a}
    \end{subfigure}
    \hspace*{\fill} % separation between the subfigures
    \begin{subfigure}{\textsubfigure}
        \includegraphics[width=\linewidth]{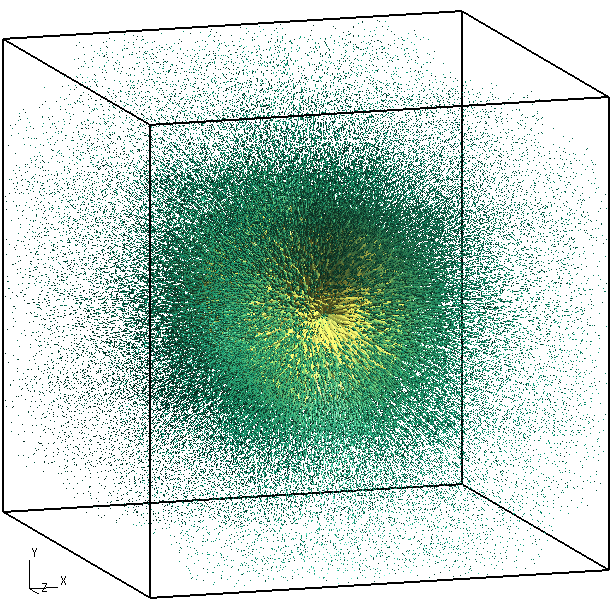}
        \captionsetup{justification=centering}
        \caption{$2\w_0$} \label{fig:9b}
    \end{subfigure}
    %\hspace*{\fill} % separation between the subfigures
    \centering
    \begin{subfigure}{\textsubfigure}
        \includegraphics[width=\linewidth]{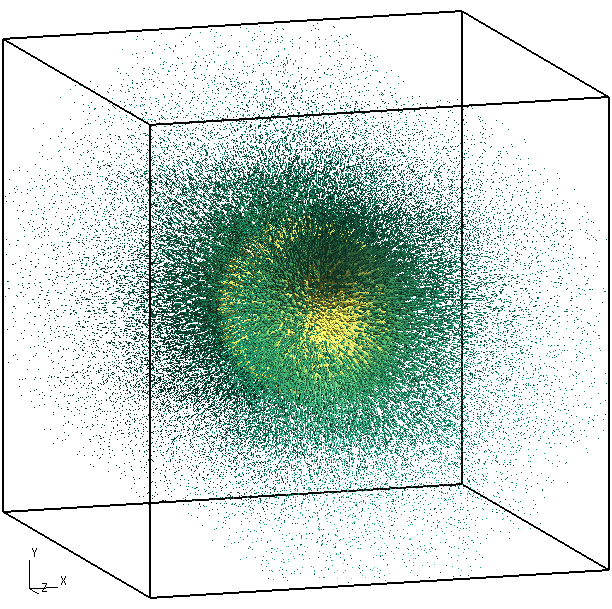}
        \captionsetup{justification=centering}
        \caption{$3\w_0$} \label{fig:9c}
    \end{subfigure}
    \hspace*{\fill} % separation between the subfigures
    \centering
    \begin{subfigure}{\textsubfigure}
        \includegraphics[width=\linewidth]{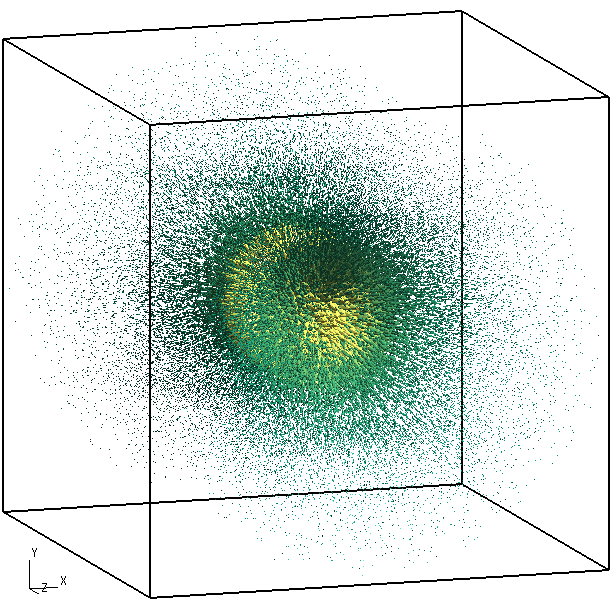}
        \captionsetup{justification=centering}
        \caption{$4\w_0$} \label{fig:9d}
    \end{subfigure}
    \captionsetup{justification=centering}
    \caption{Harmonic components of the real part of the incident Poynting vector field $\Field[l]{S}^0$ for l = $1,2,3,4$} \label{fig:9}
\end{figure}

\begin{figure}[htbp]
    \centering
    \begin{subfigure}{\textsubfigure}
        \includegraphics[width=\linewidth]{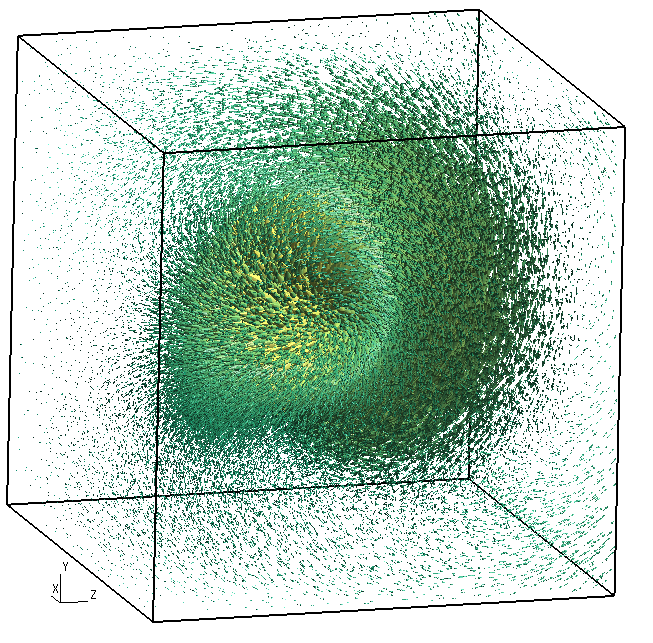}
        \captionsetup{justification=centering}
        \caption{$\w_0$} \label{fig:6a}
    \end{subfigure}
    \hspace*{\fill} % separation between the subfigures
    \begin{subfigure}{\textsubfigure}
        \includegraphics[width=\linewidth]{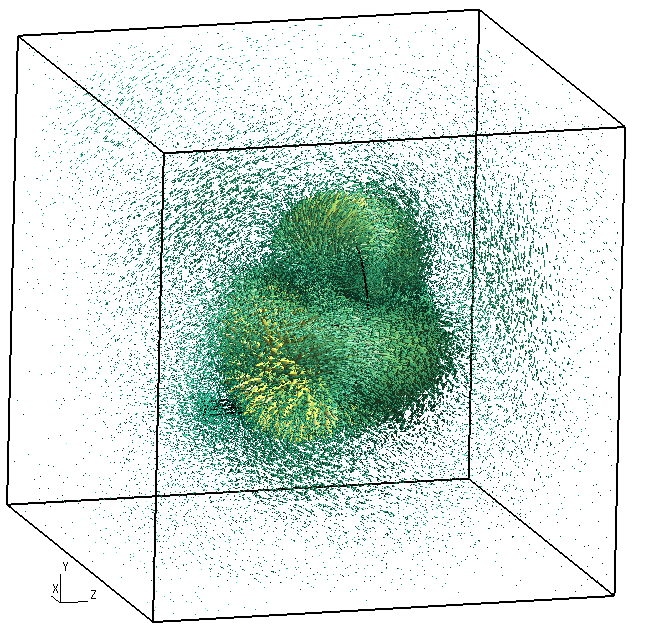}
        \captionsetup{justification=centering}
        \caption{$2\w_0$} \label{fig6b}
    \end{subfigure}
    \centering
    \begin{subfigure}{\textsubfigure}
        \includegraphics[width=\linewidth]{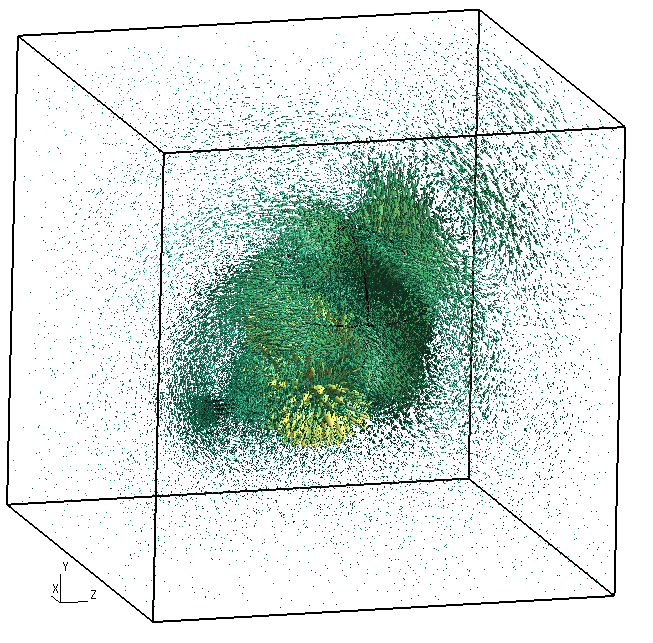}
        \captionsetup{justification=centering}
        \caption{$3\w_0$} \label{fig:6c}
    \end{subfigure}
    \hspace*{\fill} % separation between the subfigures
    \centering
    \begin{subfigure}{\textsubfigure}
        \includegraphics[width=\linewidth]{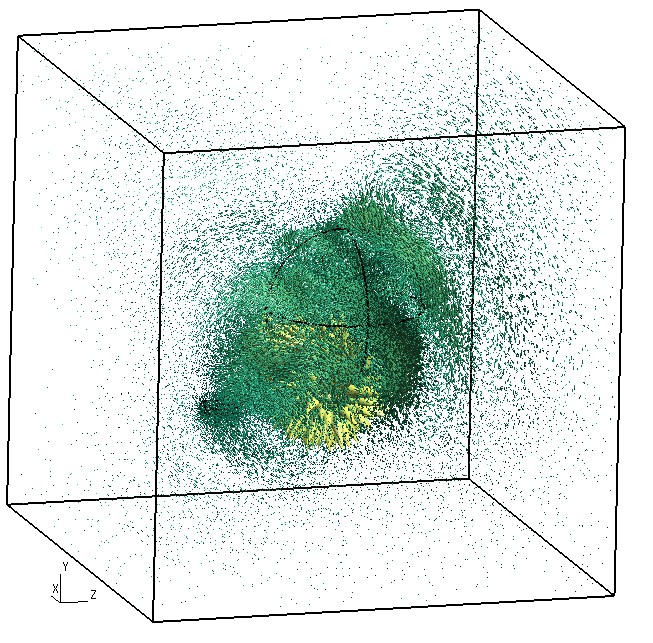}
        \captionsetup{justification=centering}
        \caption{$4\w_0$} \label{fig:6d}
    \end{subfigure}
    \captionsetup{justification=centering}
    \caption{Harmonic components of the real part of the diffracted electric field $\Field[l]{E}^1$ for l = $1,2,3,4$} \label{fig:6}
\end{figure}

\begin{figure}[htbp]
    \centering
    \begin{subfigure}{\textsubfigure}
        \includegraphics[width=\linewidth]{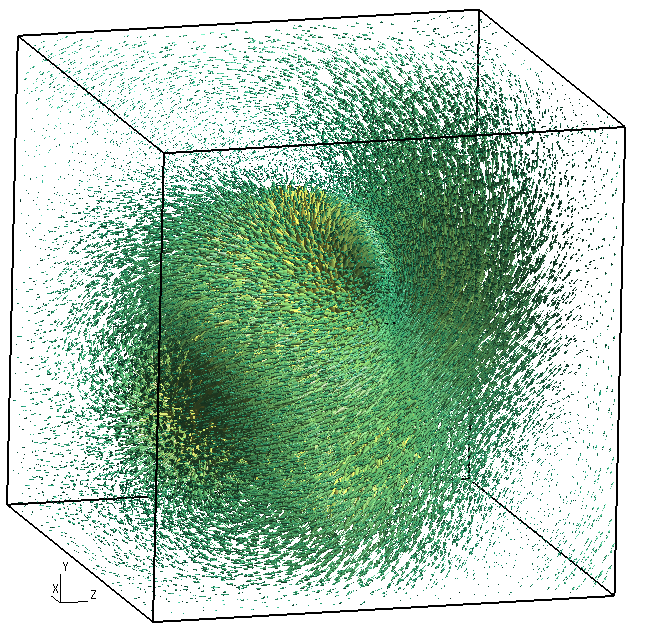}
        \captionsetup{justification=centering}
        \caption{$\w_0$} \label{fig:7a}
    \end{subfigure}
    \hspace*{\fill} % separation between the subfigures
    \begin{subfigure}{\textsubfigure}
        \includegraphics[width=\linewidth]{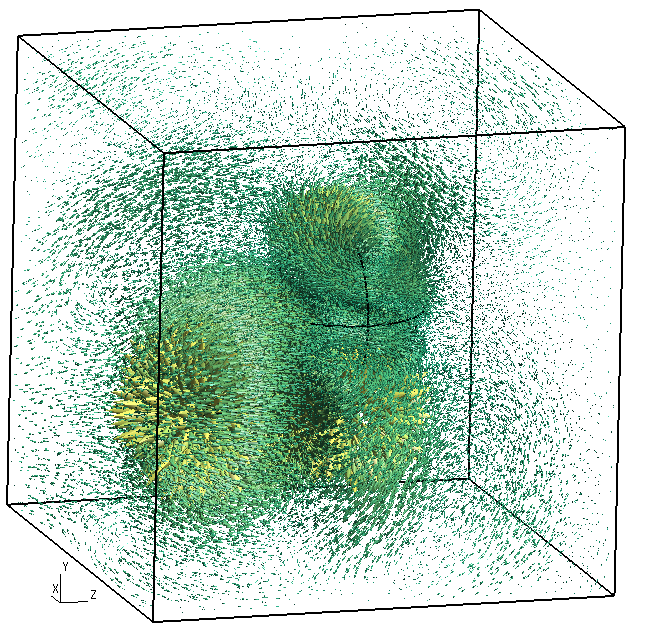}
        \captionsetup{justification=centering}
        \caption{$2\w_0$} \label{fig7b}
    \end{subfigure}
    \centering
    \begin{subfigure}{\textsubfigure}
        \includegraphics[width=\linewidth]{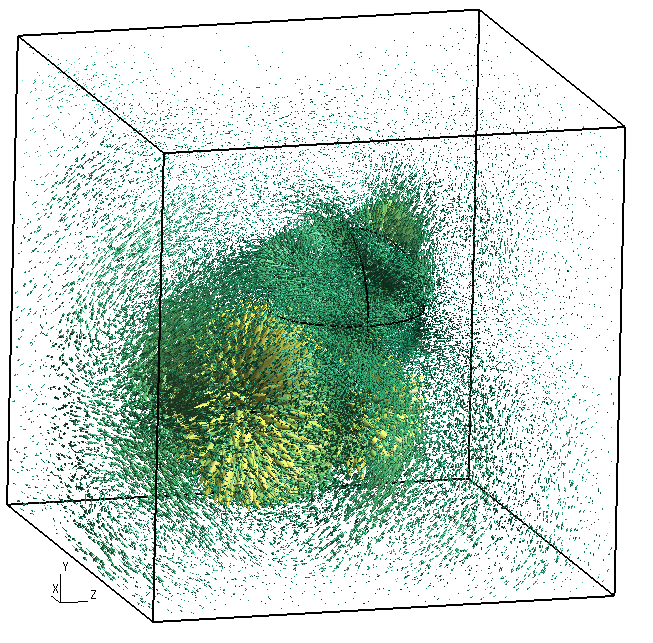}
        \captionsetup{justification=centering}
        \caption{$3\w_0$} \label{fig:7c}
    \end{subfigure}
    \hspace*{\fill} % separation between the subfigures
    \centering
    \begin{subfigure}{\textsubfigure}
        \includegraphics[width=\linewidth]{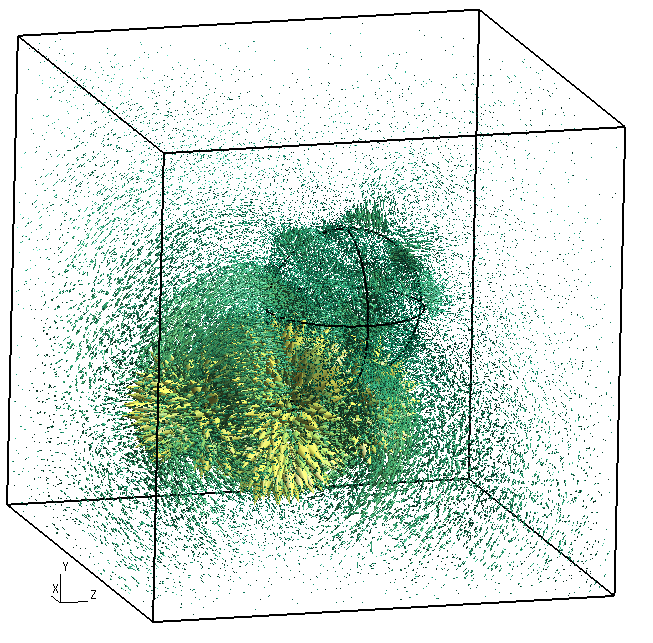}
        \captionsetup{justification=centering}
        \caption{$4\w_0$} \label{fig:7d}
    \end{subfigure}
    \captionsetup{justification=centering}
    \caption{Harmonic components of the real part of the total electric field $\Field[l]{E}$ for l = $1,2,3,4$} \label{fig:7}
\end{figure}

\begin{figure}[htbp]
    \centering
    \begin{subfigure}{\textsubfigure}
        \includegraphics[width=\linewidth]{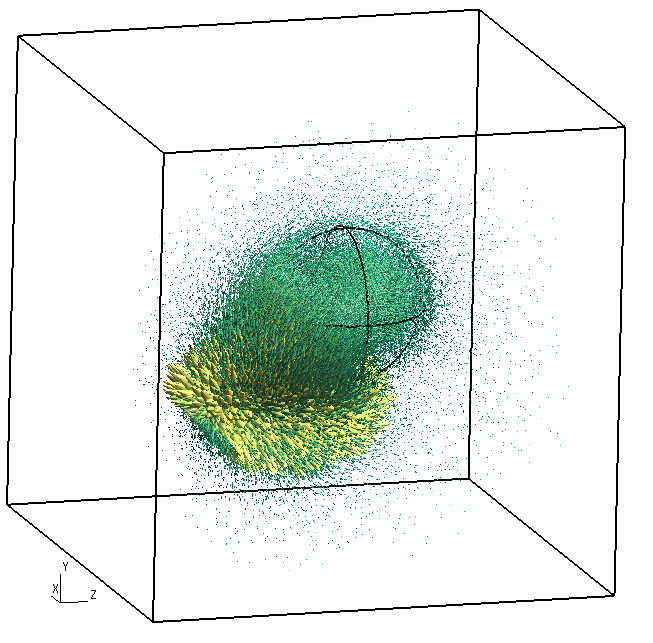}
        \captionsetup{justification=centering}
        \caption{$\w_0$} \label{fig:8a}
    \end{subfigure}
    \hspace*{\fill} % separation between the subfigures
    \begin{subfigure}{\textsubfigure}
        \includegraphics[width=\linewidth]{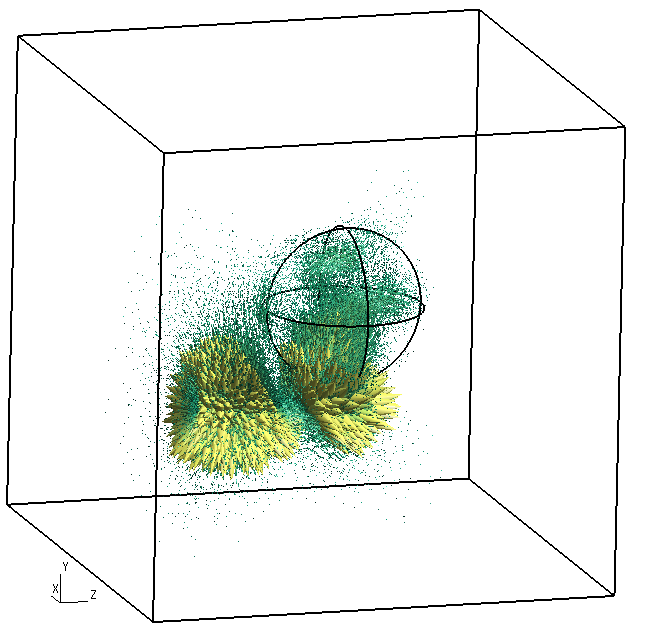}
        \captionsetup{justification=centering}
        \caption{$2\w_0$} \label{fig8b}
    \end{subfigure}
    
    \centering
    \begin{subfigure}{\textsubfigure}
        \includegraphics[width=\linewidth]{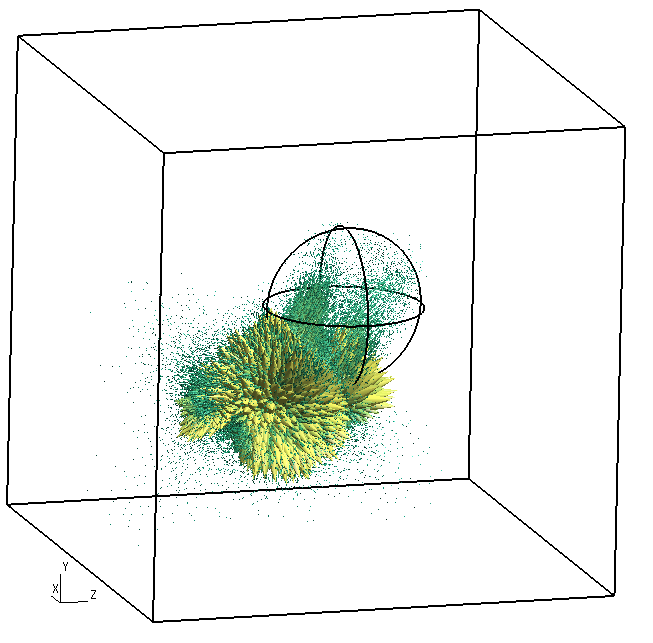}
        \captionsetup{justification=centering}
        \caption{$3\w_0$} \label{fig:8c}
    \end{subfigure}
    \hspace*{\fill} % separation between the subfigures
    \centering
    \begin{subfigure}{\textsubfigure}
        \includegraphics[width=\linewidth]{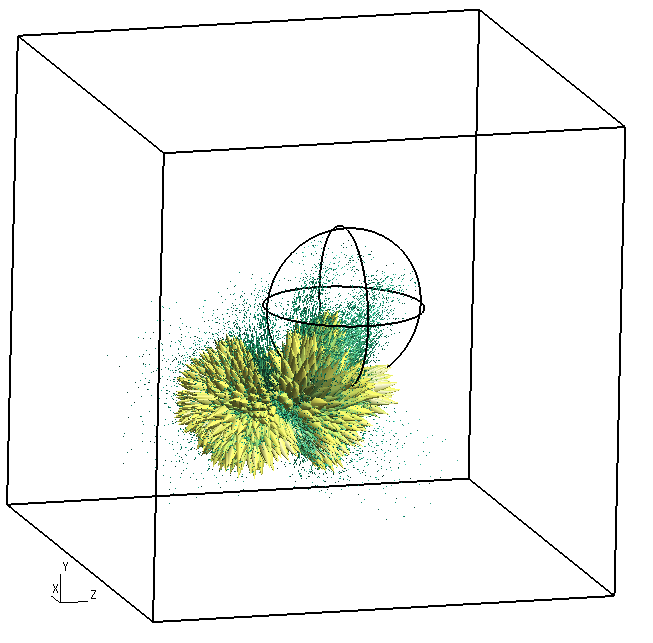}
        \captionsetup{justification=centering}
        \caption{$4\w_0$} \label{fig:8d}
    \end{subfigure}
    \captionsetup{justification=centering}
    \caption{Harmonic components of the real part of the total Poynting vector field $\Field[l]{S}$ for l = $1,2,3,4$} \label{fig:8}
\end{figure}

In order to illustrate our discussion, we present the following numerical results for the case of an oscillating charged particle close to a sphere (albeit this method can be applied for more complicated geometries). For this example it is considered that the particle oscillates at a frequency $\w_0 = \frac{2\pi c}{\lambda_0}$ with $\lambda_0 = 900$ nm. The amplitude of oscillation is $2a$ with $a = \frac{\lambda_0}{7}$. The radius of the sphere is $\frac{a}{2}$ and its center is separated from the $z$-axis by a distance of $a$, in addition the permittivity of the sphere is 9+$i$. Finally, all geometries and conformal meshes have been obtained using the  Gmsh software~\cite{gmsh} and all the finite element formulations in this article are implemented thanks to the flexibility of the finite element software GetDP~\cite{Geuzaine_getDP} The incident field was hard-coded as well in GetDP). The mesh size (see Fig.~\ref{fig:3}) is set to $\lambda_0/20$, which is very fine at $\omega_0$ and resonably coarse at $\omega_4=4\omega_0\leftrightarrow \lambda_4=\lambda_0/4$. The 3D scattering problem uses high order Webb hierarchical edge elements \cite{geuzaine1999convergence,webb1993hierarchal,jin02FEM-electromag} with 20 unknowns per tetrahedron (2 unknowns per edge, 2 unknowns per face). The direct problem described in \Eq{Eq:Driffracted_field_eq} is solved using the direct solver MUMPS \cite{mumps-userguide} interfaced in GetDP.   Next, we present our results as follows:
\begin{itemize}
\item  Figures \ref{fig:4} and \ref{fig:5} show the $\Field[l]{E}^0$ and $\Field[l]{B}^0$ fields respectively, for $l = 1,2,3,4$. The computation of these fields has been made by means of equations (\ref{Eq:E_harm_final}) and (\ref{Eq:B_harm_final}). In order to compute these integrals numerically, the following change of variables is used: $z^\prime = a \sin(\theta)$, this allows to eliminate the term $\frac{1}{\sqrt{a^2-z^{\prime 2}}}$ that comes from the definition of $w(z')$ (see \ref{Eq:w_equation}). Once the change of variable is made the formulae are coded into GetDP using a simple trapezoidal rule with 300 integration points. The only drawback we have encountered is that $\Field[l]{E}^0$ might be singular between $(0,0,-a)$ and $(0,0,+a)$. This is a problem that needs to be tackled in the future. The illustration of their Poynting vector $\Field[l]{S}^0 = \frac{1}{2 \mu_0}\Field[l]{E}^0 \times \overline{\Field[l]{B}}^0$ is shown in figure \ref{fig:9}.

\item Figure \ref{fig:6} shows the harmonic components of diffracted field $\Field[l]{E}^1$ for $l=1,2,3,4$. Notice that this field was obtained as a numerical solution of \ref{Eq:Driffracted_field_eq} by using FEM. Moreover, due to the fact that the support of source term on the right hand side of \ref{Eq:Driffracted_field_eq} is within the sphere, the possible singularity of $\Field[l]{E}^0$ does not affect our numerical results. PML's were used to truncate the surrounding free space.

\item Figure \ref{fig:7} shows the total electric field ($\Field[l]{E} = \Field[l]{E}^1 + \Field[]{E}^0 $) and its interaction with the sphere, whereas the total Poynting vector has been computed as $\Field[l]{S} = \frac{1}{2 \mu_0}\Field[l]{E} \times \overline{\Field[l]{B}}$  and can be seen in Figure \ref{fig:8} for $l= 1,2,3,4$. It is important to see in the case of the Poynting vector how this one is kind of \emph{pulled} by the sphere. This is due to the passivity of the material (remember that the permittivity of the sphere is 9+$i$).       
\end{itemize}

\begin{figure}[htbp]
\centering
\begin{subfigure}{\textsubfigure}
\includegraphics[width=\linewidth]{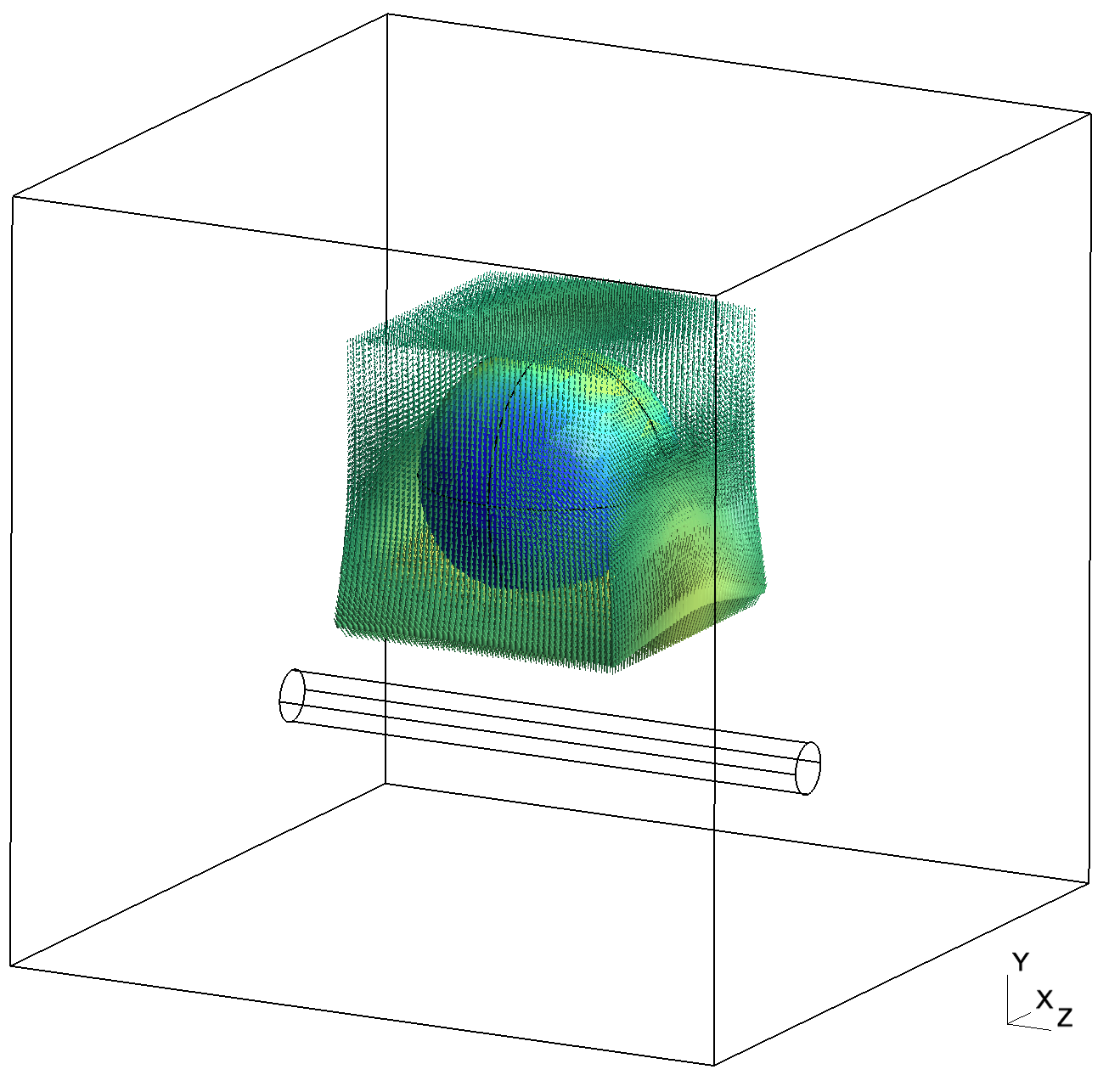}
\captionsetup{justification=centering}
\caption{Integration box that shows the} \label{fig:10a}
\end{subfigure}
\hspace*{\fill} % separation between the subfigures
\centering
\begin{subfigure}{\textsubfigure}
\includegraphics[width=\linewidth]{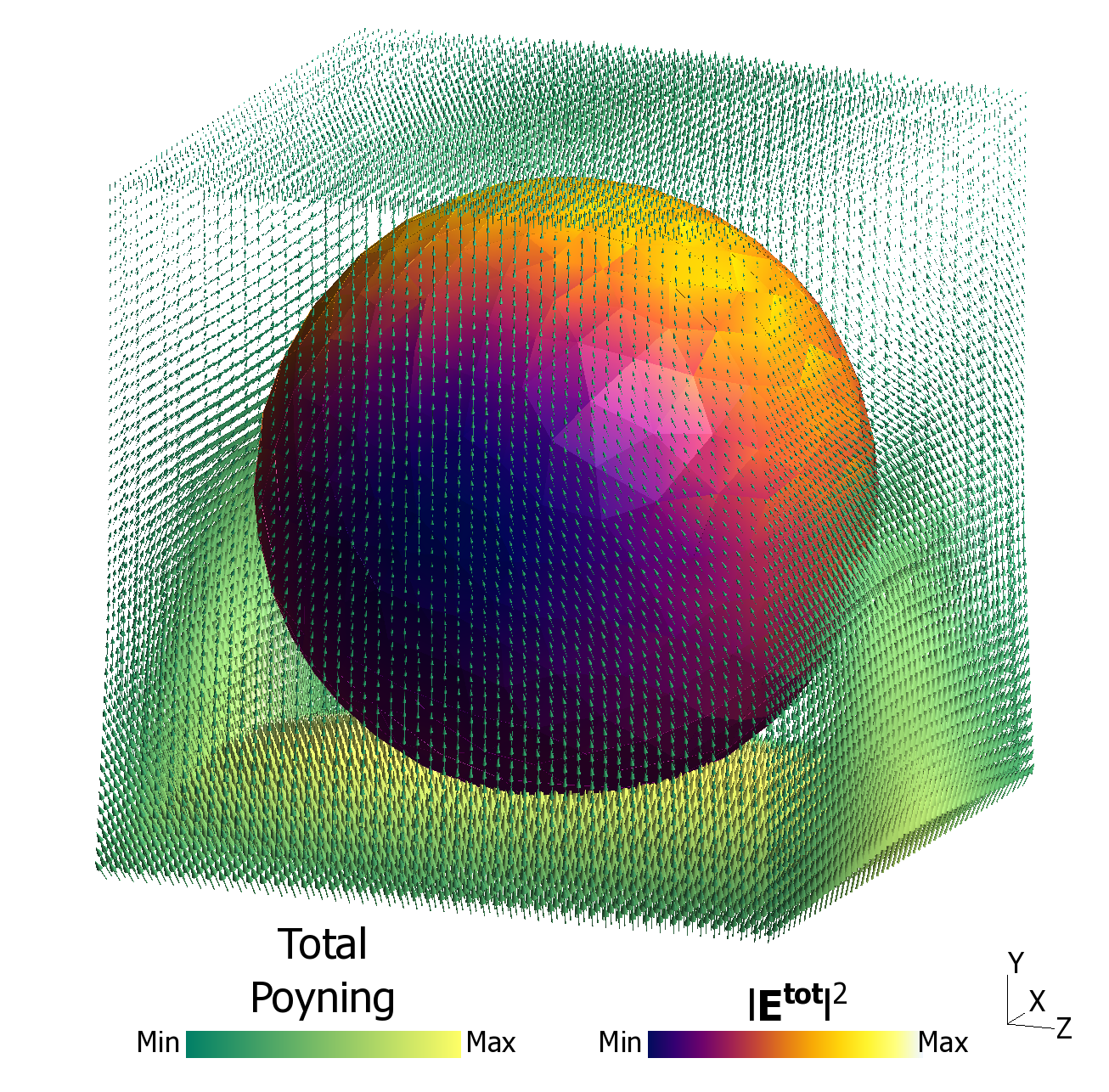}
\captionsetup{justification=centering}
\caption{Zoom enclosing the sphere} \label{fig:10b}
\end{subfigure}
\captionsetup{justification=centering}
\caption{Pill-box surface enclosing the sphere. The tube below the sphere in Figure \ref{fig:10a} is just there to refine the mesh around the trajectory of the particle (see figure \ref{fig:3b}) and to represent the total field.} \label{fig:10}
\end{figure}

Finally, all our results have been corroborated by considering an energy balance that measures the total energy flux that crosses a pill-box that surrounds the sphere. This is shown in Figure \ref{fig:10} .
%----------------------Figure-----------%
%----------------------Figure-----------%
% %----------------------Figure-----------%

% %----------------------Figure-----------%

% %----------------------Figure-----------%
% %----------------------Figure-----------%

% %----------------------Figure-----------%

%----------------------Figure-----------%

%%%%%%%%%%%%%%%%%%%%%%%%%%%%%%%%%%%%%%%%%%%%%%%%%%%%%%%%%%%%%%%%%%%%%%%%%%%%%%%%%%%%%%%%%%%%%%%%%%%%%%%%%%%%%%%%%%%%%
\section{Conclusion}\label{Sec:Osc_Conclusions}
In this \textst{chapter} work the problem of an oscillating particle near to a nano-sphere has been studied by using the diffracted field formalism. The first problem to tackle was the description of the electromagnetic field generated by an oscillating particle \emph{in vacuo}. Albeit this problem can be solved in principle by the Liénard-Wiechert fields, the resulting equations are not useful in practice. In order to overcome this difficulty we propose a harmonic decomposition of the sources $\rho$ and $\Field[]{\boldsymbol{\jmath}}$ which will allow us to find the harmonic representation of the electromagnetic fields. Next, we make an analysis of the radiated fields for the relativistic and non relativistic cases. Finally, by using the finite Elements Method, we solve the full problem that involves the interaction between the oscillating charge and the dispersive nano sphere. The solutions obtained have been validated by an energy balance.
%%%%%%%%%%%%%%%%%%%%%%%%%%%%%%%%%%%%%%%%%%%%%%%%%%%%%%%%%%%%%%%%%%%%%%%%%%%%%%%%%%%%%%%%%%%%%%%%%%%%%%%%%%%%%%%%%%%%%

% APENDICES
\begin{appendix}
\section{Annexe 0}\label{An:Fourier-T}

In this appendix we set the Fourier transform convention (in the classical sense) for this work as: 
\begin{align}
	\label{Eq:FT_Ap}
	&\hat{f}(\w)  :=\frac{1}{2\pi} \INT{t} f(t) e^{+i \w t} dt, \\
	\label{Eq:IFT_Ap}
    & f(t)  := \INT{\w} \hat{f}(\w) e^{-i \w t} \,\mathrm{d}\w.
\end{align}

If we want to extend these definitions to a wider family of mathematical objects (\emph{e.g.} $\delta$’s, unit steps, ramps, sines, cosines), it is necessary to consider the definition of the Fourier transform in the sense of distributions as per \cite{howell2016principles,Osgood_stanford_2007,asmar2005partial}:
\begin{equation}
	\label{Eq:FT_Ap_G}
	\braket{\FTw{f(t)}(\w), \varphi(\w)}_{\w \in \mathbb{R}} := \braket{f(t), \hat{\varphi}(t)}_{t \in \mathbb{R}},  
\end{equation}
accordingly the inverse Fourier transform is defined as:
\begin{equation}
	\label{Eq:IFT_Ap_G}
	\braket{\FTIw{\hat{f}(\w)}(t), \hat{\varphi}(t) }_{t \in \mathbb{R}} := \braket{\hat{f}(\w), \varphi(\w)}_{\w \in \mathbb{R}},  
\end{equation}
where the brackets denote integration in the whole real line when dealing with regular distributions and $\phi(\w)$ is a gaussian test function as it is explained in \cite{howell2016principles}.

%%%%%%%%%%%%%%%%%%%%%%%%%%%%%%%%%%%%%%%%%%%%%%%%%%%%%%%%%%%%%%%%%%%%%%%%%%%%%%%%%%%%%%%%%%%%%%%%%%%%%%%%%%%%%%%%%%%%%
\section{Annexe 1}\label{An:Compute-cl}
In this Annexe we show to obtain the poly harmonic representation of the sources.

For this, the distribution $\varrho(z,t)$ has to be considered as a $\delta$-distribution of a function $f(t):= z-a\cos(\w_0 t)$. Using the expansion given in \cite{arfken2012mathematical, jackson_classical_1998}, we obtain:
\begin{equation}\label{Eq:varrho_DiracC}
	\varrho(z,t) = \delta(f(t)) = \sum_{l\in \mathbb{Z}} \frac{1}{\dot{f}(t_l)} \delta(t-t_l),
\end{equation}
where $t_l$ are the zeros of the function $f(t)$, \emph{i.e.}
\begin{equation}
	\cos(\w_0 t) = \frac{z}{a}.
\end{equation}
Given the fact that the absolute value of the cosine is bounded by 1, there are only two solutions within the range $[-a,a]$. Then, by considering the principal branch of the arc cosine function and denoting $t_0(z):= \frac{1}{\w_0} \mathrm{arccos} ( \frac{z}{a} )$, these zeros are
\begin{align}
	t_l^+(z) &= t_0(z) + 2 l\pi/\w_0,\\
	t_l^-(z) &= -t_0(z) + 2 l\pi/\w_0.
\end{align}
In addition, it turns out that for every $t_l$ ($t^+_l$ or $t^-_l$) $|\dot{f}(t)| = \w_0 \sqrt{a^2-z^2}$ and the $\delta$-distribution reads:
\begin{equation}\label{Eq:varrho_Wpm}
	\varrho(z,t) = \delta(f(t)) =  \frac{1}{\w_0 \sqrt{a^2-z^2}}(W_+(t)+W_-(t))\chi_{[-a,a]}(z),
\end{equation}
where $W_{\pm}(t,z) = \sum_{l\in \mathbb{Z}} \delta(t-t^{\pm}(z))$. These last two series of distributions (Dirac's combs) are expandable in a Fourier series as follows \cite{Osgood_stanford_2007}:
\begin{equation}
	W_{\pm}(t,z) = \sum_{l\in \mathbb{Z}} \delta \bigg(t \mp t_0(z)-\frac{2 \pi l}{\w_0} \bigg) = \sum_{m\in \mathbb{Z}} \Phi^{\pm}_m e^{i\frac{m \pi}{T_0}t}, 
\end{equation}	
with $T_0 = \frac{\pi}{\w_0}$. Making the change of variables $t= \tau \pm t_0(z)$ one gets: 
\begin{equation}\label{Eq:Sum_delta_tau}
	\sum_{l\in \mathbb{Z}} \delta (\tau -2lT_0 ) = \sum_{m\in \mathbb{Z}} \Phi^{\pm}_m e^{i\frac{m \pi}{T_0}(\tau \pm t_0(z))}. 
\end{equation}
Multiplying this equation by $e^{-i\frac{n \pi}{T_0}(\tau \pm t_0(z))}$, taking the integral from $-T_0$ to $T_0$ and performing the change of variables $\sigma = \tau-2lT_0$, equation (\ref{Eq:Sum_delta_tau}) can be rewritten as:
\begin{equation}
	\sum_{l\in \mathbb{Z}} \int_{-(2l-1)T_0}^{-(2l+1)T_0} \delta(\sigma) e^{-im\frac{\pi \sigma}{T_0}} e^{\mp im \frac{t_0(z)}{T_0} } d\sigma = 2 T_0 \Phi^{\pm}_m.
\end{equation}
or in a more illuminating way
\begin{equation}
	 e^{\mp im \frac{t_0(z)}{T_0} } \int_{-\infty}^{+\infty} \delta(\sigma) e^{-im\frac{\pi \sigma}{T_0}}  d\sigma = 2 T_0 \Phi^{\pm}_m,
\end{equation}
which implies
\begin{equation}
	\Phi^{\pm}_m = \frac{1}{2 T_0} e^{\mp im \frac{t_0(z)}{T_0} } = \frac{\w_0}{2 \pi} e^{\mp im \w_0 t_0(z)}.
\end{equation}
Then
\begin{equation}
	W_{\pm}(t,z)  = \frac{\w_0}{2 \pi} \sum_{m\in \mathbb{Z}} e^{\mp im \w_0 t_0(z)} e^{im \w_0 t}, 
\end{equation}
and naturally (recovering the $l$ index)
\begin{align}
	W_{+}(t,z)+W_{-}(t,z) &= \frac{\w_0}{2 \pi} \sum_{l\in \mathbb{Z}}\bigg[e^{il \w_0 t_0(z)} + e^{-il \w_0 t_0(z)}\bigg]  e^{im \w_0 t}, \\
	& = \frac{\w_0}{\pi} \sum_{l\in \mathbb{Z}} \cos(l \w_0 t_0(z)) e^{il \w_0 t}.
\end{align}

Remembering the definition of $t_0$ one gets:
\begin{equation}
	\cos(l \w_0 t_0)  = \cos\bigg(l \arccos \bigg(\frac{z}{a} \bigg) \bigg) = T_l\bigg(\frac{z}{a} \bigg),  
\end{equation}
where $T_l$ are the Chebyshev polynomials of first kind. Therefore
\begin{equation}
	W_{+}(t,z)+W_{-}(t,z) = \sum_{l \in \Z } T_l\bigg(\frac{z}{a}\bigg) e^{+il\w_0 t} 
\end{equation}
Plugging this equation into Eq.~(\ref{Eq:varrho_Wpm}) and defining the function 
\begin{equation}
	\label{Eq:w_equation}
	w(z):= \frac{1}{\pi \sqrt{a^2-z^2}} \chi_{[-a,a]}(z),
\end{equation}
the function $\varrho$ reads
\begin{equation}\label{Eq:varrho_exp}
	\varrho(z,t) = \sum_{l\Z} w(z) T_l\bigg(\frac{z}{a}\bigg)e^{+il\w_0 t}.
\end{equation}
As a consequence of this, the function $\boldsymbol{\jmath}$ is given by
\begin{equation}\label{Eq:j_cos}
	\boldsymbol{\jmath}(z,t) =    -a\w_0 \sin(\w_0 t)\sum_{l\Z} w(z) T_l\bigg(\frac{z}{a}\bigg)e^{+il\w_0 t}.
\end{equation}
In short, $\varrho$ and $\boldsymbol{\jmath}$ are in the following form
\begin{equation}
	\varrho(z,t) = \sum_{l \in \Z }\varrho_l^F(z,t), \quad \boldsymbol{\jmath}(z,t) = \sum_{l \in \Z } \boldsymbol{\jmath}_l^F(z,t),  
\end{equation}
with 
\begin{equation}
	\label{Eq:rho_F_A}
	\varrho_l^F(z,t) = w(z) T_l\bigg(\frac{z}{a}\bigg)e^{+il\w_0 t} ,
\end{equation}
\begin{equation}
	\label{Eq:j_F_A}
	\boldsymbol{\jmath}_l^F(z,t) = -a\w_0 \sin(\w_0 t) w(z) T_l\bigg(\frac{z}{a}\bigg)e^{+il\w_0 t}.
\end{equation}
On the other hand, the Fourier transform $\varrho(z,t):= \delta(z-a\cos(\w_0t))$ can be derived by considering the convention:
\begin{eqnarray}
 \hat{\varrho}(k,t)& = &\mathcal{F}_{z \rightarrow k} \{ \delta(z-a\cos(\w_0t)) \}  \\
 							&=& \frac{1}{2 \pi}\int_{z \in \mathbb{R}} \delta(z-a\cos(\w_0t)) e^{+izk}dz \, .   
\end{eqnarray}
 After a suitable change of variable ($\sigma = z-a\cos(\w_0t)$) we obtain that:  
\begin{equation}\label{Eq:Final_varrho}
 \hat{\varrho}(k,t) = \frac{1}{2 \pi} e^{+ia\cos(\w_0 t)k} = \frac{1}{2 \pi} \sum_{l \in \mathbb{Z}} i^l \mathrm{J}_l(ka)  e^{+i l \w_0 t},
\end{equation}
where the last equality comes from the generating function of the Bessel's functions \cite{gray1952treatise, asmar2002applied, asmar2005partial}.
Taking the Fourier transform of Eq.~(\ref{Eq:varrho_exp}) and keeping in mind that these two Fourier series are equal term by term we arrive to this beautiful and unexpected expression: 
\begin{equation}\label{Eq:FT_zk_Bessel}
\mathcal{F}_{z \rightarrow k}\bigg\{ w(z) T_l\bigg(\frac{z}{a}\bigg) \bigg\} = \frac{i^l}{2\pi}\mathrm{ J}_l(ka)  \, ,  
\end{equation}
which will be used later.
%%%%%%%%%%%%%%%%%%%%%%%%%%%%%%%%%%%%%%%%%%%%%%%%%%%%%%%%%%%%%%%%%%%%%%%%%%%%%%%%%%%%%%%%%%%%%%%%%%%%%%%%%%%%%%%%%%%%%ù
% \section{Annexe 2}\label{An:ComputeXXX}
% \end{appendix}
\section{Annexe 2}\label{An:LW-deduction}

This Annexe is devoted to the deduction of the \emph{Liénard-Wiechert} fields.

In order to fix ideas, we are going to deal just with $\Field[int]{B}$ in \Eq{Eq:B_int_delta_int}. At first sight it would be tempting to just evaluate everything at $\Field[]{x}^{\prime} = \Field[]{u}$. However, it is important to remember that $\Field[]{u}$, and $\Field[]{n}$ are functions of the retarded time $t_r$ which is defined as:
\begin{equation}\label{Eq:t_r}
	t_r = t-\frac{|\Field[]{x}-\Field[]{u}(t_r)|}{c} \, .
\end{equation}
 A change of strategy is then necessary, instead of asking in \Eq{Eq:B_int_delta_int} which $\Field[]{x}^{\prime}$ makes $\Field[]{x}^{\prime}-\Field[]{u} = 0$ for each $t_r$, we ask which $t_r$ satisfies the transcendental equation (\ref{Eq:t_r}) for each $t$, the present time, and the given trajectory $\Field[]{u}$ \cite{heald2012classical}. Mathematically, the space integral in \ref{Eq:B_int_delta_int} is equivalent to:
\begin{equation}\label{Eq:B_int_tr}
	\Field[int]{B} =  \frac{c \mu_0}{4 \pi}\int_{\mathbb{R}} \delta\bigg(t_r - t +\frac{|\Field[]{x}-\Field[]{u}(t_r)|}{c} \bigg) \frac{\boldsymbol{\beta} \times \Field[]{n}}{R^2}   d t_r. 
\end{equation}
The next step to evaluate this integral, is to perform a change of variable with respect to the argument of the delta distribution:
\begin{equation}
	\tau = t_r - t +\frac{|\Field[]{x}-\Field[]{u}(t_r)|}{c}.
\end{equation}
Then, the differential $d\tau$ is given by:
\begin{equation}
	d \tau =d t_r\bigg[1 + \frac{1}{c}\frac{d}{dt_r}|\Field[]{x}-\Field[]{u}(t_r)|\bigg].
\end{equation}
The derivative of $|\Field[]{x}-\Field[]{u}(t_r)|$ with respect to the retarded time, can be easily computed by remembering that: 
\begin{equation}
\frac{d}{dt_r}|\Field[]{x}-\Field[]{u}(t_r)|^2 = \frac{d}{dt_r}(\Field[]{x}-\Field[]{u}(t_r))\cdot (\Field[]{x}-\Field[]{u}(t_r)),
\end{equation} 
and taking the derivative from both sides we have:
\begin{equation}
2|\Field[]{x}-\Field[]{u}(t_r)|\frac{d}{dt_r}|\Field[]{x}-\Field[]{u}(t_r)| = -2\Field[]{v}(t_r)\cdot (\Field[]{x}-\Field[]{u}(t_r)),
\end{equation} 
which after some elementary manipulations gives:
\begin{equation}
\frac{d}{dt_r}|\Field[]{x}-\Field[]{u}(t_r)| = -\Field[]{v}(t_r)\cdot \Field[]{n}(t_r),
\end{equation} 
and by defining $K :=  1 - \boldsymbol{\beta}(t_r)\cdot \Field[]{n}(t_r)$ we finally obtain:
\begin{equation}
	d \tau =d t_r\big[1 - \boldsymbol{\beta}(t_r)\cdot \Field[]{n}(t_r)\big] = dt_r K.
\end{equation}
Thus, the intermediate magnetic induction field is:
\begin{equation}\label{Eq:B_int_no_int}
	\Field[int]{B} =  \frac{c \mu_0}{4 \pi}\int_{\mathbb{R}} \delta (\tau) \frac{\boldsymbol{\beta} \times \Field[]{n}}{K R^2}   d \tau = \frac{c \mu_0}{4 \pi} \frac{\boldsymbol{\beta} \times \Field[]{n}}{K R^2} \bigg|_{\tau = 0},
\end{equation}
where due to the fact that $\tau = 0$ we have the retarded time defined implicitly as in \Eq{Eq:B_int_tr}. \emph{Mutatis mutandis}, this procedure can be repeated for the other fields in equations (\ref{Eq:B_rad_delta_int}-\ref{Eq:E_rad_delta_int}). Then, the total electric and magnetic induction fields read: 
\begin{equation}\label{Eq:B_deriv}
	\Field[]{B} = \frac{qc \mu_0}{4 \pi} \bigg[ \bigg(\frac{\boldsymbol{\beta} \times \Field[]{n}}{K R^2}\bigg)\bigg|_{t_r} + \partial_t  \bigg(\frac{ \boldsymbol{\beta} \times \Field[]{n}}{c K R}\bigg)\bigg|_{t_r} \bigg],
\end{equation}
\begin{align}\label{Eq:E_deriv}
    \nonumber
	 \Field[]{E} =& \frac{q}{4 \pi \eps_0} \bigg[\bigg( \frac{(\boldsymbol{\beta} \times \Field[]{n})}{K R^2}\times \Field[]{n}\bigg)\bigg|_{t_r} + \partial_t  \bigg(\frac{(\boldsymbol{\beta} \times \Field[]{n})}{c K R}\times \Field[]{n} \bigg)\bigg|_{t_r} \bigg]\\
    &+\frac{q}{4 \pi \eps_0}\bigg(\Field[]{n}\frac{1+\Field[]{n}\cdot\boldsymbol{\beta}}{K R^2} \bigg)\bigg|_{t_r}.
\end{align}
Now, it is necessary to compute the derivatives with respect to the present time $t$. Assuming the convention that the doted quantities denote partial derivation with respect to time $t$, the following identities will be useful:
\begin{align}
	\Field[]{\dot{R}} &= \partial_t(\Field[]{x}-\Field[]{u}(tr)) = -\PD[]{t_r}{t}\Field[]{v}(t_r) = -c\PD[]{t_r}{t}\boldsymbol{\beta}(t_r) \\
    \label{Eq:R_dot_1}
    \dt{R} &= \partial_t(\sqrt[]{\Field[]{R}\cdot \Field[]{R}}) = \frac{1}{R} \Field[]{R} \cdot \Field[]{\dt{R}} = -c(\Field[]{n}\cdot \boldsymbol{\beta})\PD[]{t_r}{t}\\
    \label{Eq:R_dot_2}
    &= c\partial_t(t-t_r) = c\bigg(1-\PD[]{t_r}{t} \bigg)
\end{align} 
from equations \eqref{Eq:R_dot_1} and \eqref{Eq:R_dot_2} one can solve for:
\begin{equation}\label{Eq:dtr}
	\PD[]{t_r}{t} = \frac{1}{1-\Field[]{n}\cdot \boldsymbol{\beta}} = \frac{1}{K},
\end{equation}
and upon substitution of \Eq{Eq:dtr} into equations \eqref{Eq:R_dot_1} and \eqref{Eq:R_dot_2} one gets:
\begin{align}
	\Field[]{\dt{R}} &= \partial_t(\Field[]{x}-\Field[]{u}(t_r)) = -\PD[]{t_r}{t}\Field[]{v}(t_r) = -\frac{c}{K}\boldsymbol{\beta} \\
    \label{Eq:R_dot_1_K}
    \dot{R} &= \partial_t(\sqrt[]{\Field[]{R}\cdot \Field[]{R}}) = \frac{1}{R} \Field[]{R} \cdot \Field[]{\dt{R}} = -\frac{c}{K}(\Field[]{n}\cdot \boldsymbol{\beta})\\
    \label{Eq:R_dot_2_K}
    &= c\partial_t(t-t_r) = c\bigg(1-\frac{1}{K}\bigg)
\end{align} 
From the expression $R\Field[]{n}=\Field[]{R}$ we can derive with respect to $t$ and after some manipulations we obtain:
\begin{eqnarray}\label{Eq:n_dot}
	\Field[]{\dt{n}} &=&\frac{1}{R}\big[\Field[]{\dt{R}}- \dot{R}\Field[]{n} \big]  \nonumber \\
							&=& \frac{1}{R}\bigg[-\frac{c}{K}\boldsymbol{\beta} - c\bigg(1-\frac{1}{K}\bigg)\Field[]{n} \bigg] \nonumber \\
							&=& -\frac{c}{KR}\big[(K-1)\Field[]{n}+\boldsymbol{\beta} \big].
\end{eqnarray}
In addition, we have:
\begin{equation}
	\boldsymbol{\dt{\beta}} = \frac{\Field[]{a}}{Kc}
\end{equation}
and
\begin{align}	
    \nonumber
	\dt{K} =& -\partial_t(\boldsymbol{\beta}\cdot \Field[]{n})= -\boldsymbol{\dt{\beta}}\cdot\Field[]{n}-\boldsymbol{\beta}\cdot\Field[]{\dt{n}}\\ 
    \nonumber
	=& -\frac{\Field[]{a}\cdot \Field[]{n}}{Kc}+\frac{c}{KR}\big[(K-1)\boldsymbol{\beta}\cdot\Field[]{n}+\beta^2 \big]\\
    =& - \frac{\Field[]{a}\cdot \Field[]{n}}{Kc}+\frac{c}{KR}\big[\beta^2 - (K-1)^2 \big]
\end{align}
Using these identities, we will first deal with $\Field[]{B}$ in \Eq{Eq:B_deriv}:
\begin{equation}\label{Eq:B_M}
	\Field[]{B} = \frac{qc\mu_0}{4 \pi} \bigg[\frac{\Field[]{M}}{R} + \frac{\Field[]{\dt{M}}}{c} \bigg], 
\end{equation}
where we have defined:
\begin{equation}\label{Eq:M_def}
 \Field[]{M} := \frac{\boldsymbol{\beta}\times \Field[]{n}}{KR}.
\end{equation}
It is then necessary to calculate the derivative of $\Field[]{M}$:
\begin{equation}\label{Eq:M_dot}
 \Field[]{\dt{M}} = \frac{1}{KR}\big[\partial_t(\boldsymbol{\beta}\times \Field[]{n})-\Field[]{M}\partial_t(KR)\big].
\end{equation}
The first derivative in the right hand side of \Eq{Eq:M_dot} is: 
\begin{align}\label{Eq:M_dot_1}
	\nonumber
	\partial_t(\boldsymbol{\beta}\times \Field[]{n}) = & \boldsymbol{\dt{\beta}}\times \Field[]{n} + \boldsymbol{\beta}\times \Field[]{\dt{n}},\\
	\nonumber
    =& \frac{\Field[]{a}\times \Field[]{n}}{Kc} -\frac{c}{KR}\big[(K-1)\boldsymbol{\beta}\times\Field[]{n}+\boldsymbol{\beta}\times \boldsymbol{\beta} \big], \\
    =&  \frac{\Field[]{a}\times \Field[]{n}}{Kc} -c(K-1)\Field[]{M},
\end{align}
and the second one is given by:
\begin{align}\label{Eq:M_dot_2}
	\nonumber
	\partial_t(KR) = & \dt{K}R+\dt{R}K, \\
    \nonumber
	=&- \frac{\Field[]{a}\cdot \Field[]{n}}{Kc}R+\frac{c}{K}\big[\beta^2 - (K-1)^2 \big]+c(K-1),\\
    &- \frac{\Field[]{a}\cdot \Field[]{n}}{Kc}R+\frac{c}{K}\big[\beta^2 - 1]+c.
\end{align}
Plugging \Eq{Eq:M_dot_1} and \Eq{Eq:M_dot_2} into \Eq{Eq:M_dot} we obtain:
\begin{equation}\label{Eq:M_dot_explicit}
 \Field[]{\dt{M}} = \bigg[\frac{\Field[]{a}\times \Field[]{n}}{K^2Rc} -c\frac{\Field[]{M}}{R}+\Field[]{M}\frac{\Field[]{a}\cdot \Field[]{n}}{cK^2}+\frac{c}{K^2R}\big[1-\beta^2]\Field[]{M}\bigg],
\end{equation}
and substituting \Eq{Eq:M_dot_explicit} into \Eq{Eq:B_M} we obtain the Liénard-Wiechert magnetic induction field: 
\begin{equation}\label{Eq:B_LW}
	 \Field[]{B} = \frac{qc\mu_0}{4 \pi} \bigg[\frac{\Field[]{a}\times \Field[]{n}}{K^2Rc^2}+\frac{\Field[]{a}\cdot \Field[]{n}(\boldsymbol{\beta}\times \Field[]{n})}{c^2K^3R} +\frac{(1-\beta^2)(\boldsymbol{\beta}\times \Field[]{n})}{K^3R^2} \bigg]_{t_r}.
\end{equation}

For the case of the electric field, we start again by expressing \Eq{Eq:E_deriv} in terms of \Eq{Eq:M_def}: 
\begin{equation}\label{Eq:E_M} 
	\Field[]{E} = \frac{q}{4 \pi \eps_0}\bigg[ \bigg( \frac{\Field[]{M}}{R} + \frac{\Field[]{\dt{M}}}{c} \bigg)\times \Field[]{n} +  \frac{\Field[]{M}}{c} \times \Field[]{\dt{n}}+\frac{2-K}{K R^2}\Field[]{n}\bigg]_{t_r}.
\end{equation}
Then, we consider the second term in the right hand side of \ref{Eq:E_M}:
\begin{align}
	\nonumber
	\frac{\Field[]{M}}{c} \times \Field[]{\dt{n}} &= \frac{-\big[(K-1)(\boldsymbol{\beta}\times \Field[]{n})\times \Field[]{n} + (\boldsymbol{\beta}\times \Field[]{n})\times \boldsymbol{\beta} \big]}{K^2R^2},\\
	\nonumber
    &= \frac{-\big[(K-1)(\Field[]{n}(\boldsymbol{\beta}\cdot \Field[]{n})-\boldsymbol{\beta}) - (\boldsymbol{\beta}(\boldsymbol{\beta}\cdot \Field[]{n})-\beta^2\Field[]{n}) \big]}{K^2R^2},
\end{align}
and using $1-K = \boldsymbol{\beta}\cdot \Field[]{n}$
\begin{align}\label{Eq:E_M_1}
	\nonumber
	\frac{\Field[]{M}}{c} \times \Field[]{\dt{n}} 
    &= \frac{-\big[(K-1)(\Field[]{n}(1-K)-\boldsymbol{\beta}) - (\boldsymbol{\beta}(1-K)-\beta^2\Field[]{n}) \big]}{K^2R^2},\\
    \nonumber
    &=\frac{-\Field[]{n}}{K^2R^2}(\beta^2-(K-1)^2),\\
    &= -\frac{2-K}{KR^2}\Field[]{n} + \frac{\Field[]{n}}{K^2R^2}(1-\beta^2).
\end{align}
Notice that the first term in \Eq{Eq:E_M_1} is going to cancel with the third term in \Eq{Eq:E_M}. Then, we can focus our attention in the first term of \Eq{Eq:E_M}:
\begin{eqnarray}\label{Eq:E_M_First}
\bigg( \frac{\Field[]{M}}{R} + \frac{\Field[]{\dt{M}}}{c} \bigg)\times \Field[]{n}  
& =&  \nonumber \\
 \bigg[\frac{\Field[]{a}\times \Field[]{n}}{K^2Rc^2}+\frac{\Field[]{a}\cdot \Field[]{n}(\boldsymbol{\beta}\times \Field[]{n})}{c^2K^3R} +\frac{(1-\beta^2)(\boldsymbol{\beta}\times \Field[]{n})}{K^3R^2} \bigg]\times \Field[]{n} 
\end{eqnarray}
Let us work with the first two terms of \Eq{Eq:E_M_First}:
%\begin{widetext}
\begin{align}
\nonumber
\bigg[\frac{\Field[]{a}\times \Field[]{n}}{K^2Rc^2}+\frac{\Field[]{a}\cdot \Field[]{n}(\boldsymbol{\beta}\times \Field[]{n})}{c^2K^3R}\bigg]\times \Field[]{n} 
=& \frac{\big[K(\Field[]{a}\times \Field[]{n}) +\Field[]{a}\cdot \Field[]{n}(\boldsymbol{\beta}\times \Field[]{n}) \big]}{c^2K^3R}\times \Field[]{n},\\
\nonumber
=& \frac{\big[(1-\boldsymbol{\beta}\cdot \Field[]{n})(\Field[]{a}\times \Field[]{n}) +\Field[]{a}\cdot \Field[]{n}(\boldsymbol{\beta}\times \Field[]{n}) \big]}{c^2K^3R}\times \Field[]{n}.
\end{align}
%\end{widetext}
Now, by means of the vector identity $\Field[]{A}\cdot(\Field[]{B}\times\Field[]{C})\Field[]{D}=(\Field[]{A}\cdot \Field[]{D})(\Field[]{B}\times \Field[]{C}) + (\Field[]{B}\cdot \Field[]{D})(\Field[]{C}\times \Field[]{A})+ (\Field[]{C}\cdot \Field[]{D})(\Field[]{A}\times \Field[]{D})$ \cite{ruiz1995calculo} and letting $\Field[]{A} = \Field[]{a}$, $\Field[]{B} = \boldsymbol{\beta}$ and $\Field[]{C} = \Field[]{D} = \Field[]{n}$ we have:
\begin{align}\label{Eq:E_M_2}
\nonumber
\bigg[\frac{\Field[]{a}\times \Field[]{n}}{K^2Rc^2}+\frac{\Field[]{a}\cdot \Field[]{n}(\boldsymbol{\beta}\times \Field[]{n})}{c^2K^3R} \bigg]\times \Field[]{n} 
=& \frac{[\Field[]{a}\times\Field[]{n}-\boldsymbol{\beta}\times\Field[]{n}+\Field[]{a}\cdot(\boldsymbol{\beta}\times\Field[]{n})\Field[]{n}]}{c^2K^3R}\times \Field[]{n},\\
=& \frac{\Field[]{a}\times(\Field[]{n}-\boldsymbol{\beta})\times \Field[]{n} }{c^2K^3R}.
\end{align}
On the other hand, the third term on \ref{Eq:E_M_First} can be seen as: 
\begin{align}\label{Eq:E_M_3}
	\nonumber
	\frac{(1-\beta^2)(\boldsymbol{\beta}\times \Field[]{n})}{K^3R^2}\times \Field[]{n} 
   &= \frac{(1-\beta^2)( \Field[]{n}(1-K)-\boldsymbol{\beta})}{K^3R^2}, \\
   &= \frac{(1-\beta^2)( \Field[]{n}-\boldsymbol{\beta})}{K^3R^2} - \frac{(1-\beta^2) \Field[]{n}}{K^2R^2}
\end{align}
Putting  \Eq{Eq:E_M_2}, and \Eq{Eq:E_M_3} into \Eq{Eq:E_M_First} and then plugging this result and \ref{Eq:E_M_1} into \ref{Eq:E_M}, we finally obtain the electric Liénard-Wiechert field:
\begin{equation}\label{Eq:E_LW}
 \Field[]{E} = \frac{q}{4 \pi \eps_0}\bigg[\frac{(1-\beta^2)( \Field[]{n}-\boldsymbol{\beta})}{K^3R^2} + \frac{\Field[]{a}\times(\Field[]{n}-\boldsymbol{\beta})\times \Field[]{n} }{c^2K^3R} \bigg]_{t_r}.
\end{equation}
 We can see that, along expression \Eq{Eq:B_LW}, these are the same results as obtained by Griffiths \cite{griffiths1999introduction} and Heald and Marion \cite{heald2012classical}.
%%%%%%%%%%%%%%%%%%%%%%%%%%%%%%%%%%%%%%%%%%%%%%%%%%%%%%%%%%%%%%%%%%%%%%%%%%%%%%%%%%%%%%%%%%%%%%%%%%%%%%%%%%%%%%%%%%%%%
\section{Annexe 4}\label{An:Bessel identities}
Here the goal is to express $P_l$ with the minimum of \emph{ad hoc} special functions. For this reason, we are going to derive the integrals used in Section \ref{Sec:Far_Field_Power}. As we know, the Bessel functions of first 
kind are non diverging solutions at the origin of the differential equation \cite{simmons2016differential, gray1952treatise, asmar2005partial}:
\begin{equation}\label{Eq:Bessel_eq}
	r^2 \frac{d^2}{dr^2} J_{l}(r) + r\frac{d}{dr} J_{l}(r) + J_{l}(r)\bigg(r^2 -l^2\bigg)= 0 
\end{equation}
which satisfy the identities \cite{asmar2005partial,arfken2012mathematical, abramowitz1964handbook}:
\begin{align}
\label{Eq:Identity_1}
 &J_{l-1}(r) + J_{l+1}(r)  = \frac{2l}{r}J_{l}(r),\\
\label{Eq:Identity_2} 
 &J_{l-1}(r) - J_{l+1}(r)  = 2\frac{d}{dr}J_{l}(r),%\\
%\label{Eq:Identity_3}
% &\frac{d}{dr} J_{l}(r) = J_{l-1}(r) - \frac{l}{r}J_{l}(r),\\
% \label{Eq:Identity_4}
% &\frac{d}{dr} J_{l}(r) = J_{l+1}(r) + \frac{l}{r}J_{l}(r),
\end{align}
Equipped with these tools, our first goal is to compute the integral
\begin{equation}
	\label{Eq:I_ls}
	\mathcal{I}_l^{S}(A) = \int_{0}^{A} \frac{J_l^2}{r^2}dr. 
\end{equation}
In order to do that, consider \eq{Eq:Bessel_eq} and divide it over $r$, after some elementary manipulations this equation reads:
\begin{equation}\label{Eq:J_l}
	\frac{d}{dr} \bigg[r \frac{d J_{l}}{dr}\bigg] + J_{l}\bigg(r -\frac{l^2}{r} \bigg)= 0,	
\end{equation}
and analogously for $J_{k}$
\begin{equation}\label{Eq:J_k}
	\frac{d}{dr} \bigg[r \frac{d J_{k}}{dr}\bigg] + J_{k}\bigg(r -\frac{k^2}{r} \bigg)= 0.	
\end{equation}
Multiplying \eq{Eq:J_l} by $ J_k$, \eq{Eq:J_k} by $J_l$ and taking its difference one gets:
\begin{equation}
	J_k \frac{d}{dr} \bigg[r \frac{d J_{l}}{dr}\bigg]-J_l \frac{d}{dr} \bigg[r \frac{d J_{k}}{dr}\bigg] + (k^2-l^2)\frac{1}{r}J_k J_l = 0.
\end{equation}
Next, we take the integral from 0 to $A$ to obtain:
\begin{equation}
	J_k r \frac{dJ_{l}}{dr}\bigg|_{0}^{A} - J_l r \frac{d J_{k}}{dr}\bigg|_{0}^{A} = (l^2-k^2)\int_{0}^{A} \frac{1}{r}J_k J_l dr, 
\end{equation}
and then: 
\begin{equation}\label{Eq:J_l_J_k_o_r}
	\int_{0}^{A} \frac{1}{r}J_k J_l dr = \frac{A}{(l^2-k^2)} \bigg[  J_k\frac{dJ_{l}}{dr}- J_l\frac{dJ_{k}}{dr} \bigg]_{A}. 
\end{equation}
Now, by means of the identity in (\ref{Eq:Identity_1}), it is easy to see \eq{Eq:I_ls} as: 
\begin{equation}\label{Eq:J_o_r_2}
 \int_{0}^{A} \frac{1}{r^2} J_l^2 dr = \frac{1}{2l} \bigg[  \int_{0}^{A} \frac{1}{r}J_{l-1} J_l dr +  \int_{0}^{A} \frac{1}{r}J_{l+1} J_l dr \bigg]. 
\end{equation}
These last two integrals in the right hand side of \eq{Eq:J_o_r_2} can be easily obtained via (\ref{Eq:J_l_J_k_o_r}) and finally
\begin{equation}\label{Eq:Final_J_o_r_2}
	\boxed{\mathcal{I}_l^S(A) = \frac{A}{2l} \bigg[\frac{J_{l}J^{\prime}_{l+1}- J_{l+1}J^{\prime}_{l} }{1+2l} + \frac{J_{l}J^{\prime}_{l-1}- J_{l-1}J^{\prime}_{l} }{1-2l}\bigg]_{A}.}
\end{equation}
%%%%%%%%%%%%%%%%%%%%%%%%%%%%%%
The second goal, is to compute the integral
\begin{equation}
	\label{Eq:I_lR}
	\mathcal{I}_l^{R}(A) = \int_{0}^{A} J_l^2  dr. 
\end{equation}
Despite its harmless appearance, this integral is in general not easy to integrate and there is no analytical expression in the consulted references \cite{simmons2016differential, gray1952treatise,asmar2005partial,arfken2012mathematical, abramowitz1964handbook}. Thus, we propose here a semi analytical approach that give us a way to obtain this integral in a recursive manner. 

Starting by assuming $l > 1$, taking the product of equations (\ref{Eq:Identity_1}) and (\ref{Eq:Identity_2}) and integrating this resulting equation from 0 to $A$ one gets:
\begin{equation}
	 \int_{0}^{A} J_{l-1}^2 dr - \int_{0}^{A} J_{l+1}^2 dr  = 4l \int_{0}^{A} \frac{1}{r} J_{l} \frac{d J_{l}}{dr} dr,
\end{equation}
and after performing integration by parts in the right hand side of this equation one arrives to the expression:
\begin{equation}\label{Eq:recurrence}
	\boxed{\mathcal{I}_{l+1}^{R}(A) = \mathcal{I}_{l-1}^{R}(A)   - \frac{2l}{A} J_l^2(A) -2l \mathcal{I}_l^{S}(A).}
\end{equation}
Notice that the last integral in the right hand side of \eq{Eq:recurrence} is given by (\ref{Eq:Final_J_o_r_2}). This  establishes a two step recurrence relation between 
the integrals $I_{l+1}^{R}(A)$ and $I_{l-1}^{R}(A)$ as defined in \eq{Eq:I_lR}. Next, we are going to get an expression for $I_{1}^{R}(A)$ by considering \eq{Eq:J_l} with $l = 0$
\begin{equation}
	\frac{d}{dr} \bigg[r \frac{d}{dr}J_{0}\bigg] + J_{0}r = 0.	
\end{equation}
 Remembering that $\frac{d}{dr}J_{0} = - J_{1}$ we get:
 \begin{equation}
	-\frac{d}{dr} \bigg[r J_{1}\bigg] + J_{0}r = 0,	
\end{equation}
and multiplying by $\frac{dJ_{0}}{dr} = - J_{1}$ 
 \begin{equation}
	J_1\frac{d}{dr} \bigg[r J_{1}\bigg] + \frac{dJ_{0}}{dr}J_{0}r = 0	
\end{equation}
which implies
\begin{equation}
	J_{1}^2 + \frac{1}{2}\frac{d}{dr}[J_{1}^2 + J_{0}^2]r = 0	
\end{equation}
and after performing integration by parts
\begin{equation}
	\int_{0}^{A} J_{1}^2 dr - \int_{0}^{A} J_{0}^2 dr + A[J_{1}^2 + J_{0}^2]_{A} = 0.
\end{equation}
This expression can be arranged in a more illuminating way as per:
\begin{equation}
	\boxed{\mathcal{I}_1^{R}(A) = \mathcal{I}_0^{R}(A) - A[J_{1}^2 - J_{0}^2]_{A}}.
\end{equation}
Therefore \eq{Eq:I_lR} depends at the end only of 
\begin{equation}\label{Eq:I_R_0}
	\boxed{\mathcal{I}_0^{R}(A) = \int_{0}^{A} J_{0}^2 dr,}
\end{equation}
which can be evaluated numerically in a very precise way by the Periodisation method described for instance in \cite{helluy1998integration}.
\end{appendix}

\bibliographystyle{ieeetr}
\bibliography{biblio}

\begin{thebibliography}{10}

\bibitem{harish2007antennas}
A.~Harish and M.~Sachidananda, {\em Antennas and wave propagation}.
\newblock Oxford University Press, USA, 2007.

\bibitem{huang2008antennas}
Y.~Huang and K.~Boyle, {\em Antennas: from theory to practice}.
\newblock John Wiley \& Sons, 2008.

\bibitem{sommerfeld1960partial}
A.~Sommerfeld, {\em Partial Differential Equations In Physics: Lectures On
  Theoretical Physics}.
\newblock No.~v. 6 in Pure and applied mathematics, 1, Sarat Book House, 1960.

\bibitem{Lassalle:17}
E.~Lassalle, A.~Devilez, N.~Bonod, T.~Durt, and B.~Stout, ``Lamb shift
  multipolar analysis,'' {\em J. Opt. Soc. Am. B}, vol.~34, pp.~1348--1355, Jul
  2017.

\bibitem{klimov_1996}
V.~V. Klimov, M.~Ducloy, and V.~S. Letokhov, ``Radiative frequency shift and
  linewidth of an atom dipole in the vicinity of a dielectric microsphere,''
  {\em Journal of Modern Optics}, vol.~43, no.~11, pp.~2251--2267, 1996.

\bibitem{jackson_classical_1998}
J.~D. Jackson, {\em Classical {Electrodynamics}}.
\newblock Wiley, Aug. 1998.

\bibitem{griffiths1999introduction}
D.~Griffiths, {\em Introduction to Electrodynamics}.
\newblock Prentice Hall, 1999.

\bibitem{novotny2012principles}
L.~Novotny and B.~Hecht, {\em Principles of nano-optics}.
\newblock Cambridge university press, 2012.

\bibitem{Raimond_PMC_2016}
J.-M. Raimond, {\em {M1} {Th\'eorie} {Classique} des {Champs}, {Notes} de
  {Cours}}.
\newblock (Universit\'e Pierre et Marie Curie, Laboratoire Kastler Brossel
  2016) [retrieved 8 Jun 2017],
  http://www.lkb.upmc.fr/cqed/wp-content/uploads/sites/14/2016/09/notes-de-cours.pdf,
  2016.

\bibitem{landau1975classical}
L.~Landau and E.~Lifshitz, {\em The Classical Theory of Fields}.
\newblock Course of theoretical physics, Butterworth-Heinemann, 1975.

\bibitem{panofsky2005classical}
W.~K. Panofsky and M.~Phillips, {\em Classical electricity and magnetism}.
\newblock Courier Corporation, 2005.

\bibitem{jefimenko1966electricity}
O.~Jefimenko, {\em Electricity and Magnetism: An Introduction to the Theory of
  Electric and Magnetic Fields}.
\newblock Series in physics, Appleton-Century-Crofts, 1966.

\bibitem{petit_outil_1991}
R.~Petit, {\em L'outil mathématique: distributions, convolution,
  transformations de {Fourier} et de {Laplace}, fonctions d'une variable
  complexe, fonctions eulériennes}.
\newblock Masson, 1991.

\bibitem{ruiz1995calculo}
C.~Ruiz, {\em C{\'a}lculo vectorial}.
\newblock Prentice Hall Hispanoamericana, S.A., 1995.

\bibitem{spohn2004dynamics}
H.~Spohn, {\em Dynamics of charged particles and their radiation field}.
\newblock Cambridge university press, 2004.

\bibitem{heald2012classical}
M.~A. Heald and J.~B. Marion, {\em Classical electromagnetic radiation}.
\newblock Courier Corporation, 2012.

\bibitem{jentschura2017advanced}
U.~D. Jentschura, {\em Advanced classical electrodynamics: green functions,
  regularizations, multipole decompositions}.
\newblock World Scientific Publishing Company, 2017.

\bibitem{stratton1941electromagnetic}
J.~Stratton, {\em Electromagnetic theory}.
\newblock International series in pure and applied physics, McGraw-Hill book
  company, inc., 1941.

\bibitem{kreyszig1989introductory}
E.~Kreyszig, {\em Introductory functional analysis with applications}, vol.~1.
\newblock wiley New York, 1989.

\bibitem{reddy2013introductory}
B.~D. Reddy, {\em Introductory functional analysis: with applications to
  boundary value problems and finite elements}, vol.~27.
\newblock Springer Science \& Business Media, 2013.

\bibitem{saichev1997distributions}
A.~I. Saichev and W.~A. Woyczynski, {\em Distributions in the Physical and
  Engineering Sciences. Volume I}.
\newblock Springer, 1997.

\bibitem{luo2003cerenkov}
C.~Luo, M.~Ibanescu, S.~G. Johnson, and J.~Joannopoulos, ``Cerenkov radiation
  in photonic crystals,'' {\em science}, vol.~299, no.~5605, pp.~368--371,
  2003.

\bibitem{lin2018controlling}
X.~Lin, S.~Easo, Y.~Shen, H.~Chen, B.~Zhang, J.~D. Joannopoulos,
  M.~Solja{\v{c}}i{\'c}, and I.~Kaminer, ``Controlling cherenkov angles with
  resonance transition radiation,'' {\em Nature Physics}, vol.~14, no.~8,
  pp.~816--821, 2018.

\bibitem{arfken2012mathematical}
G.~Arfken, H.~Weber, and F.~Harris, {\em Mathematical Methods for Physicists: A
  Comprehensive Guide}.
\newblock Elsevier, 2012.

\bibitem{Geuzaine_getDP}
C.~Geuzaine and D.~Patrik, {\em {GetDP} reference manual: the documentation for
  a {General} {Environment} for the {Treatment} of {Discrete} {Problems}}.
\newblock (Université de Liège 1997) [retrieved 9 Nov 2014],
  http://getdp.info, 2017.

\bibitem{abramowitz1964handbook}
M.~Abramowitz and I.~A. Stegun, {\em Handbook of mathematical functions: with
  formulas, graphs, and mathematical tables}, vol.~55.
\newblock Courier Corporation, 1964.

\bibitem{zolla_foundations_2005}
F.~Zolla, G.~Renversez, A.~Nicolet, B.~Kuhlmey, S.~Guenneau, D.~Felbacq,
  A.~Argyros, and S.~Leon-Saval, {\em Foundations of {Photonic} {Crystal}
  {Fibres}}.
\newblock Imperial College Press, Jan. 2005.
\newblock Google-Books-ID: iVZXwXDswv0C.

\bibitem{demesy2010all}
G.~Dem{\'e}sy, F.~Zolla, A.~Nicolet, and M.~Commandr{\'e}, ``All-purpose finite
  element formulation for arbitrarily shaped crossed-gratings embedded in a
  multilayered stack,'' {\em JOSA A}, vol.~27, no.~4, pp.~878--889, 2010.

\bibitem{gmsh}
C.~Geuzaine and J.~F. Remacle, ``Gmsh: a three-dimensional finite element mesh
  generator with built-in pre- and post-processing facilities,'' {\em
  International Journal for Numerical Methods in Engineering}, vol.~79, no.~11,
  pp.~1309--1331, 2009.

\bibitem{geuzaine1999convergence}
C.~Geuzaine, B.~Meys, P.~Dular, and W.~Legros, ``{ Convergence of high order
  curl-conforming finite elements [for {EM} field calculations]},'' {\em IEEE
  Transactions on Magnetics}, vol.~35, no.~3, pp.~1442--1445, 1999.

\bibitem{webb1993hierarchal}
J.~Webb and B.~Forgahani, ``{ Hierarchal scalar and vector tetrahedra},'' {\em
  IEEE Transactions on Magnetics}, vol.~29, no.~2, pp.~1495--1498, 1993.

\bibitem{jin02FEM-electromag}
J.~Jin, {\em The {F}inite {E}lement {M}ethod in {E}lectromagnetics}.
\newblock John Wiley \& Sons Inc., 3rd~ed., 2014.

\bibitem{mumps-userguide}
P.~Amestoy, I.~Duff, A.~Guermouche, J.~Koster, J.-Y. L'Excellent, and
  S.~Pralet, {\em MUltifrontal Massively Parallel Solver, (MUMPS 4.8.4), Users'
  guide}.
\newblock CERFACS, ENSEEIHT-IRIT, and INRIA, December 2008.
\newblock {h}ttp://mumps.enseeiht.fr and http://graal.ens-lyon.fr/MUMPS.

\bibitem{howell2016principles}
K.~B. Howell, {\em Principles of Fourier analysis}.
\newblock CRC Press, 2016.

\bibitem{Osgood_stanford_2007}
B.~Osgood, {\em Stanford {Engineering} {Everywhere} {\textbar} {EE}261 - {The}
  {Fourier} {Transform} and its {Applications}}.
\newblock (Stanford 2007) [retrieved 6 Jan 2017],
  https://see.stanford.edu/Course/EE261, 2007.

\bibitem{asmar2005partial}
N.~Asmar, {\em Partial Differential Equations with Fourier Series and Boundary
  Value Problems}.
\newblock Pearson Prentice Hall, 2005.

\bibitem{gray1952treatise}
A.~Gray, G.~Mathews, and T.~MacRobert, {\em A treatise on Bessel functions and
  their applications to physics}.
\newblock Macmillan and co., limited, 1952.

\bibitem{asmar2002applied}
N.~Asmar and G.~Jones, {\em Applied Complex Analysis with Partial Differential
  Equations}.
\newblock Prentice Hall, 2002.

\bibitem{simmons2016differential}
G.~F. Simmons, {\em Differential equations with applications and historical
  notes}.
\newblock CRC Press, 2016.

\bibitem{helluy1998integration}
P.~Helluy, S.~Maire, and P.~Ravel, ``Int{\'e}gration num{\'e}rique d'ordre
  {\'e}lev{\'e} de fonctions r{\'e}guli{\`e}res ou singuli{\`e}res sur un
  intervalle,'' {\em Comptes Rendus de l'Acad{\'e}mie des Sciences-Series
  I-Mathematics}, vol.~327, no.~9, pp.~843--848, 1998.

\end{thebibliography}

%\bibliographyfullrefs{biblio}

\end{document}